\documentclass{aa} %\documentclass{aa}

\usepackage{graphicx}
\usepackage{txfonts}
\usepackage[colorlinks=true, urlcolor=blue, linkcolor=blue, citecolor=blue]{hyperref}
\usepackage{academicons}
    \def\orcid#1{\kern .08em \href{https://orcid.org/#1}{\includegraphics[keepaspectratio,width=0.7em]{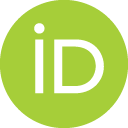}}}
\usepackage{amsmath,amssymb}
\usepackage{mathtools}
\usepackage{multirow}
\usepackage{pdflscape}
\usepackage{color}
\usepackage{textcomp}
\usepackage{gensymb}
\usepackage{placeins}

\begin{document} 

    \title{Multi-frequency analysis of the ALMA and VLA high resolution continuum observations of the substructured disc around CI~Tau}
    \subtitle{Preference for submillimetre-sized low-porosity amorphous carbon grains}
    
    \titlerunning{Multi-frequency analysis of CI~Tau's (sub-)millimetre continuum observations}
    
    \author{
        Francesco Zagaria\inst{1,2}\thanks{E-mail: \url{frzagaria@mpia.de}} \orcid{0000-0001-6417-7380}
        \and
        Stefano Facchini\inst{3} \orcid{0000-0003-4689-2684}
        \and
        Pietro Curone\inst{3,4,5} \orcid{0000-0003-2045-2154}
        \and
        Jonathan P. Williams\inst{6} \orcid{0000-0001-5058-695X} 
        \and
        Cathie J. Clarke\inst{2} \orcid{0000-0003-4288-0248} 
        \and
        Álvaro Ribas\inst{2} \orcid{0000-0003-3133-3580} 
        \and
        Marco Tazzari\inst{2,7} \orcid{0000-0003-3590-5814}
        \and
        Enrique Macías\inst{4} \orcid{0000-0003-1283-6262} 
        \and
        Richard A. Booth\inst{8,9} \orcid{0000-0002-0364-937X} 
        \and
        Giovanni P. Rosotti\inst{10,11,3} \orcid{0000-0003-4853-5736} 
        \and
        Leonardo Testi\inst{4,12} \orcid{0000-0003-1859-3070} 
    }
    \authorrunning{F. Zagaria et al.}
    \institute{
        Max Planck Institute for Astronomy, Königstuhl 17, 69117 Heidelberg, Germany
        \and
        Institute of Astronomy, University of Cambridge, Madingley Road, Cambridge CB3 OHA, UK
        \and
        Università degli Studi di Milano, Via Giovanni Celoria 16, 20133 Milano, Italy
        \and
        European Southern Observatory, Karl-Schwarzschild-Str. 2, 85748 Garching bei München, Germany
        \and
        Departamento de Astronomía, Universidad de Chile, Camino El Observatorio 1515, Las Condes, Santiago, Chile
        \and
        Institute for Astronomy, University of Hawaii, Honolulu, HI 96822, USA
        \and
        INAF – Osservatorio Astrofisico di Arcetri, L.go E. Fermi 5, 50125 Firenze, Italy
        \and
        Astrophysics Group, Department of Physics, Imperial College London, Prince Consort Rd, London SW7 2A2, UK
        \and
        School of Physics and Astronomy, University of Leeds, Leeds, LS2 9JT, UK
        \and
        Leiden Observatory, Leiden University, P.O. Box 9513, 2300 RA Leiden, The Netherlands
        \and
        School of Physics \& Astronomy, University of Leicester, University Road, Leicester LE1 7RH, UK
        \and
        Dipartimento di Fisica e Astronomia, Università di Bologna, Via Gobetti 93/2, I-40122 Bologna, Italy
    }
    
    \date{Received ; accepted }

    \abstract
    {
    We present high angular resolution ($50\,{\rm mas}$) and sensitivity Atacama Large Millimeter/submillimeter Array (ALMA) Band~3 ($3.1\,{\rm mm}$) and Very Large Array (VLA) Ka~band ($9.1\,{\rm mm}$) observations of the multi-ringed disc around the $3\,{\rm Myr}$-old solar-mass star CI~Tau. These new data were combined with similar-resolution archival ALMA Band~7 ($0.9\,{\rm mm}$) and 6 ($1.3\,{\rm mm}$) observations and new and archival VLA Q ($7.1\,{\rm mm}$), Ku ($2.0\,{\rm cm}$), X ($3.0\,{\rm cm}$) and C~band ($6.0\,{\rm cm}$) photometry to study the properties of dust in this system. At wavelengths $\leq3.1\,{\rm mm}$, the continuum emission from CI~Tau is very extended ($\geq200\,{\rm au}$) and highly substructured (with three gaps, four rings, and two additional gap-ring pairs identified by non-parametric visibility modelling). In contrast, the VLA Ka~band data are dominated by a centrally peaked bright component, only partially ($\leq 50\%$) due to dust emission, surrounded by a marginally detected faint and smooth halo. We fitted the ALMA and VLA Ka~band data together, adopting a physical model that accounts for the effects of dust absorption and scattering. For our fiducial dust composition (`Ricci' opacities), we retrieved a flat maximum grain size distribution across the disc radius, with $a_{\rm max}=(7.1\pm0.8)\times10^{-2}\,{\rm cm}$, that we tentatively attributed to fragmentation of fragile dust or bouncing. We tested, for the first time, the dependence of our results on the adopted dust composition model to assess which mixture can best reproduce the observations. We found that `Ricci' opacities work better than the traditionally adopted `DSHARP' ones, while graphite-rich mixtures perform significantly worse. We also show that for our fiducial composition, the data prefer low porosity ($\leq70\%$) grains. This is in contrast with recent claims of highly porous aggregates in younger sources, which we tentatively justified by time-dependent compaction at the fragmentation or bouncing barrier. Our results on composition and porosity are in line with constraints from disc population synthesis models and naturally arise from CI~Tau's peculiar spectral behaviour (i.e. the abrupt steepening of its spectral index at wavelengths longer than $3.1\,{\rm mm}$), making this disc a unique target to characterise the properties of disc solids and thus ideal for deeper centimetre-wavelength observations and follow-up dust polarisation studies.
    }


    \maketitle
    
\section{Introduction}\label{sec:1}
Dust is an essential ingredient of every planet-formation model. First and foremost, it is the material planet(esimal)s are made~of. By dominating the disc opacity, solids also set the temperature structure and thus the excitation conditions and reaction rates of volatiles \citep[e.g.][]{Gavino2021,Oberg2023}. In addition, electron-ion recombination and chemical reactions on the surface of grains affect the disc ionisation structure \citep[e.g.][]{Desch&Turner2015} and prompt the formation of complex organic molecules \citep[e.g.][]{Oberg&Bergin2021}. Dust collisional evolution (growth and fragmentation) and transport (vertical settling~and radial drift) not only determine the availability of solids to form planetary cores \citep{Brauer2008,Birnstiel2010} but also affect the dust-to-gas thermal coupling \citep{Facchini2017} and the abundance of gas-phase chemical species. As dust migrates radially, the volatiles frozen onto grains undergo thermal desorption upon transition across their snowlines \citep[e.g.][]{Booth2017,Booth&Ilee2019,Krijt2020,vanClepper2022}, thus locally enriching the feedstock of exoplanet atmospheres \citep{Madhusudhan2019}. Therefore, unsurprisingly, determining the physical and chemical properties of protoplanetary disc solids is an essential first step to understanding planet formation \citep[e.g.][]{Miotello2023,Birnstiel2024}. For this~reason, attempts have been made to infer the density, size, distribution, and temperature of disc solids by modelling the continuum emission from planet-forming discs at multiple wavelengths.

Early works, which either focused on a few targets of interest or surveyed tens of discs in nearby star-formation regions,~measured integrated spectral indices, first in the (sub-)millimetre and then at progressively longer wavelengths significantly smaller than in the interstellar medium (ISM). While the contribution of optically thick regions could not be ruled out \citep{Ricci2012}, especially at (sub-)millimetre wavelengths, these results were generally interpreted \citep{Beckwith&Sargent1991,Andrews&Williams2005,Rodmann2006}, sometimes in combination with information on the (wavelength dependence of the) disc radius \citep[e.g.][]{Testi2003,Wilner2005,Ricci2010_Oph,Ricci2010,Isella2010,Banzatti2011,Guilloteau2011}, as being a consequence of substantial grain growth up to millimetre or centimetre sizes, depending on the adopted dust optical properties (see \citealt{Testi2014} and \citealt{Andrews2020} for a summary picture). In addition, the first multi-wavelength studies able to resolve a few bright sources from (sub-)millimetre to centimetre wavelengths  \citep{Perez2012,Perez2015,Trotta2013,Tazzari2016} inferred radially decreasing grain size profiles, in qualitative agreement with the predictions of dust evolution models \citep{Brauer2008,Birnstiel2010}.

The advent of the Atacama Large Millimeter/submillimeter Array (ALMA) substantially revolutionised this picture. One of ALMA's major breakthroughs was the almost ubiquitous detection of small scale structures in the continuum emission of (the largest and brightest) protoplanetary discs \citep{Andrews2020,Bae2023}. At (sub-)millimetre wavelengths, substructures are most commonly detected in the form of axisymmetric troughs and peaks (colloquially referred to as `gaps' and `rings'; e.g. \citealt{Andrews2016,Long2018,Andrews2018,Huang2018}) or cavities \citep[e.g.][]{Keppler2019,Facchini2020}, but spirals \citep{Perez2016,Huang2018_spirals,Kurtovic2018,Paneque-Carreno2021} and crescents \citep[e.g.][]{vanderMarel2013,Isella2018,Perez2018,Dong2018} are sometimes also present. On the one hand, the most tantalising interpretation of these structures is that they trace dynamical interactions between massive (0.1 to $10.0\,M_{\rm Jup}$, \citealt{Lodato2019,Bae2023,Ruzza2024}) protoplanets and their hosting disc \citep[e.g.][]{Zhang2018_sims}. However, despite many attempts using infrared (IR) high-contrast imaging \citep[e.g.][]{Testi2015,Guidi2018}, photospheric emission of accreting planets was detected only in one case (PDS~70, e.g. \citealt{Keppler2018,Haffert2019}). On the other hand, a plethora of alternative mechanisms have been put forward to explain the wealth of observed substructures \citep[e.g.][]{Bae2023}, such as (magneto-)hydrodynamic instabilities (\citealt{Lesur2023} and references therein), thermal and magnetic winds (\citealt{Pascucci2023} and references therein), eccentric modes (sometimes triggered by gravitationally bound companions or flybys, e.g. \citealt{Ragusa2017,Cuello2023}), or physical-chemical processes that alter the properties of dust at the snowlines of the major volatile species \citep[e.g.][]{Zhang2015,Okuzumi2016}.

Determining the properties of disc substructures, such as dust density and size (see below), gap-to-ring contrast \citep{Huang2018}, and trapping ability \citep{Dullemond2018,Rosotti2020,Doi&Kataoka2023}, is of paramount importance. This is crucial not only to discriminate between their possible origins \citep{Bae2023} but also to assess their potential to form (second generation, see the discussion of \citealt{Bae2023}) planet(esimal)s by streaming instability \citep[][see \citealt{Carrera2021,Carrera2022,Xu&Bai2022_rings} for the specific case of pressure bumps]{Youdin&Goodman2005,Johansen2007} or gravitational instability \citep{Goldreich&Ward1973} and the subsequent growth by pebble accretion \citep{Lau2022,Jiang&Ormel2023}.

The substructured discs observed with ALMA at high angular resolution \citep[e.g.][]{Andrews2018} showed clear indications of moderate-to-high absorption optical depths in the inner regions and at the position of the bright rings \citep{Huang2018,Dullemond2018}, which is considered to be an indication of unresolved optically thick substructures \citep{Dullemond2018}, self-regulating planetesimal formation \citep{Stammler2019},~or high extinction in the presence of self-scattering by highly reflective grains \citep[e.g.][although \citealt{Isella2018}~and \citealt{Guzman2018} showed that dust rings do not completely suppress line emission]{Zhu2019}. The spectral index radial profile modulations at the position of gaps and rings inferred from high angular resolution multi-frequency continuum observations \citep[e.g.][]{ALMAPartnership2015,Tsukagoshi2016,Macias2019,Huang2020} and the `anomalously low' (i.e. $\alpha<2$) spectral indices in the innermost regions of a few sources \citep[e.g.][]{Huang2018_TWHya,Dent2019,Paneque-Carreno2021,Ribas2023,Houge2024}, which are most often attributed to dust self-scattering \citep{Liu2019}, go well in line with this picture of moderate-to-high optical depths in the inner disc and the bright rings. These clues suggest that the classically adopted assumption of optically thin continuum emission is rather inaccurate when analysing (sub-)millimetre observations and leads to overestimation of the grain size and underestimation of the dust density \citep[e.g.][]{Zhu2019,Carrasco-Gonzalez2019}.

Broadband high angular resolution multi-frequency continuum observations with ALMA \citep{Macias2021,Sierra2021,Mauco2021,Ueda2022,Ohashi2023,Carvalho2024,Sierra2025}, complemented with Very Large Array (VLA) data in some cases \citep{Carrasco-Gonzalez2019,Guidi2022,Zhang2023,Sierra2024,Guerra-Alvarado2024}, allowed the degeneracies between intrinsic dust properties and the wavelength dependence of the optical depth to be broken, thus opening up the possibility of thoroughly characterising the temperature, density, and size of solids in gaps and rings. Not surprisingly, most of these works obtained rather different constraints. While partly due to the intrinsic diversity of the analysed sources, the different modelling hypotheses (e.g. the adopted dust composition and porosity or the fitting methods) might have impacted their results substantially. In this paper, we combine ALMA~and VLA high angular resolution and sensitivity (sub-)millimetre continuum observations and centimetre-wavelength photometry of CI~Tau (whose stellar parameters are summarised in \autoref{tab:app1}) to provide constraints on the properties of solids and simultaneously explore different assumptions on dust composition and porosity.

\subsection{The case of CI~Tau}\label{sec:1.1}

CI Tau is the youngest pre-main-sequence star where periodic variability in multi-epoch IR radial velocity measurements, also supported by optical spectroscopy and photometric monitoring, was attributed to a moderately eccentric ($0.28\pm0.16$) massive ($11.29\pm2.13\,M_{\rm Jup}$) candidate planet on a nine-day period~orbit \citep{Johns-Krull2016}. This claim was further backed by the detection of a similar-period variability in \textit{K2} (the extended mission of the \textit{Kepler Space Telescope}) photometry \citep{Biddle2018,Biddle2021}, consistent with accretion modulation on the planet's orbital period (see \citealt{Teyssandier&Lai2020}), and of CO emission, proposed to be associated with the candidate hot Jupiter \citep{Flagg2019}.

Since the in situ formation of this candidate planet was considered unlikely (see \citealt{Rosotti2017_HJ} and references therein), planet migration and interactions with or scattering by an unseen companion were put forward as possible explanations for the origin of this hot Jupiter. On the one hand, while early simulations suggested that the planet eccentricity can be pumped up while migrating in a low-mass disc over long timescales \citep{Rosotti2017_HJ,Ragusa2018}, it was recently shown that type~II~migration preferably leads to the pile-up of planets between 1 and $10\,{\rm au}$, precluding the formation of hot Jupiters by migration in a (non-self-gravitating) disc \citep{Scardoni2022}. On the other hand, high-contrast IR angular differential imaging ruled out the presence of companions more massive than 2 to $4\,M_{\rm Jup}$ external to $30\,{\rm au}$ \citep{Shimizu2023}, in line with the prediction~that eccentricity-driving lower-mass companions are ejected or accreted after dynamical interactions \citep{Rosotti2017_HJ}.

Most recently, (sub-)au resolution VLTI/GRAVITY observations \citep{Soulain2023} revealed that CI~Tau's inner disc emission is non-axisymmetric, in line with the predictions of planet-disc interaction simulations \citep[e.g.][]{Muley&Dong2021}, and misaligned by $\approx70^\circ$ with the outer disc (see \citealt{Clarke2018}), potentially due to the gravitational torque induced by a massive close-in companion. Moreover, the inner dust disc rim is located two to four times further out than the expected position of the sublimation radius, as expected from the presence of an inner companion carving a gap in the disc \citep{Muley&Dong2021}. This hypothesis was further supported by the shape of the CO $\nu=1-0$ ro-vibrational transitions targeted by IRTF/iSHELL and Keck/NIRSPEC \citep{Kozdon2023}. These lines show~a~blueshifted core and redshifted wings, that can be explained by two eccentric disc components with oppositely aligned arguments of periapsis, separated by a massive eccentric planet, albeit with much lower eccentricity ($\approx0.05$) than estimated by \citet{Johns-Krull2016}.

However, the spectro-polarimetric monitoring of CI~Tau with ESPaDOnS and SPIRou/CFHT \citep{Donati2020,Donati2024} recently showed that the nine-day periodicity detected by \citet{Johns-Krull2016} is most-likely driven by the stellar rotation, and the CO emission detected by \citet{Flagg2019} might be connected to the accretion funnels onto the star, thus suggesting that the presence of a hot Jupiter in the system needs to be reconsidered. Likewise, the most recent detection of a 25.2-day periodicity in ESPaDOnS and SPIRou/CFHT multi-epoch radial velocity monitoring, as well as \textit{K2} and LCOGT photometric time series, attributed to a highly eccentric ($0.58\pm0.05$) massive ($3.6\pm0.3\,M_{\rm Jup}$) planet \citep{Manick2024}, was disputed by \citet{Donati2024}, who proposed that it can be explained by an asymmetry in the inner disc. Whether the disc around CI~Tau hosts a hot Jupiter or not remains debated.

Being bright and large at (sub-)millimetre wavelengths, CI~Tau~was extensively targeted in the pre-ALMA era, both in single dish \citep[e.g.][]{Beckwith1990,Beckwith&Sargent1991,Andrews&Williams2005} and interferometric observations \citep[e.g.][]{Dutrey1996,Andrews&Williams2007,Andrews2013,Kwon2015}. Early multi-wavelength analyses measured a millimetre spectral index of $\alpha_{\rm 1.3-2.7mm}=2.5\pm0.2$ \citep{Dutrey1996,Ricci2010,Guilloteau2011} that steepens to $\alpha_{\rm 1.3-7.0mm}=3.0\pm0.3$ at longer wavelengths \citep{Rodmann2006}, and a $\beta^{\rm abs}_{\rm 1.3-2.7mm}$ profile radially increasing between $\lesssim0.5$ and $\gtrsim1.0$ (see Figure 10 and 11 of \citealt{Guilloteau2011}).

High-angular resolution and sensitivity ALMA observations of CI~Tau at $1.3\,{\rm mm}$ revealed a very extended ($\geq200\,{\rm au}$) disc characterised by a sequence of four rings and three gaps, compatible with the presence of three 0.75, 0.15, and $0.40\,M_{\rm Jup}$~planets in the system \citep{Clarke2018,Long2018}, or the snowlines of CO and clathrate-hydrated CO and N$_2$ for the two innermost gaps \citep{Long2018}. Adopting a super-resolution imaging technique, \citet{Jennings2022b} identified two additional gaps, making CI~Tau one of the most substructured discs observed to date, together with AS~209 \citep{Guzman2018}, HL~Tau \citep{ALMAPartnership2015}, TW~Hya \citep{Andrews2016}, HD~163296 \citep{Isella2018}, and RU~Lup \citep{Huang2020_RULup}. Follow-up observations at $0.9\,{\rm mm}$ showed a similar continuum morphology \citep{Rosotti2021}. On the contrary, the scattered light emission in SPHERE polarimetric images shows no substructures, but a bright $\approx100\,{\rm au}$ inner disc surrounded by a faint halo extending to $\gtrsim200\,{\rm au}$ \citep{Garufi2022_CITau}.

As for the lines, although both $^{12}{\rm CO}\ J=2-1$ and $J=3-2$ emission is heavily absorbed towards CI~Tau \citep{Rosotti2021,Semenov2024}, a kinematic planetary signature external to the dusty disc was proposed in one of the non-absorbed channels \citep{Rosotti2021}. Instead, $^{13}{\rm CO}\ J=2-1$ and $J=3-2$ emission is not absorbed. When observed at high-angular resolution, the latter shows a low-contrast gap co-located with the $\approx60\,{\rm au}$ continuum ring, attributed by \citet{Rosotti2021} to shadowing by the inner disc. A shallow dip in the $^{12}{\rm CO}\ J=3-2$ emission height at the same position of the $\approx50\,{\rm au}$ gap, and a change of slope co-located with the $\approx100\,{\rm au}$ ring, were detected by \citet{Law2022}. Similarly, the $^{12}{\rm CO}$ temperature profile shows two dips at $\approx70$ and $\approx120\,{\rm au}$ and a bump at $\approx90\,{\rm au}$ co-located with similar features in the continuum and $^{13}{\rm CO}\ J=3-2$ emission \citep{Law2022}. The ${\rm CS}\ J=7-6$ emission is ring-like, while the $J=5-4$ one is smooth, most likely due to chemical effects \citep[e.g.][]{LeGal2019,Rosotti2021}. CCH $N=3-2$ and C$^{18}$O $J=2-1$ emission are relatively bright and centrally peaked \citep{Bergner2019,Semenov2024}.

This paper is organised as follows. In \autoref{sec:2} we introduce our datasets, discuss their (self-)calibration and imaging. CI~Tau's (integrated and resolved) spectral properties are discussed in \autoref{sec:3}. In \autoref{sec:4} we introduce our analysis procedure, while in \autoref{sec:5} we present our results, that are discussed in \autoref{sec:6}. In \autoref{sec:8} we draw our conclusions.

\section{Observations, calibration, and imaging}\label{sec:2}

\subsection{Observations}\label{sec:2.1}
We introduce, for the first time, ALMA Band~3, VLA Ka, Ku, X, and C band observations of CI~Tau. We make use of archival ALMA Band~6 \citep{Konishi2018,Clarke2018} and 7 \citep{Rosotti2021} observations to study the properties of solids in this disc at high angular resolution and sensitivity. Information about these datasets is summarised in \autoref{tab:app2}.

\paragraph*{ALMA Band~3 observations --} ALMA Band~3 observations were conducted in Cycle~6 and 7 between June~2019 and July~2021 as part of the program 2018.1.00900.S (PI: M.~Tazzari). The data were correlated from four spectral windows (SPWs) in dual polarisation mode. All the SPWs were set in frequency division mode (FDM), each with 1920 channels $976.562\,{\rm kHz}$ wide, spanning a total bandwidth of $1.875\,{\rm GHz}$, and centred at 90.5, 92.5, 102.6, and $104.5\,{\rm GHz}$. The target was observed in the more compact C43-6 configuration (short baselines, SBs), and the more extended C43-9 configuration (long baselines, LBs).

The SB observations used an array with baseline lengths from $15.3\,{\rm m}$ to $2.4\,{\rm km}$ ($\approx400\,{\rm mas}$ resolution) and 41 antennas. The total integration time on the science target was $\approx38\,{\rm min}$. The phase calibrator J0426+2327 was observed in an alternating sequence with the science target, every $10\,{\rm min}$. The bandpass and flux calibrator J0237+2848 was observed at the beginning of the observing block. The LB observations used an array with baseline lengths from $83.1\,{\rm m}$ to $16.2\,{\rm km}$ ($\approx50\,{\rm mas}$ resolution). Two execution blocks (EBs) were scheduled. The first used 42 antennas and the second 46. The total integration time on the science target was $\approx44\,{\rm min}$ for each EB. The phase calibrator J0426+2327 was observed in an alternating sequence with the science target, every minute. The bandpass and flux calibrator, J0423--0120 and J0510+1800 for different EBs, respectively, was observed at the beginning of the observing block. An additional `check' calibrator, J0425+2235, was observed every $\approx15\,{\rm min}$ to assess the quality of phase transfer.

\paragraph*{ALMA Band~6 observations --} The publicly available archival ALMA Band~6~observations (that we took into account) were conducted in Cycle~3 in August 2016 as part of the program 2015.1.01207.S (PI: H.~Nomura) and in Cycle~4 in September 2017 as part of the program 2016.1.01370.S (PI: C.~J.~Clarke). The data were correlated from four SPWs in dual polarisation mode. For the program 2016.1.01370.S, all the SPWs were set in time division mode (TDM), each with 128 channels $15.625\,{\rm MHz}$ wide, spanning a total bandwidth of $2.0\,{\rm GHz}$, and centred at 224.0, 226.0, 240.0, and $242.0\,{\rm GHz}$. Instead, for the program 2015.1.01207.S, the SPWs were set in FDM and centred at: (i) $220.0\,{\rm GHz}$ with a bandwidth of $234.0\,{\rm MHz}$ in 480 channels of $488.3\,{\rm kHz}$; (ii--iii) 216.7 and $231.2\,{\rm GHz}$ with a bandwidth of $937.5\,{\rm MHz}$ in 1920 channels of $488.3\,{\rm kHz}$; (iv) $234.4\,{\rm GHz}$ with a bandwidth of $937.5\,{\rm MHz}$ in 960 channels of $976.6\,{\rm kHz}$.

The 2016.1.01370.S program observations were scheduled in two EBs, both relied on an array with baseline lengths from $41.4/21.0\,{\rm m}$ to $12.1\,{\rm km}$ ($\approx35\,{\rm mas}$ resolution) in the C40-8/9 configuration (LBs). The first EB used 40 antennas and the second 41. The total integration time on the science target was $\approx33\,{\rm min}$ for each EB. The phase calibrator J0426+2327 was observed in an alternating sequence with the science target, every minute. The bandpass and flux calibrator J0510+1800 was observed at the beginning of the observing block. An additional “check” calibrator J0438+2153 was observed every $\approx12\,{\rm min}$.

The 2015.1.01207.S program observations used an array with baseline lengths from $15.1\,{\rm m}$ to $1.6\,{\rm km}$ ($\approx250\,{\rm mas}$ resolution) and 44 antennas in the C40-6 configuration (SBs). The total integration time on the science targets was $\approx18\,{\rm min}$ (half of which on CI~Tau). The phase calibrator J0431+2037 was observed in an alternating sequence with the science targets, every $6\,{\rm min}$. The bandpass and flux calibrator J0510+1800 was observed at the beginning of the observing block.

\paragraph*{ALMA Band~7 observations --} ALMA Band~7 observations were conducted in Cycle~5 in December 2017 as part of the DDT program 2017.A.00014.S (PI: G.~P.~Rosotti). The data were correlated from four SPWs in dual polarisation mode. Three SPWs were set in FDM each with 1920 channels $448.3\,{\rm kHz}$-wide centred at 330.7, 343.0, and $345.7\,{\rm GHz}$, and spanning a total bandwidth of $937.5\,{\rm MHz}$. The remaining SPW, centred at $333.2\,{\rm GHz}$, was set in TDM, with 128 channels $15.625\,{\rm MHz}$-wide, and spanning a total bandwidth of $2\,{\rm GHz}$. The observations used an array with baseline lengths from $15.1\,{\rm m}$ to $3.3\,{\rm km}$ ($\approx90\,{\rm mas}$ resolution) and 43 antennas in the C43-7 configuration. Two EBs were scheduled. The total integration time on the science target was $\approx39\,{\rm min}$ for each EB. The phase calibrator J0426+2327 was observed in an alternating sequence with the science target, every minute. The bandpass and flux calibrator J0510+1800 was observed at the beginning of the observing block. An additional “check” calibrator J0435+2532 was observed every $\approx 13\,{\rm min}$ to assess the quality of phase transfer.

\paragraph*{VLA Ka, Ku, X, and C band observations --} CI~Tau was observed by the VLA in Ka~band in two configurations as part of the project 19A-440 (PI: M.~Tazzari). The SPWs covered the frequency range between 29 and $37\,{\rm GHz}$ (wavelengths between 8.1 and $10.3\,{\rm mm}$). Observations in B~configuration, with a maximum baseline of $11.1\,{\rm km}$, were performed during March and April 2019 and September and October 2020, for a total on-source integration time of $9.8\,{\rm hrs}$. Observations in A~configuration, with a maximum baseline of $36.6\,{\rm km}$, were executed between September and October~2019, resulting in a total on-source integration time of $8.6\,{\rm hrs}$. For both configurations, 3C147 served as the flux and bandpass calibrator, J0403+2600 was used for pointing calibration, and J0431+2037 for phase calibration. 

Under the same VLA project 19A-440, X~band observations in A~configuration (maximum baseline of $36.6\,{\rm km}$) were conducted in August 2019, for a total integration time on the science target of $1.3\,{\rm hrs}$. The bandwidth extended from 8 to $12\,{\rm GHz}$ (corresponding to wavelengths from 2.5 to $3.7\,{\rm cm}$). 3C147 was used for flux and bandpass calibration, J0403+2600 for phase calibration, while no pointing calibrator was observed. 

As part of the VLA project 20A-373 (PI:~M.~Tazzari), CI~Tau was observed in the Ku and C~band, both in C~configuration, with a maximum baseline of $3.4\,{\rm km}$. Ku~band observations were performed between February and March 2020, with a spectral coverage from 12 to $18\,{\rm GHz}$ (1.7 to $2.5\,{\rm cm}$), for a total on-source integration time of $44.5\,{\rm min}$. C~band observations were carried out in 2020, with a bandwidth from 4 to $8\,{\rm GHz}$ (3.7 to $7.5\,{\rm cm}$), and a total on-source integration time of $14.9\,{\rm min}$. For both bands, 3C147 was employed for the flux and bandpass calibration, while J0431+2037 for the phase calibration. J0431+2037 also served as pointing calibrator for Ku~band, whereas C~band did not require any pointing calibration.

\subsection{(Self-)calibration}\label{sec:2.2}
The standard pipeline calibration was performed manually for ALMA Band~3, Band~6 SB and the VLA observations. The other datasets were pipeline-calibrated by the European ARC at ESO. Self-calibration was performed using the software \texttt{CASA} (Common Astronomy Software Applications, \citealt{McMullin2007,CASAteam2022}) \texttt{v6.4.3.27} for ALMA observations (\textit{ex novo} also for the archival Band~6 and 7 data) and \texttt{v6.2.1.7} for the VLA data, as explained hereafter.

\paragraph*{ALMA observations --} We followed the standard methodology adopted by the DSHARP collaboration \citep{Andrews2018}, with some minor changes. To begin with, channel-averaging was performed ensuring the same number of channels per SPW in each EB. To avoid bandwidth smearing, we adopted the criterion of \citet{Bridle&Schwab1989} for a reduction of $<1$\% in peak response to a point source at the edge of the primary beam. For Band~7, a pseudo-continuum measurement set (MS) was first created, flagging data within $-5\leq v/({\rm km}\, {\rm s}^{-1}) \leq 15$ (to account for CI~Tau's systematic velocity) of the $^{13}$CO $J = 3-2$ ($\nu_0=330.588\,{\rm GHz}$), CS $J = 7-6$ ($\nu_0=342.883\,{\rm GHz}$), and $^{12}$CO $J = 3-2$ ($\nu_0=345.796\,{\rm GHz}$) molecular line transition centres.

When available, the SB data were self-calibrated first, initially in phase-only mode. The gain solutions were computed using the task \texttt{gaincal}\footnote{\url{https://casadocs.readthedocs.io/en/v6.4.4/api/tt/casatasks.calibration.gaincal.html}}. For the first run, this was done separately for each polarisation (\texttt{gaintype=`G'}) for Band~3 and combining both polarisations (\texttt{gaintype=`T'}) for Band~6, combining different scans, with an infinite solution interval. From the second run on, for both Band~3 and 6 data, the gain solutions were computed with \texttt{gaintype=`T'}, combining different scans and spectral windows, with solution intervals progressively decreasing from 240, 120, 60, 30, to $15\,{\rm s}$. Images were reconstructed after each iteration using the \texttt{tclean} task. We set the CLEANing threshold to three times the noise level, used 10 pixels per beam, a multi-scale multi-frequency synthesis deconvolver (\texttt{mtmfs}, see \citealt{Rau&Cornwell2011}) with \texttt{nterms=2}, and adopted a \texttt{briggs} weighting scheme \citep{Briggs1995} with \texttt{robust=0.5}. We estimated an improvement of the peak signal-to-noise ratio (S/N) of $\approx28$\% and $\approx40$\% for Band~3 and Band~6 data, respectively. Finally, a round of phase and amplitude self-calibration was performed. The gain solutions were computed with \texttt{gaintype=`T'}, combining different scans and spectral windows, with an infinite solution interval. However, we did not observe any improvement of the peak S/N (as already reported by \citealt{Konishi2018} for Band~6 data).

Then, all the EBs in the same band were concatenated and self-calibrated together. To remove spatial offsets, we measured the emission centroid positions of each dataset with a Gaussian fit in the image plane. Subsequently, the \texttt{phaseshift} task was used to move the disc centre to the fitted phase centre position, and the \texttt{fixplanets} task to align each EB to the phase centre of the LBs with the highest peak S/N (i.e. the EB with the smallest astrometric uncertainty and lowest noise). Before these steps, we performed a run of phase-only self-calibration on the LB data to increase their peak S/N ratio. The gain solutions were computed with \texttt{gaintype=`G'}, combining different scans with an infinite solution interval. This additional step helped to improve the alignment procedure. Subsequently, the visibilities of each dataset were deprojected and azimuthally averaged. To do so, we used the median of the posterior distribution of the inclination ($i=49.24\, {\rm deg}$) and position angle (${\rm PA}=11.28\, {\rm deg}$) that \citet{Clarke2018} determined by fitting the Band~6 LB data in the Fourier space parametrically with \texttt{galario} \citep{Tazzari2018}. These deprojected visibilities were then inspected to identify and correct for mismatches in the amplitude scales. This step is crucial to ensure an accurate flux scaling with wavelength, and proved to be particularly important for Band~6 data, where flux density variations larger than 10\% over a single day were measured. We refer to \autoref{sec:app1} for a detailed discussion of our correction procedure, and the time variability of the flux calibrators.

The short and long baseline data in the same ALMA bands were then combined using the \texttt{concat} task. We performed a number of phase-only self-calibration iterations on the concatenated MS. For the first run, the gain solutions were computed with \texttt{gaintype=`G'} for Band~3 and 7, and \texttt{gaintype=`T'} for Band~6, combining different scans with an infinite solution interval. From the second run on, the gain solutions were computed with \texttt{gaintype=`T'}, combining different scans and spectral windows, with solution intervals progressively decreasing from 360, 180, to 60 for Band~3, and to 30 and $15\,{\rm s}$ for Band~6 and 7. Image reconstruction was performed as for the SBs, but with \texttt{robust=1.0}. We estimated an improvement of the peak S/N of $\approx21$\% for Band~6, $\approx26$\% for Band~7, and only marginal for Band~3. Finally, we performed a round of phase and amplitude self-calibration with \texttt{gaintype=`T'}, combining different scans and spectral windows with an infinite solution interval. We flagged the most extreme gain solutions (with gains $<0.8$ or $>1.2$), that are associated with long-baseline antennas and artificially reduced the beam size. After this iteration, the peak S/N ratio improved substantially only in Band~7 by an additional factor of $\approx20$\% (i.e. a total improvement by 56\%).

\paragraph*{VLA observations --} We conducted spectral averaging on each dataset, considering the necessary precautions to prevent bandwidth smearing\footnote{\url{https://casadocs.readthedocs.io/en/v6.3.0/notebooks/synthesis_imaging.html}}. Time averaging was not performed. In the imaging process, we adopted a \texttt{briggs} weighting scheme with \texttt{robust=1.0}, representing the optimal trade-off between S/N and side lobe effects. The \texttt{mtmfs} deconvolver with \texttt{nterms=2} was employed to accurately account for the large bandwidth-over-observing-frequency ratio in the VLA data.

For the Ka~band observations, we initially aligned the data from different EBs to ensure that the disc centre was consistent across all the datasets. Then, we applied a single round of phase-only self-calibration only to the B~configuration (lower angular resolution) data, leading to a slight improvement of~the peak S/N by $\approx5\%$. We adopted the following \texttt{gaincal} parameters: \texttt{gaintype='G'}, \texttt{combine='scan'}, \texttt{solint='inf'}. Subsequently, the B~configuration data were combined with the A~configuration (higher angular resolution) ones. Similarly, a round of phase-only self-calibration was performed, resulting in a further increase in the peak S/N by $\approx10\%$. Additional self-calibration iterations  with shorter solution intervals did not yield any further improvement. For the Ku~band data, we applied one round of phase-only self-calibration (again with \texttt{gaintype='G'}, \texttt{combine='scan'}, \texttt{solint='inf'}) resulting in a marginal improvement of the S/N by $\approx10\%$. Due to the initially (too) low S/N, no self-calibration was performed on~the X and C~band data to avoid introducing spurious effects.

\subsection{Fiducial images and radial profiles}\label{sec:2.3}
We reconstructed the ALMA and VLA Ka~band fiducial images CLEANing over an elliptical mask with a semi-major axis of $1\farcs5$, a semi-minor axis of $0\farcs9$, and ${\rm PA} = 11.28\, {\rm deg}$ \citep{Clarke2018}, down to the estimated noise level (i.e. a $1\sigma$ RMS noise threshold), using 8 pixels per beam semi-minor axis. Different \texttt{robust} parameters were tested, progressively increasing by 0.5 from $-1.0$ to 1.5. We adopted a \texttt{mtmfs} deconvolver with \texttt{nterms=2}, and a set of (Gaussian) deconvolution scales \citep{Cornwell2008}, different for each band and \texttt{robust} parameter, including a point source, scales corresponding to one and two beams, and, for ALMA data, to the ring locations (from \citealt{Clarke2018}), and the outer disc radius. The image noise was estimated over a circular annulus centred on and larger than the target with inner and outer radius of $2\farcs25$ and $6\farcs00$. The results of the imaging process are summarised in Columns (1)--(9) of \autoref{tab:app3}. The reconstruction of the VLA Ku, X, and C~band images is discussed in \autoref{sec:app2} and their properties are also summarised in \autoref{tab:app3}. The continuum flux densities in \autoref{tab:app3} were measured integrating emission within the CLEANing mask (for ALMA observations and the VLA Ka~band data) and a synthesised beam around the target (for the unresolved Ku, X, and C~band emission) and its uncertainty was determined as the standard deviation of the flux densities measured over 34 (for the ALMA observations and the VLA Ka~band data) or 80 (for Ku, X, and C~band data) masks of the same size away from the source.

For ALMA data, we also reconstructed a set of images with $4\sigma$ threshold and applied the Jorsater and van Moorsel's (JvM) correction \citep{Jorsater&vanMoorsel1995,Czekala2021}. Their radial profiles agree well (within 10\%) with those imaged with a deeper, $1\sigma$ threshold in most cases. Notable exceptions are (i) the dark rings in the Band~6 images reconstructed with a \texttt{robust} parameter $<0.0$, due to their low S/N, and (ii) the outermost dark and bright rings in the Band~3 images reconstructed with a \texttt{robust} parameter $>0.5$, due to the elongated shape of their point-spread function (PSF), whose first null lies external to $2\farcs5$. However, in both cases, the radial profiles are off by less than 30\%. In our analysis, we used the deep-CLEANed images for consistency with the VLA observations, where we chose not to apply the JvM correction altogether because of the prominent beam sidelobes.

\begin{figure*}
    \centering
    \includegraphics[width=0.98\textwidth]{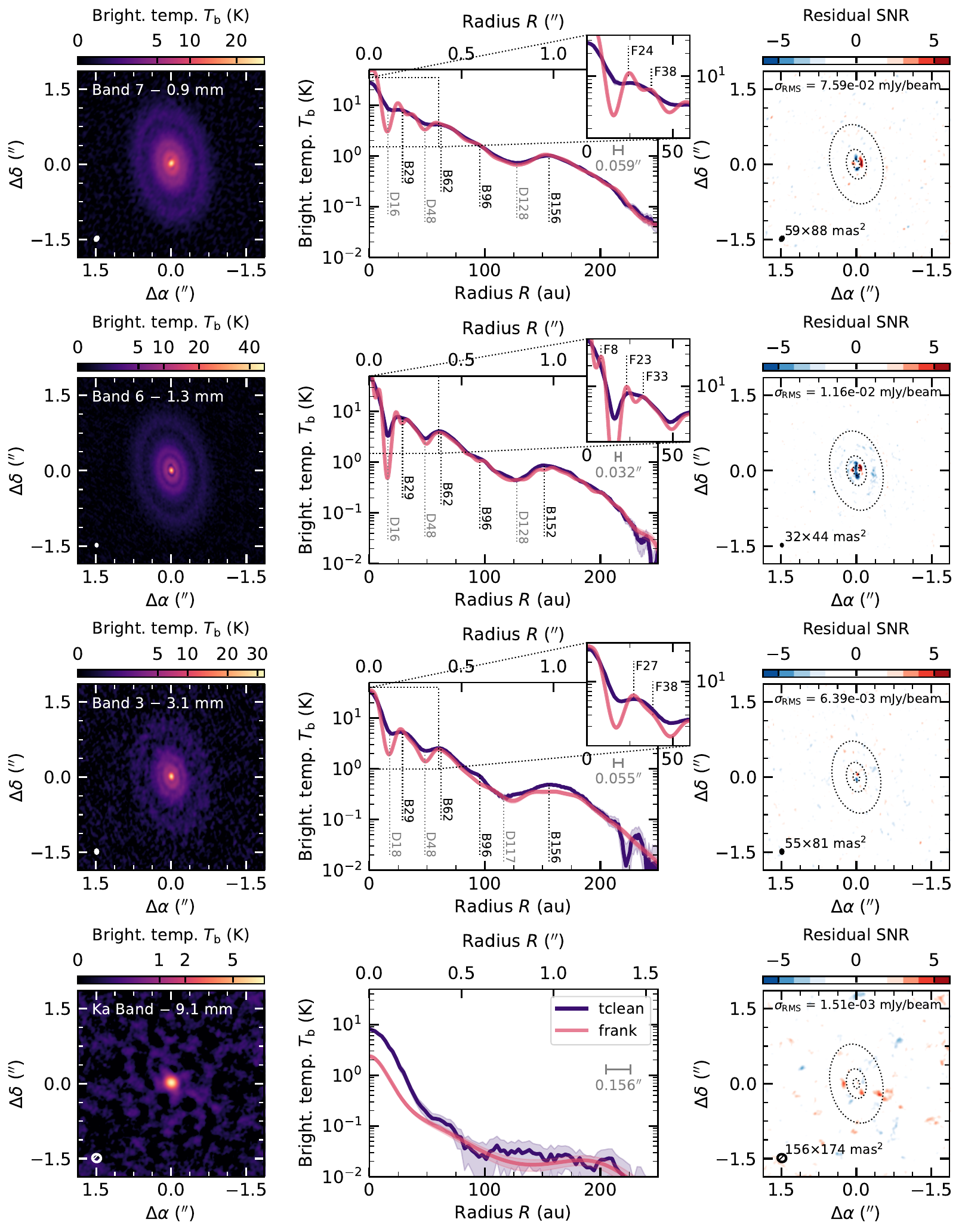}
    \caption{From top to bottom: CI~Tau's ALMA Band~7, 6, 3, and VLA Ka~band continuum emission. Left column: CLEAN images. Central column: Azimuthally averaged surface brightness radial profiles. Those obtained from the \texttt{tclean} images are in violet, purple is used for the best-fit \texttt{frank} profiles (a point-source component was subtracted from the 3.1 and $9.1\,{\rm mm}$ visibilities before fitting). Right column: Residual images of the \texttt{frank} fit. Dotted ellipses mark the location of the dark rings in the CLEAN images. The synthesised CLEAN beam is shown as an ellipse in the bottom-left corner of each image and as a segment with full width half maximum equal to the beam minor axis in each radial profile subplot.}
    \label{fig:1}
\end{figure*}

\paragraph*{Image analysis --} The left and central columns of \autoref{fig:1} display a montage of the ALMA Band~7, 6, 3, and VLA Ka~band \texttt{tclean} images, and their radial profiles, plotted in violet, obtained deprojecting and azimuthally averaging their corresponding images using \texttt{GoFish} \citep{Teague2019} with the best-fit inclination and position angle of \citet{Clarke2018}. For both images and radial profiles, the brightness temperature was computed in the Rayleigh-Jeans approximation. The ALMA images show a similar morphology, with three dark and four bright rings, as previously detected by \citet{Clarke2018} and \citet{Long2018} in Band~6 and \citet{Rosotti2021} in Band~7. These substructures are labelled by the prefix `D' and `B' for dark and bright rings followed by their radial location in au \citep[see][]{Huang2018}. On the contrary, the VLA Ka~band data display a smoothly declining surface brightness, with no clear substructures. Although this image could be reconstructed at a significantly higher angular resolution, comparable to that of our ALMA datasets, for \texttt{robust} parameters $<1.0$ its surface brightness radial profile is centrally peaked and noise dominated further out. For this reason, we adopted a conservative synthesised beam that allows for the detection, albeit with a low ${\rm S/N}\lesssim3$, of emission extending up to $1\farcs2$ in the radial profile. 

We measured continuum disc sizes using a curve of growth method. The azimuthally averaged surface brightness radial profiles were integrated up to the disc radius where their S/N (measured weighting the image RMS noise in \autoref{tab:app3} by the number of beams per annulus at each radius) falls below 3. Uncertainties were obtained likewise, adopting the fiducial brightness profiles plus or minus their standard deviation at each radius. Our results for the 68\% and 95\% disc radius are summarised in Columns (10)--(11) of \autoref{tab:app3}. A clear trend of decreasing continuum sizes with wavelength can be seen, especially in the case of VLA Ka~band data, where the disc is unresolved in all the images reconstructed with a \texttt{robust} parameter $<1.0$. Emission at VLA Ku, X, and C~band wavelengths is also unresolved, regardless of the imaging parameters.

\paragraph*{Visibility plane fits --} We searched for additional continuum substructures using \texttt{frank} \citep{Jennings2020}, a tool that performs non-parametric fits of the data in the visibility space and can achieve sub-beam resolution. Our fitting procedure is described in \autoref{sec:app3}. We highlight that, at long baselines, the ALMA Band~3 and the VLA Ka~band visibilities do not flatten out to zero, as expected from fully resolved continuum emission, but plateau to some non-null amplitude offset, indicative of the presence of an additional point-source (PS) component. This visibility offset corresponds to $<2\%$ of the total flux density in Band~3, but it accounts for $\approx50\%$ of the Ka~band continuum emission. This PS component was subtracted from the visibilities prior to the fit both at 3.1 and $9.1\,{\rm mm}$. In the latter case, this step is essential to reach convergence (see \autoref{sec:app4.1} for more details).

The best-fit \texttt{frank} brightness profiles are displayed in the central column of \autoref{fig:1} in purple. For all the ALMA bands, the \texttt{frank} profiles show that the bright ring at $\approx 29\,{\rm au}$ in the CLEAN images is in fact a blend of two smaller-scale bright rings (see the insert in the top-right corner of the central column plots, zooming in the inner disc region). In Band~6, an additional bright ring is detected at $\approx8\,{\rm au}$, as was previously suggested by \citet{Jennings2022b}, who modelled only the Band~6 LB data. The detection of these new structures is supported by the presence of similar features in the CLEAN model and, for the $\approx 29\,{\rm au}$ structure, also in the CLEAN images reconstructed with \texttt{robust} parameters $\leq0.0$. Using a similar notation to that of \citet{Huang2018}, we labelled these sub-beam bright rings using the prefix `F', to mark that they were identified using \texttt{frank}. 

For the VLA Ka~band data, instead, the best-fit \texttt{frank} profile is very similar to the \texttt{tclean} one (albeit fainter in the inner $50\,{\rm au}$ due to PS subtraction) and shows no clear substructures. Most importantly, it recovers the extended emission marginally detected in the CLEAN profile reconstructed from the same data. This can be also seen comparing the disc sizes in \autoref{tab:app3}, from \texttt{tclean}, and those in \autoref{tab:app4}, from \texttt{frank}, roughly three-times larger due to a combination of the subtracted visibility offset and the fact that a more extended emission profile is recovered.

Finally, we imaged the \texttt{frank} fit residuals using \texttt{tclean} and the same parameters adopted for the data but \texttt{niter=0} \citep{Jennings2022a,Jennings2022b}. Our results are shown in the right column of \autoref{fig:1}. The dotted ellipses in these subplots display the location of the three dark rings detected in the CLEAN ALMA images. No significant residuals can be identified consistently across our datasets other than the fourfold structure in the inner disc, clearly visible in Band~7 and 6 data, that cannot be readily explained by the adoption of a wrong disc geometry when deprojecting the visibilities \citep[see Appendix A of][]{Andrews2021}, and is instead most likely connected to the presence of sub-beam structures along to the $29\,{\rm au}$ ring \citep[see][]{Scardoni2024}.

\section{Spectral properties}\label{sec:3}
\subsection{Spectral flux density distribution}\label{sec:3.1}

\begin{figure*}
    \sidecaption
    \includegraphics[width=12cm]{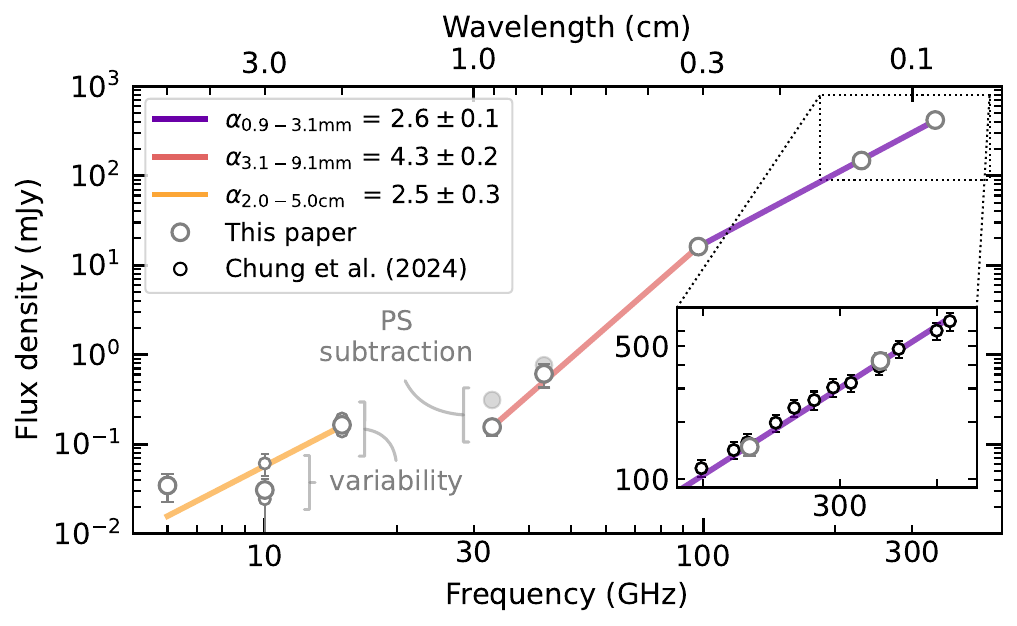}
    \caption{CI~Tau's spectral flux density distribution. The grey dots display CI~Tau's photometry from this paper's data and those of \citet{Rodmann2006}. Photometry by \citet{Chung2024} is over-plotted with black dots in the insert. Full markers show the total flux density (i.e. before point-source subtraction) for the ALMA Band~3 and the VLA Q and Ka~band data. The maximum and minimum integrated flux densities across different scans are displayed as smaller dots for the Ku and X~band data to highlight their short timescale variability.}
    \label{fig:2}
\end{figure*}

\autoref{fig:2} displays CI~Tau's spectral flux density distribution from 6.0 to $340.0\,{\rm GHz}$ (i.e. $5.0\,{\rm cm}$ to $0.9\,{\rm mm}$), built from the flux densities in \autoref{tab:app3}\footnote{When different \texttt{robust} parameters or weighting schemes were used in the imaging process, the flux densities corresponding to the largest beam were plotted. A point-source offset was subtracted from the ALMA Band~3 and VLA Ka~band flux densities (see \autoref{sec:app4.1}).} and the VLA Q~band ($43.3\,{\rm GHz}$ or $6.9\,{\rm mm}$) one ($0.76\pm0.17\,{\rm mJy}$) published by \citet{Rodmann2006}.~The spectral flux density distribution can be divided in three branches. To measure their slope (or `spectral index'), we fitted a line to the data in log-space solving the least-squares problem by adopting the Levenberg-Marquardt algorithm implemented in \texttt{scipy.optimize.curve\_fit} \citep{Virtanen2020}.

The spectral index between 0.9 and $3.1\,{\rm mm}$ is $\alpha_{\rm 0.9-3.1mm}=2.6\pm0.1$, consistent with marginally optically thin dust emission or optically thick dust emission with albedo increasing with wavelength \citep{Zhu2019}. This value is in excellent agreement with that measured by \citet{Chung2024} using 12 independent SMA flux density samplings between 200 and $400\,{\rm GHz}$ (black dots in the bottom-right insert of \autoref{fig:2}), and broadly compatible with similar measurements in other bright discs observed at high resolution and sensitivity with ALMA at multiple wavelengths, such as HL~Tau ($\alpha_{\rm 0.9-2.9mm}\approx2.8$, \citealt{Carrasco-Gonzalez2019}), HD~163296 ($\alpha_{\rm 0.9-3.2mm}\approx2.6$, \citealt{Guidi2022}), and TW~Hya ($\alpha_{\rm 0.9-3.1mm}\approx2.6$, \citealt{Macias2021}). 

The spectral index between 3.1 and $9.1\,{\rm mm}$ is $\alpha_{\rm 3.1-9.1mm}=3.7\pm0.1$, that steepens to $\alpha_{\rm 3.1-9.1mm}=4.3\pm0.1$ upon subtraction of the VLA Ka~band PS component (see \autoref{sec:app4.1}), suggesting that dust emission is optically thin at these wavelengths. Emission at $6.9\,{\rm mm}$ (not considered in the fit) is slightly brighter than expected from $\alpha_{\rm 3.1-9.1mm}$ due to contamination by non-dust emission (expected to be $\approx20\%$ and thus consistent with the offset in \autoref{fig:2}, according to \citealt{Rodmann2006}, who attributed it to free-free emission). We stress that $\alpha_{\rm 3.1-9.1mm}$ is steeper in CI~Tau compared with the spectral indices at similar wavelengths in HL~Tau ($\alpha_{\rm 2.9-9.1mm}\approx3.4$, \citealt{Carrasco-Gonzalez2019}), HD~163296 ($\alpha_{\rm 3.2-9.7mm}\approx2.6$, \citealt{Guidi2022}), and TW~Hya ($\alpha_{\rm 3.1-9.3mm}\approx3.0$, \citealt{Menu2014,Macias2021}), that we do not expect to steepen substantially when the contribution of non dust emission is subtracted, since this was estimated to account for $20\%$ in TW~Hya \citep{Macias2021}, 10\% in HL~Tau \citep{Carrasco-Gonzalez2019}, and only $5\%$ in HD~163296 \citep{Guidi2022} at the longest wavelength in these ranges (9.1 to $9.7\,{\rm mm}$), compared to 50\% in CI~Tau.
The rather unusual nature of CI~Tau is confirmed by the $7.0\,{\rm mm}$ survey of \citet[][see also \citealt{Chung2025}]{Rodmann2006}: of a larger sample of 14 discs in Taurus only GM~Aur and DG~Tau~B have a spectral index steeper than 3.0 (although the contribution of non-dust emission to their centimetre-wavelength luminosity cannot be precisely assessed due to the lack of longer wavelength photometry for most of the sample).

Finally, between 2.0 and $5.0\,{\rm cm}$, the spectral index is difficult to determine due to the low S/N of our X and C~band data, as well as the flux density variability in Ku and (especially) X~band observations (highlighted by the smaller dots in \autoref{fig:2}, that display the maximum and minimum flux densities measured across different scans for 2.0 and $3.0\,{\rm cm}$ data; see \autoref{sec:app4} for more details). When taken at face value, the flux densities give $\alpha_{\rm 2.0-5.0cm}=1.8\pm0.3$, flatter than expected from dust emission. Indeed, under a conservative extrapolation from the VLA Ka~band flux density with slope $\alpha_{\rm 3.1-9.1mm}$, it can be seen that dust emission contributes less than $3\%$ at wavelengths of $2.0\,{\rm cm}$ and longer. The possible origins of this emission and its variability are discussed hereafter.

\subsection{Centimetre-wavelength emission}\label{sec:3.2}
A number of physical processes were invoked to explain the centimetre-wavelength continuum emission in excess to thermal dust emission from circumstellar dust \citep{Dulk1985,Gudel2002}.

Free electrons interacting with circumstellar ionised gas produce bremsstrahlung (or free-free) thermal radiation. While in the case of emission from a homogeneous plasma, the spectral index is either $\alpha=2.0$, in the optically thick limit, or $\alpha=-0.1$, in the optically thin one \citep{Mezger1967}, for an ionised, isothermal, spherically symmetric stellar wind, both for constant velocity and accelerated flows, emission scales with $\alpha=0.6$ \citep{Wright&Barlow1975,Panagia&Felli1975,Olnon1975}. Instead, for a wider class of collimated ionised winds, the spectral index can be $-0.1\leq\alpha\leq2.0$ \citep{Reynolds1986}. While jets are typical of young Class 0/I sources \citep{Anglada2018}, free-free emission was attributed to collimated outflows also in a handful of more evolved systems, such as AB~Aur \citep{Rodriguez2014} and GM~Aur \citep{Macias2016}, where elongated structures were detected in $3.0\,{\rm cm}$ images with 3 to $4\sigma$ significance. Optically thin free-free emission could well explain the slightly decreasing spectral index between the VLA Ku and (the visibility offset subtracted from the) Ka~band data, and would also be consistent with the PS contribution to the ALMA Band~3 data. However, even in the optically thick regime, free-free emission is not fully consistent with the slope of the flux density distribution at wavelengths longer than~$2.0\,{\rm cm}$.

Other than circumstellar plasma, it was proposed that thermal free-free radiation could also originate from EUV-driven photoevaporative winds or the bound hot disc atmosphere close to the star heated up by X-ray irradiation \citep{Pascucci2012,Owen2013}. At wavelengths shorter than $10.0\,{\rm cm}$, as is the case for our data, numerical models and radiative transfer calculations predict that free-free emission becomes optically thin, and its spectrum much flatter. As an example, for a $0.7M_\odot$ star and at $3.0\,{\rm cm}$, $\alpha\lesssim0.3$ and 0.7 for the EUV and X-ray irradiation models \citep{Owen2013}, both too flat to explain by themselves the spectral indices measured in CI~Tau between 2.0 and $5.0\,{\rm cm}$.

Electrons whirling around the stellar magnetic field lines radiate non-thermal gyromagnetic emission. Depending on the particle velocity (relativistic or not) and electron energy distribution (thermal or power-law) gyromagnetic emission is characterised by different spectral indices \citep{Dulk1985,Gudel2002}. Radiation from highly relativistic electrons (synchrotron emission) is well described by a power-law electron distribution and a spectral index $\alpha=2.5$ in the optically thick regime and $\alpha\leq-0.5$ in the optically thin regime. In the case of mildly relativistic electrons (gyrosynchrotron radiation), both thermal and power-law electron distributions are possible scenarios. In the former case, $\alpha=2.0$ in the optically thick limit and $\alpha=-8.0$ in the optically thin one, while in the latter, $\alpha=2.5$ in the optically thick limit and $\alpha\leq-0.6$ in the optically thin one \citep{Dulk&Marsh1982,Gudel2002}. Gyromagnetic radiation could broadly explain the excess radio emission between VLA Ku and C~band in CI~Tau, as well as the unresolved flux density offset in Ka~band, provided that a sharp transition between optically thick and thin emission occurs at about $1.0\,{\rm cm}$. However, its steep negative spectral index in the optically thin limit is not consistent with the unresolved flux density offset in ALMA Band~3 data.

It is well known that both free-free emission and gyromagnetic emission can be variable, but their characteristic timescales are expected to differ \citep{Lommen2009,Ubach2012,Ubach2017,Dzib2013,Dzib2015,Liu2014,Coutens2019,Curone2023}. While the former might change by 20 up to 40\% over a few weeks to years, the latter can vary by a factor of two or more over a timescale of minutes to days (e.g. \citealt{Ubach2012,Ubach2017} and references therein). In \autoref{sec:app4} we show how CI~Tau's continuum luminosity changes by $\lesssim30\%$ over a timescale of a few minutes to weeks in the Ku~band but is highly variable (by more than a factor of two) in the first $15\,{\rm min}$ of observations in the X~band, as expected from gyromagnetic emission. No information on C~band variability can be extracted due to the prohibitively low S/N of our data.

Therefore, we propose that optically thick gyrosynchrotron radiation most likely contributes to the excess emission (and connected variability) the most at wavelengths longer than $2.0\,{\rm cm}$. Here, upon gyrosynchrotron emission transitioning from being optically thick to thin, optically thin free-free radiation becomes the dominant mechanism of non-dust emission down to wavelengths as short as $3.1\,{\rm mm}$. Monitoring CI~Tau over weeks to months with deeper follow-up multi-wavelength (between $9.1\,{\rm mm}$ to $6.0\,{\rm cm}$) and multi-epoch observations is necessary to conclusively assess the origin of centimetre-wavelength continuum emission in this source.

\subsection{Spectral index radial profiles}\label{sec:3.3}

Our high-resolution observations can be used to study how the spectral index changes as a function of the disc radius. To do~so, we first subtracted the point-source visibility offset measured in \autoref{sec:app4.1} to the real part of the ALMA Band~3 (for~both XX and YY polarisations) and the VLA Ka~band (only for the RR and LL polarisations) visibilities. Then we imaged each dataset using two different synthesised beams: (i) we reconstructed the ALMA and VLA Ka~band observations using a circular beam with $0\farcs195$ radius. This conservative choice allowed us to recover the VLA Ka~band outer-disc emission displayed in \autoref{fig:1} with ${\rm S/N}>3$; (ii) we also imaged the ALMA data with a smaller, $0\farcs058\times0\farcs087$, ${\rm PA}=-33.20\,{\rm deg}$ beam. This is the best trade off between angular resolution and sensitivity for Band~7 data (those with shortest maximum baseline). In this case, we chose not to circularise the beam to keep the maximum angular resolution compatible with our data.

To minimise any uncertainty due to convolution in the image plane, we first CLEANed each dataset adopting different \texttt{uvtaper} parameters to reach a similar beam shape (i.e. beam axes within $\leq1.5\,{\rm mas}$ and position angles within $\leq0.02\,{\rm deg}$) to our target ones. Only then, these images were smoothed, using the \texttt{imsmooth} task, to our target beam sizes. The optimal tapering parameters were found by trial and error, simulating the \texttt{CASA} PSF reconstruction routine. For each target beam, we explored different combinations of the \texttt{robust} and \texttt{uvtaper} parameters, but no notable differences were found. Finally, these smoothed intensity maps were deprojected as explained in \autoref{sec:2.3} and their azimuthally averaged radial profiles were used to compute the spectral index radius by radius as
\begin{equation}\label{eq:3.1}
    \alpha_{\nu_1-\nu_2} = \dfrac{\log I_{\nu_1} - \log I_{\nu_2}}{\log \nu_1-\log \nu_2},
\end{equation}
where $I_\nu$ is the azimuthally averaged surface brightness profile at frequency $\nu$. We preferred this method over the alternative strategy of computing 2D spectral index maps from CLEAN images at different wavelengths using the \texttt{immath} task. In fact, when deprojected and azimuthally averaged, such maps generally provide much noisier spectral index radial profiles.

\begin{figure*}
    \centering
    \includegraphics[width=\textwidth]{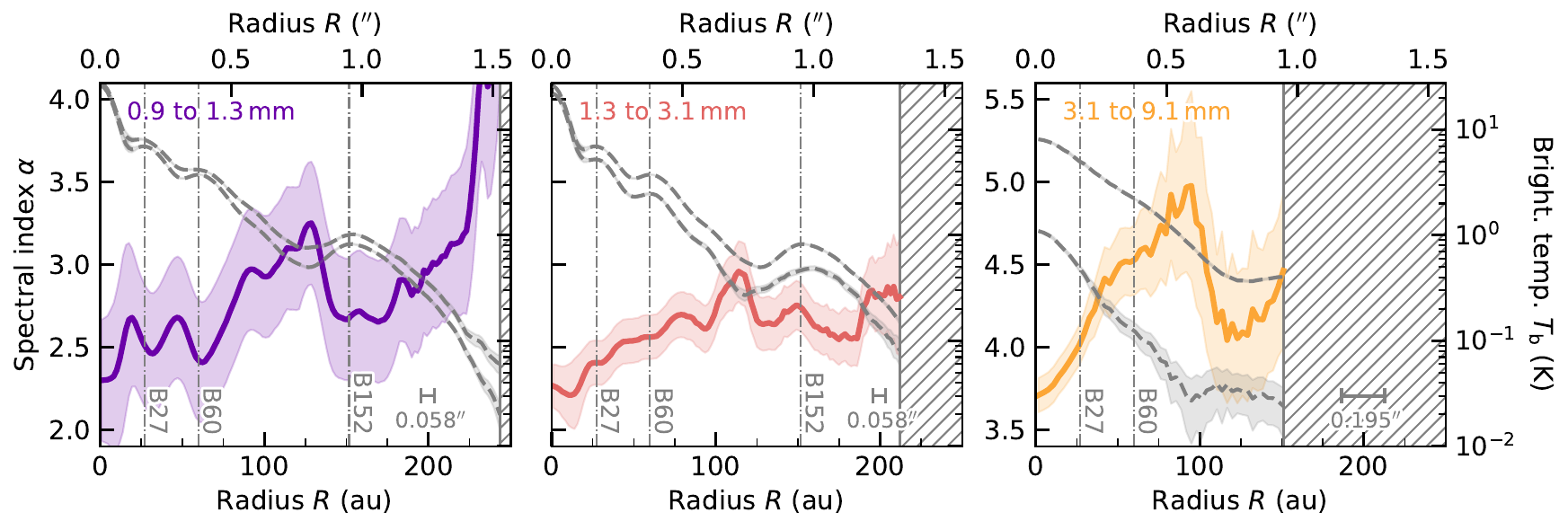}
    \caption{Spectral index radial profiles (solid lines) and their $1\sigma$ uncertainty (shaded areas). The hatched regions mark those locations where ${\rm S/N}\leq5$ (left and central panel) and 3 (right panel), for at least one of the emission profiles. The dashed grey lines in each panel display the surface brightness radial profiles combined to determine the spectral index.}
    \label{fig:3}
\end{figure*}

CI~Tau's spectral index profiles measured between pairs of progressively longer wavelengths are plotted as solid lines in \autoref{fig:3}. The shaded regions, instead, display their uncertainty, obtained propagating the quadrature sum of the error of the mean of each surface brightness radial profile and their absolute flux calibration uncertainty. The hatched regions mark the locations where ${\rm S/N}\leq5$ (left and central panel) and ${\rm S/N}\leq3$ (right panel) for at least one of the emission profiles. These profiles are plotted as dashed lines in the background. 

The spectral indices between ALMA Band~7 and 6, $\alpha_{\rm B7-B6}$, and between ALMA Band~6 and 3, $\alpha_{\rm B6-B3}$, data increase towards the outer disc, suggesting that emission is optically thinner at large radii or tracing radial variations of the dust opacity owing most likely to (a larger fraction of) smaller grains further away from the central star. At the position of the bright rings, marked by the vertical dashed-dotted lines and labelled as in \autoref{fig:1}, the spectral index radial profiles reach local minima, as expected in the case of more optically thick emission or the presence of (a larger fraction of) larger grains. This trend is more easily seen~for $\alpha_{\rm B7-B6}$ than $\alpha_{\rm B6-B3}$, whose profile shows much shallower modulations. This difference can be explained by: (i) the different ring-to-gap contrast in the surface brightness radial profiles at different wavelengths. Indeed, while this contrast increases between ALMA Band~7 and 6 observations, it is very similar for ALMA Band~6 and 3 data (see the grey dashed lines in \autoref{fig:3}). This trend can be interpreted as a consequence of emission being optically thin(ner) at 1.3 and $3.1\,{\rm mm}$. In this regime, the less substructured spectral index radial profile between ALMA Band~6 and 3 data suggests that only small modulations of the grain size radial profile at the location of gaps and rings might be expected; (ii) the different gap and ring centres in Band~6 and 3 data (e.g. the outer disc gap in the central panel of \autoref{fig:3}).

As already discussed in \autoref{sec:3.1}, compared with $\alpha_{\rm B7-B6}$ and $\alpha_{\rm B6-B3}$, the spectral index between ALMA Band~3 and VLA Ka~band data, $\alpha_{\rm B3-Ka}$, is larger (notice the different scale in the right panel of \autoref{fig:3}), suggesting that emission is optically thin at these wavelengths. Also $\alpha_{\rm B3-Ka}$ increases towards the outer disc, albeit only until the VLA Ka~band brightness profile flatten out due to its lower S/N. No modulations due to substructures in the emission profiles can be seen.

\section{Analysis method}\label{sec:4}
\subsection{Model}\label{sec:4.1}
The spectral dependence of dust emission can inform us about dust properties such as their temperature ($T_{\rm dust}$), density ($\Sigma_{\rm dust}$), size, composition, and morphology. To try and constrain (at least some of) these quantities, we fitted CI Tau's surface brightness profiles between 0.9 and $9.1\,{\rm mm}$, radius by radius, adopting a physical model that takes into account the effects of optical depth and dust scattering \citep[e.g.][]{Carrasco-Gonzalez2019,Macias2021,Sierra2021,Guidi2022}. According to this model, in the case of an azimuthally symmetric, vertically isothermal, and razor-thin disc, dust thermal emission from the disc mid-plane can be computed as \citep{Rybicki&Lightman1986,Miyake&Nakagawa1993,Carrasco-Gonzalez2019}
\begin{equation}\label{eq:4.1}
    S_\nu(\tau_\nu,\mu)=B_\nu(T_{\rm dust})\Bigl[1-\exp\left(-\tau_\nu/\mu\right)+\omega_\nu F(\tau_\nu,\omega_\nu,\mu)\Bigr],
\end{equation}
where $B_\nu(T_{\rm dust})$ is the black body emission at temperature $T_{\rm dust}$ and frequency $\nu$, $\tau_\nu=\Sigma_{\rm dust}\chi_\nu$ is the disc optical depth, $\chi_\nu=\kappa_\nu^{\rm abs}+\kappa_\nu^{\rm sca}$ is the total dust opacity, $\kappa_\nu^{\rm abs}$ and $\kappa_\nu^{\rm sca}$ are the absorption and scattering opacities, $\mu=\cos i$ is the disc inclination (that we fixed to the best-fit value of \citealt{Clarke2018}), $\omega_\nu=\kappa_\nu^{\rm sca}/\chi_\nu$ is the single-scattering albedo, and
\begin{equation}\label{eq:4.2}
    \begin{gathered}
        F(\tau_\nu,\omega_\nu,\mu) = \dfrac{1}{\exp\left(-\sqrt{3}\epsilon_\nu\tau_\nu\right)\left(\epsilon_\nu-1\right)-\left(\epsilon_\nu+1\right)}\times \\[2ex]
        \hspace{-15ex}\times\left[\dfrac{1-\exp\left(-\left(\sqrt{3}\epsilon_\nu+1/\mu\right)\tau_\nu\right)}{\sqrt{3}\epsilon_\nu\mu+1} + \right.\\
        \hspace{+20ex}\left. + \dfrac{\exp\left(-\tau_\nu/\mu\right)-\exp\left(-\sqrt{3}\epsilon_\nu\tau_\nu\right)}{\sqrt{3}\epsilon_\nu\mu-1}\right],
    \end{gathered}
\end{equation}
where $\epsilon_\nu=\sqrt{1-\omega_\nu}$. In the simplest possible assumption that the absorption and scattering opacities scale with frequency as a single-power law (i.e. $\kappa_\nu^{\rm abs}=\kappa_{\nu_0}^{\rm abs}\nu^{\beta^{\rm abs}}$ and $\kappa_\nu^{\rm sca}=\kappa_{\nu_0}^{\rm sca}\nu^{\beta^{\rm sca}}$, where $\kappa_{\nu_0}^{\rm abs}$ and $\kappa_{\nu_0}^{\rm sca}$ are normalisation coefficients), \autoref{eq:4.1} depends on six free parameters ($T_{\rm dust}$, $\Sigma_{\rm dust}$, $\kappa_{\nu_0}^{\rm abs}$, $\beta^{\rm abs}$, $\kappa_{\nu_0}^{\rm sca}$, and $\beta^{\rm sca}$). Thus, at least six datasets at different frequencies would be required to observationally constrain such parameters. Since this is currently not feasible in the case of CI~Tau, where only four high angular resolution and sensitivity datasets are available, we adopted a different strategy \citep[e.g.][]{Carrasco-Gonzalez2019,Macias2021}. Instead of fitting for the absorption and scattering opacities self-consistently, we assumed a bulk dust composition and computed its optical properties, hence $\kappa^{\rm abs}_\nu$ and $\kappa^{\rm sca}_\nu$, as a function of the maximum grain size ($a_{\rm max}$) and for a grain size power law distribution ($n(a)da\propto a^{-q}da$, where $n(a)$ is the dust number density in a small size interval $da$). Thus the number of free parameters was reduced from six to four ($T_{\rm dust}$, $\Sigma_{\rm dust}$, $a_{\rm max}$, and $q$). Adopting a set of dust bulk properties also allowed us to correct for the assumption of isotropic scattering in \autoref{eq:4.1}, compute the forward-scattering coefficient ($g_\nu$, \citealt{Henyey&Greenstein1941}), and rescale the scattering opacity as $\kappa_\nu^{\rm sca,eff}=(1-g_\nu)\kappa_\nu^{\rm sca}$.

In this scenario, the composition, porosity, and mixing rules adopted to determine the dust optical properties are essential. Indeed, it is well known that different opacities can lead to significantly different estimates of the temperature, density, and size of dust \citep[see e.g.][]{Banzatti2011,Ohashi2023,Sierra2025}. Most of the previously published analyses of high angular resolution multi-frequency continuum observations similar to ours considered grains to be compact and spherical and adopted the default dust mixture proposed by \citet{Birnstiel2018} as a reference for the DSHARP survey \citep{Andrews2018}. In this paper, instead, we chose a different fiducial composition: we considered dust to be made of 60\% water ice \citep{Warren&Brandt2008}, 30\% amorphous carbon (\citealt{Zubko1996}, ACH2 sample), and 10\% astronomical silicates \citep{Draine2003} by volume. This mixture differs from that proposed by \citet{Ricci2010} only for their higher porosity (30\%)\footnote{The volume fractions quoted here differ from those in \citet{Ricci2010} that contain some typos. We follow instead those adopted in the \texttt{dsharp\_opac} package \citep{Birnstiel2018}, by \citet{Banzatti2011} and \citet{Trotta2013}, using updated refractive indices to determine the optical properties of the mixture, with only marginal differences. See Part 3 of the reference notebook \href{https://github.com/birnstiel/dsharp_opac/blob/master/notebooks/opacity_examples.ipynb}{here}.}. For this reason, we refer to our fiducial opacities as `Ricci (compact)'. Although, at the current stage, our choice might seem arbitrary, in \autoref{sec:6} we extensively discuss how our results depend on the adopted bulk dust properties, exploring different compositions and porosity filling factors, and providing a justification for our fiducial opacities. Furthermore, in recent years indirect evidence in favour of mixtures rich in amorphous carbon, such as the `Ricci (compact)' one, came from demographic disc studies in nearby star-formation regions such as those modelling the size-luminosity correlation \citep{Rosotti2019,Zormpas2022} and the spectral index distribution \citep{Stadler2022,Delussu2024}. We used the \texttt{dsharp\_opac} package \citep{Birnstiel2018} to determine the dielectric constants of our fiducial mixture\footnote{For all the dust mixtures explored in this paper, we compared the opacity tables obtained with \texttt{dsharp\_opac} \citep{Birnstiel2018} and \texttt{optool} \citep{Dominik2021}, when the optical constants of the material components of a mixture were implemented in both codes. Their absorption and scattering opacities are in excellent agreement, when plotted as a function of wavelength, for all the different maximum grain sizes ($100\,\mu{\rm m}$, $1\,{\rm mm}$, and $1\,{\rm cm}$) and dust density distributions ($q=3.5$ and $2.5$) we tested.} from the refractive indices of the aforementioned materials adopting the Bruggeman rule \citep[][see Appendix~A of \citealt{Zhang2023} for a justification]{Bruggeman1935}, and its optical properties using Mie theory for spherical grains \citep{Bohren&Huffman1998}. 

\subsection{Fitting procedure}\label{sec:4.2}
We adopted a Bayesian approach and used \texttt{emcee} \citep{Foreman-Mackey2013,Foreman-Mackey2019}, a pure-Python implementation of the affine-invariant Markov chain Monte Carlo (MCMC) ensemble sampler of \citet{Goodman&Weare2010} to estimate the posterior distribution of the model parameters at each radius. Our log-likelihood function reads as
\begin{equation}
    \ln p\left(I_\nu\middle|\vec{\theta}\right)=-\dfrac{1}{2}\sum_{\nu}\left[\left(\dfrac{I_\nu - S_\nu}{\sigma_{\nu,{\rm tot}}}\right)^2 + \ln(2\pi\sigma_{\nu,{\rm tot}}^2)\right],
\end{equation}
where $\vec{\theta}=\{T_{\rm dust},\Sigma_{\rm dust},a_{\rm max},q\}$ is the parameter vector, $I_\nu$ is the azimuthally averaged dust surface brightness radial profile at frequency $\nu$ and radius $R$ (from the observations), $S_\nu$ is the model surface brightness radial profile (from \autoref{eq:4.1}), and
\begin{equation}
    \sigma^2_{\nu,{\rm tot}} = \sigma_\nu^2 + (\delta I_\nu)^2,
\end{equation}
where $\sigma_\nu$ is the uncertainty of the azimuthally averaged brightness profile at frequency $\nu$ and radius $R$, while $\delta I_\nu$ indicates the systematic flux calibration uncertainty. Following the recommendations of the ALMA Technical Handbook\footnote{See Sect.~10.2.6 of the \href{https://almascience.nrao.edu/proposing/technical-handbook/}{ALMA Technical Handbook}.} and the Guide to Observing with the VLA\footnote{See Sect.~6.1 of the \href{https://science.nrao.edu/facilities/vla/docs/manuals/obsguide/topical-guides/hifreq/}{Guide to Observing with the VLA} and Sect.~3.11 of the \href{https://science.nrao.edu/facilities/vla/docs/manuals/oss/performance/fdscale/}{VLA Observational Status Summary}.}, we assumed $\delta=10\%$ for ALMA Band~7, 6, and the VLA Ka~band observations, and $\delta=5\%$ for the ALMA Band~3 data.

We benchmarked our fitting procedure against the results of \citet{Macias2021} in TW~Hya, using their publicly available self-calibrated high angular resolution multi-frequency continuum observations, the same dust composition, model priors, and parameter ranges. Our results are in excellent agreement with their marginalised posterior distributions for each of the fitting parameters, both in the case of a fixed ISM-like particle size distribution and when $q$ is allowed to change.

To assess convergence we estimated the integrated autocorrelation time, $\tau_{\rm e}$, using the method of \citet{Sokal1997}. Since $\tau_{\rm e}$ often increases with the number of MCMC steps, we first adopted a restrictive convergence criterion: we considered our chains to be converged if they are longer than 100 times the maximum estimated integrated autocorrelation time for each parameter, $\max(\tau_{\rm e})$, and if this quantity changes by less than 1\% over the last $10^3$ steps. To speed-up convergence, we first fitted \autoref{eq:4.1} to our spectral flux density distribution, using the Trust Region Reflective minimisation algorithm for bound problems implemented in \texttt{scipy.optimize.curve\_fit} and initialised our walkers in a 4D sphere in the parameter space around these best-fit values. Moreover, instead of the default `stretch move' of \citet{Goodman&Weare2010}, we adopted a weighted mixture of the `differential evolution proposal' \citep{TerBraak2006} and the `snooker proposal' \citep{TerBraak&Vrugt2008} moves, randomly selected with 80\% and 20\% probability, achieving a twice as fast convergence.

We tested this convergence criterion using TW~Hya data for $R=25$, 30, and $45\,{\rm au}$ (because at these locations our posterior distributions can be compared with the corner-plots published by \citealt{Macias2021}). When exploring the parameter space with $10^2$ walkers, this criterion requires chains longer than $1.6\times10^4$ steps and gives $\max(\tau_{\rm e})\leq1.4\times10^2$. Since this is impractically long to run for hundreds of fits, for the rest of the paper we used $10^2$ walkers and $10^3$ steps, $\approx7.5$ times longer than $\max(\tau_{\rm e})$, and sampled the posterior distribution function discarding, on average, the initial $2\ {\rm to}\ 4\times10^2$ `burn-in' steps (corresponding to five times the maximum estimated integrated autocorrelation time of these shorter chains). This choice gives results in remarkable agreement with those obtained with our more restrictive convergence criterion and the corner plots of \citet{Macias2021}. Our acceptance fractions were, on average, $\leq0.3$.

\subsection{Two-step methodology}\label{sec:4.3}
The main issue to deal with when fitting \autoref{eq:4.1} to the data is the lower S/N of the VLA Ka~band observations compared to the ALMA ones. Fitting all our high-resolution data together (i.e. data at the three ALMA bands and the VLA Ka~band) requires either (i) adopting a large-enough beam to recover the outer disc emission in the VLA Ka~band data, thus giving up the possibility of constraining the properties of dust on the scale of the gaps and rings in the system, or (ii) considering only the higher quality ALMA data. However, this generally leads to a bimodal dust size distribution and generally poorer constraints on dust properties (as was discussed by e.g. \citealt{Macias2021,Sierra2021,Ueda2022}). To take the best out of these two scenarios (tight constraints on dust properties from the VLA data, on the scale of gaps and rings in the ALMA images) we adopted a different method.

We first considered the low angular resolution surface brightness radial profiles (i.e. those reconstructed with a $0\farcs195$ circular beam) and fitted \autoref{eq:4.1} to the spectral flux density distribution from $0.9\,{\rm mm}$ to $9.1\,{\rm mm}$ in each radial bin, $I_\nu(R_i+\Delta R)$, independently (i.e. averaging the emission intensity in each pixel of width $\Delta R = 19.5\,{\rm mas} \approx 3.1\,{\rm au}$, for a progressively larger radius $R_{i+1}=R_i+\Delta R$). We adopted uninformative priors for all the parameters, except the temperature. In this case, following \citet{Macias2021}, our prior is based on the expected temperature profile of a passively irradiated disc as summarised in \autoref{sec:app5}. The other parameters were free to vary in the following ranges: $-3\leq\log(\Sigma_{\rm dust}/{\rm g}\,{\rm cm}^{-2})\leq3$, $-3\leq\log(a_{\rm max}/{\rm cm})\leq3$, and $1\leq q\leq4$. The posterior distributions of this `low resolution' fit were then used as priors for a `high resolution' fit to the ALMA-only surface brightness radial profiles reconstructed with a $0\farcs058\times0\farcs087$, ${\rm PA} = -33.20\,{\rm deg}$ beam. We fitted \autoref{eq:4.1} to the spectral flux density distribution from $0.9\,{\rm mm}$ to $3.1\,{\rm mm}$ in each radial bin (with $\Delta R = 11.6\,{\rm mas} \approx 1.9\,{\rm au}$) independently.

To do so, for each selected radius in the high resolution profiles, we first identified the closest (`reference') radial bin in the low resolution profiles. Then, we estimated non-parametrically the probability density function of the marginalised posterior distribution of each parameter fitting the posterior samples using the \texttt{scipy.stats.gaussian\_kde} function \citep{Virtanen2020}. We only fitted the ($>7\times10^3$) samples left after discarding the first $5\times\max(\tau_e)$ `burn-in' steps and thinning the sample set by a factor of $0.5\times\max(\tau_e)\approx15$. Finally, for every MCMC step and fitting parameter, $\theta_i$, we determined the marginalised prior through a one-dimensional piecewise linear interpolation to the KDE marginalised posterior probability density function, $p_{\rm KDE}(\theta_i|I_\nu)$. Our final prior reads as
\begin{equation}
    \begin{gathered}
        \hspace{-10ex}\ln p(\vec{\theta}) = \ln p(T_{\rm dust}) +\\
        \hspace{5ex} + \ln p_{\rm KDE}(\Sigma_{\rm dust}|I_\nu) + \ln p_{\rm KDE}(a_{\rm max}|I_\nu) + \ln p_{\rm KDE}(q|I_\nu),
    \end{gathered}
\end{equation}
where $p(T_{\rm dust})$ is the temperature prior from \autoref{sec:app5}, while $I_\nu$ indicates the $0\farcs195$ resolution ALMA and VLA Ka~band surface brightness profiles at the reference radius. We initialised our walkers around the median of the posterior distribution of the low resolution fit in the same reference radial bin to achieve a faster convergence. While writing this paper, we became aware that a similar double-fit analysis method had been independently developed by \citet{Viscardi2025}, who also benchmarked it against collisional growth and dust transport \texttt{DustPy} \citep{Stammler&Birnstiel2022} models. The analysis of \citet{Viscardi2025} lends further support to the validity of our results and we refer to their paper for further insights and validation.

\subsection{Cross-composition comparison}\label{sec:4.4} 
We explored the dependence of our results on dust composition by fitting CI~Tau's low-resolution ($0\farcs195$ circular beam) radial profiles with a range of different optical properties and comparing their posterior distributions.

In Bayesian statistics, two models, $m_1$ and $m_2$, can be compared computing the so-called Bayes factor. This is the ratio of the marginalised likelihoods, $p(I_\nu|m_1)p(m_1)$ and $p(I_\nu|m_2)p(m_2)$, where
\begin{equation}
    p(I_\nu|m_i)=\int p(I_\nu|\vec{\theta},m_i)p(\vec{\theta}|m_i)d\vec{\theta},
\end{equation}
for $i=1,2$, is the posterior normalisation or model evidence. Unfortunately, \texttt{emcee} does not provide such a marginalised likelihood and estimating it from the posterior samples is non-trivial \citep[e.g.][]{Hou2014}. For this reason, we adopted the $\chi^2$ test as a measure of the goodness of our fit and postponed a more thorough and fully Bayesian analysis to a future paper. Since models with the same number of free parameters are compared, we expect this metric to be a sufficiently robust indicator of the quality of our models.

For each adopted composition, we provide two estimates of the reduced $\chi^2$, based on the surface brightness and spectral index radial profiles. As for the surface brightness, first, for each radius, we used our MCMC samples to determine the posterior distributions of the model intensity, $S_\nu$, and estimated the reduced $\chi^2$ posterior distribution as
\begin{equation}
    \chi^2_{\rm SB}=\dfrac{1}{4}\sum_{\nu}\left(\dfrac{I_\nu-S_\nu}{\sigma_{\nu,{\rm tot}}}\right)^2,
\end{equation}
where the sum runs over all the four frequencies our data were taken at, $I_\nu$ and $\sigma_{\nu,{\rm tot}}$ have the same meaning as in \autoref{sec:4.2}, while $S_\nu$ is a model realisation from each sample of the parameter posteriors. We then determined a disc-averaged $\chi^2$ posterior distribution concatenating the last $n$ samples of the $\chi^2_{\rm SB}$ posterior distribution for each radius smaller than some maximum disc size $R_{\rm out}$. Since the number of available $\chi^2_{\rm SB}$ samples is radius-dependent, we chose $n$ as the minimum sample length among all radii smaller than $R_{\rm out}$. As for $R_{\rm out}$, instead, we adopted three different definitions: $R_{\rm out}\approx200\,{\rm au}$, the maximum radius we could fit our data for (at larger radii the low surface brightness S/N leads to unphysical results), $R_{\rm out}=151$ and $82\,{\rm au}$, the disc locations where the S/N of our VLA Ka~band and brightness profile drops below 3 and 5, respectively.

We adopted a similar strategy for the $\chi^2$ posterior distribution estimated from the spectral index radial profiles. In this case, at a given radius, the reduced $\chi^2$ reads
\begin{equation}
    \chi^2_{\rm SI}=\dfrac{1}{3}\sum_{i}\left(\dfrac{\alpha_{i,{\rm obs}}-\alpha_{i,{\rm mod}}}{\sigma_{\alpha_i,{\rm tot}}}\right)^2,
\end{equation}
where the subscripts `obs' and `mod' stand for the observed spectral index and a model realisation from each sample of the parameter posteriors, while the index $i$ runs over pairs of subsequent frequencies (i.e. $\alpha_{\rm B7-B6}$, $\alpha_{\rm B6-B3}$, and $\alpha_{\rm B3-Ka}$). $\sigma_{\alpha_i,{\rm tot}}$, instead, is the observed spectral index uncertainty, estimated propagating the errors on each surface brightness radial profile, that also include the absolute flux calibration uncertainty.

\section{Results}\label{sec:5}

\begin{figure*}
    \centering
    \includegraphics[width=\textwidth]{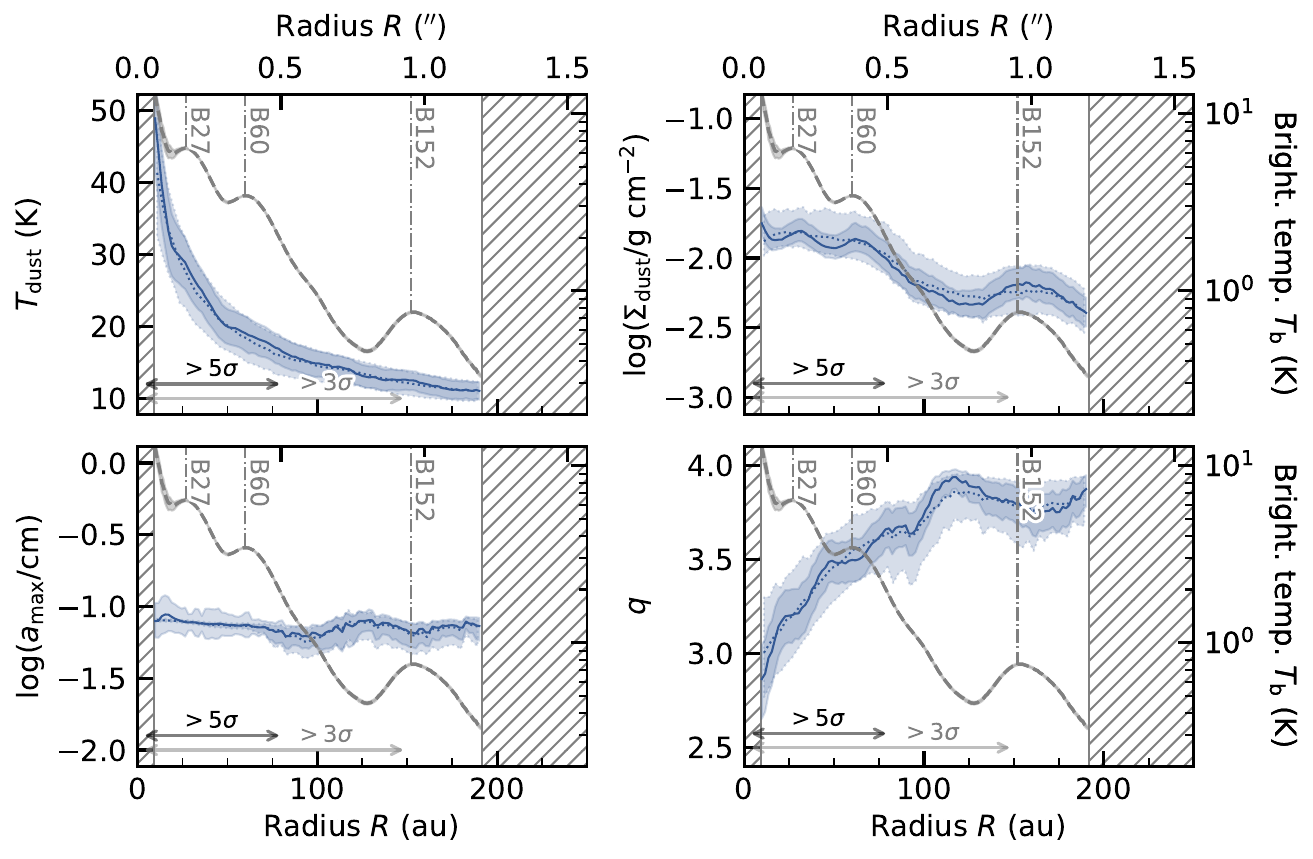}
    \caption{Dust temperature, density, maximum size, and density distribution power-law index posterior distributions for our fiducial `Ricci (compact)' composition. The blue solid lines and shaded regions display the median and $1\sigma$ uncertainty of each parameter. For comparison, the dotted lines show the results of the low resolution fit. The hatched areas mark the disc region within the synthesised beam minor axis ($0\farcs058$) and, in the outer disc, the region where our fit does not provide robust results due to the low S/N of the data. The grey dashed line displays our ALMA Band~6 surface brightness radial profile (at a resolution of $0\farcs058$), the bright ring position is indicated and labelled as in the previous plots. The black and grey horizontal arrows mark the regions where the S/N of VLA Ka~band surface brightness radial profile is $>5$ and 3.}
    \label{fig:4}
\end{figure*}

\subsection{Best-fit radial profiles, fiducial case}\label{sec:5.1}
Our results are shown in \autoref{fig:4}, while we defer to \autoref{sec:app6} a discussion on the goodness of our fit. The blue solid lines and shaded regions display the best-fit profiles and their $1\sigma$ uncertainty (defined as the median and the area between the 16\textsuperscript{th} and 84\textsuperscript{th} percentile of the marginalised posterior distribution of each parameter). For the sake of comparison, the median and $1\sigma$ spread are plotted also for the low resolution fit as dotted lines and paler shaded areas of the same colour. As can be seen, the low resolution fit is able to constrain remarkably well the four fitting parameters. However, their radial profiles are only sensitive to global variations and low-amplitude modulations at the location of the outer ring (the only substructure that is marginally resolved with an angular resolution of $0\farcs195$). Thanks to their smaller beam, the high resolution best-fit radial profiles, instead, show clear modulations at the position of all the rings (see e.g. the ALMA Band~6 surface brightness radial profile over-plotted as a grey dashed line, where the ring positions are indicated by the vertical dashed-dotted lines and labelled as in \autoref{fig:1}). 

Starting from the top left panel, the best-fit temperature profile reflects our passively irradiated disc prior and shows no clear substructures. Instead, the dust surface density radial profile locally peaks at the position of the bright rings. In particular, the $27\,{\rm au}$, $60\,{\rm au}$, and $152\,{\rm au}$ rings are $9\%$, $14\%$, and $37\%$ denser than their preceding gaps. Similar features were already observed in some rings, for example, in HL~Tau \citep{Carrasco-Gonzalez2019,Guerra-Alvarado2024}, TW~Hya \citep{Macias2021}, HD~163296 \citep{Guidi2022}, and the MAPS discs \citep{Sierra2021}. We do not detect the sharp dust surface density outer edge predicted by \citet{Birnstiel&Andrews2014} to be a signpost of radial drift. However, this is likely due the low S/N of our data in the outer disc ($R\gtrsim200\,{\rm au}$). Our best-fit maximum grain size radial profile is flat within the uncertainties and $a_{\rm max}\approx7.4\times10^{-2}\,{\rm cm}$. Moreover, it shows no clear substructures, with the exception of a marginal ($11\%$) increase at the position of the outer gap. However, this modulation lies well within the uncertainties and is not statistically significant. Indeed, we do not consider it to be physical, but an artefact of the fit, owing to the low ${\rm S/N}\gtrsim3$ of the VLA Ka~band radial profile at this location (see the black and grey arrows at the bottom-left corner of each panel that mark the detection significance of the VLA Ka~band surface brightness radial profile). Deeper data are needed to confirm this hypothesis. The absence of substructures in the maximum grain size radial profile is also a common feature to many of the sources targeted at multiple wavelengths and high angular resolution by ALMA \citep[see e.g.][]{Jiang2024}. Finally, our density distribution power-law index radial profile increases from $q\approx3.0$ in the inner disc to $q>3.5$ in the outer disc, with local troughs at the position of the bright rings, similarly to the case TW~Hya \citep{Macias2021}.

\subsection{Optical properties, fiducial case}\label{sec:5.2}

\begin{figure*}
    \centering
    \includegraphics[width=\textwidth]{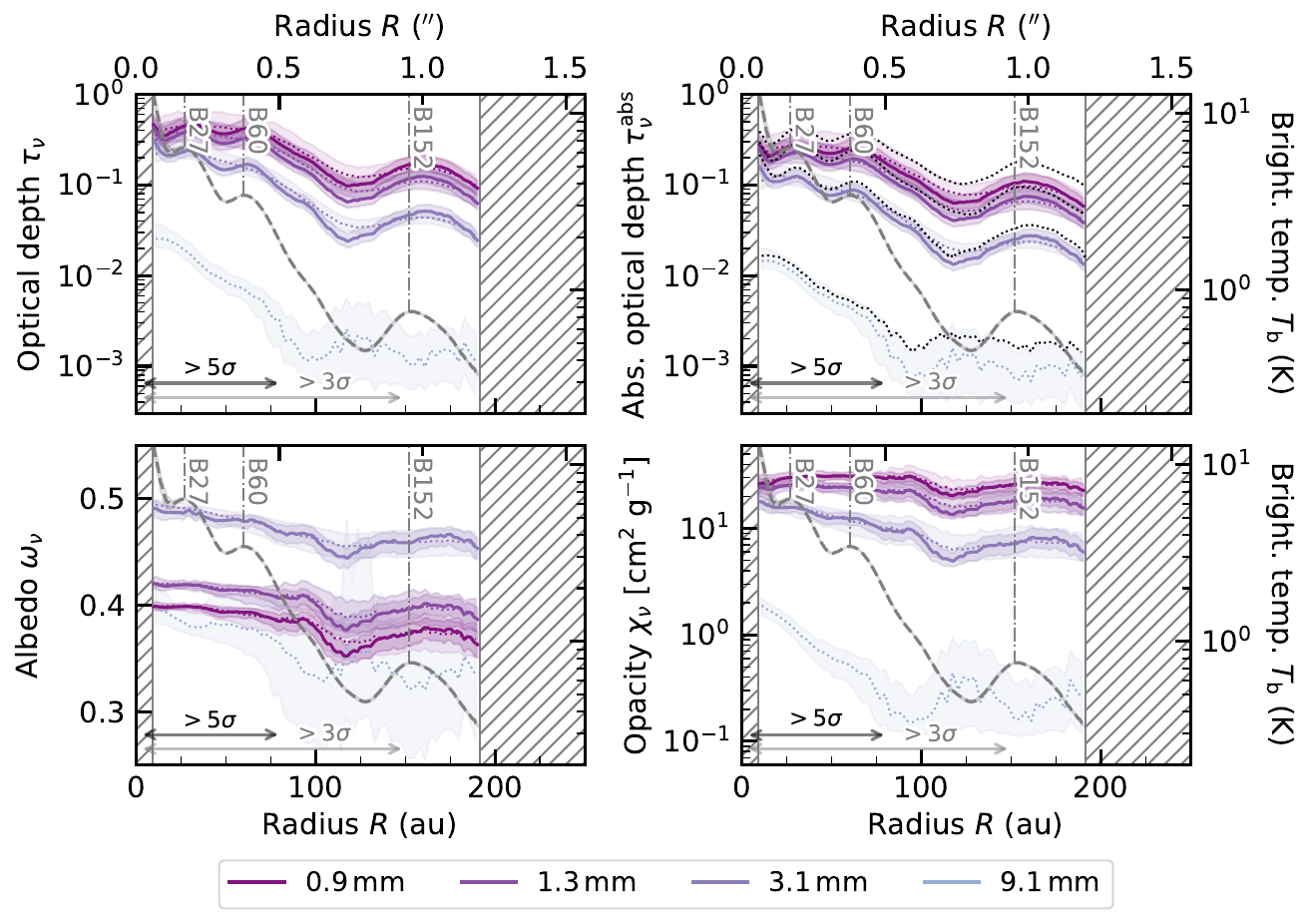}
    \caption{Optical depth, absorption optical depth, albedo, and total dust opacity posterior distributions at 0.9, 1.3, 3.1, and $9.1\,{\rm mm}$ for our fiducial `Ricci (compact)' composition. The solid lines and shaded regions display the median and $1\sigma$ uncertainty of each parameter. For comparison, the dotted lines show the results of the low resolution fit. The black dotted lines display the absorption optical depth estimated at each wavelength separately in the optically thin approximation from \autoref{eq:5.1}. Our ALMA Band~6 surface brightness radial profile is plotted with grey dashed lines. , the bright ring position is indicated and labelled as in the previous plots. The black and grey horizontal arrows mark the regions where the S/N of VLA Ka~band surface brightness radial profile is $>5$ and 3. Dust emission is marginally optically thin at all the wavelengths.}
    \label{fig:5}
\end{figure*}

\autoref{fig:5} shows the best-fit radial profiles and the $1\sigma$ uncertainties of the dust optical properties, computed from the posterior distributions discussed in \autoref{sec:4.1}. The solid lines display the ALMA-only high resolution fit results, while the dotted ones are used for the combined ALMA and VLA Ka~band low resolution fit, colour-coded by the reference wavelength of our datasets.

First and foremost, the upper-left panel shows that the optical depth is moderate to low between 0.9 and $9.1\,{\rm mm}$ ($\tau_\nu\lesssim1$), suggesting that CI~Tau's continuum emission is (marginally) optically thin at all the radii at these wavelengths. Our low and high resolution fits display similar trends, with $\tau_\nu$ decreasing towards the outer disc and longer wavelengths. Such a decrease becomes even sharper when the disc radius or the observational wavelength increases, suggesting that dust emission is progressively optically thinner in the outer disc and at longer wavelengths. The results of the high resolution fit also show that $\tau_\nu$ locally increases at the position of the bright rings. These profiles closely match the spectral index behaviour discussed in \autoref{sec:3.3}.

The upper-right panel displays the absorption optical depth, $\tau_\nu^{\rm abs}=\Sigma_{\rm dust}\kappa_\nu^{\rm abs}$. On top of the results from our high and low resolution fits, the black dotted lines show the absorption optical depth profiles inferred from each dataset separately in the optically thin limit \citep[see e.g.][]{Huang2018,Dullemond2018} as
\begin{equation}\label{eq:5.1}
    \dfrac{I_\nu}{B_\nu}=1-\exp\left(-\dfrac{\tau^{\rm abs}_\nu}{\cos i}\right)\rightarrow\tau^{\rm abs}_\nu=-\cos i\ln\left(1-\dfrac{I_\nu}{B_\nu}\right),
\end{equation}
where the dust temperature was estimated as in \autoref{eq:app5.1} fixing $\phi=0.02$ and $L_\star=1.04\,L_\odot$. Unsurprisingly, since continuum emission is (marginally) optically thin, the absorption optical depth profiles inferred from \autoref{eq:5.1} are in excellent agreement (less than a factor of two off) with those estimated from the posterior distribution of our multi-frequency analysis. By direct comparison with the profiles plotted in the upper-left panel, it can be seen that absorption contributes by about 50\% (ALMA Band~3) to 60\% (ALMA Band~7, 6, and VLA Ka~band) to the total dust opacity. This is a consequence of our fiducial dust composition, whose single-scattering albedo, plotted in the bottom-left panel, is by construction $\lesssim0.5$ at all the wavelengths.

CI~Tau's low optical depth stands out when compared with that of the other sources targeted in previous multi-frequency high resolution studies. In particular, only GM~Aur \citep{Sierra2021} shows a comparably low optical depth at (sub-)millimetre wavelengths. All the other sources, instead, are optically thick down to a few millimetres (ALMA Band~3 or 4) either in the inner 20 to $50\,{\rm au}$, such as TW~Hya \citep{Macias2021}, HD~163296 \citep{Sierra2021,Guidi2022}, AS~209 and MWC~480 \citep{Sierra2021}, or at all the radii, such as HL~Tau \citep{Carrasco-Gonzalez2019,Zhang2023} and IM~Lup \citep{Sierra2021}. Although this comparison is most likely affected by the different spectral coverage of these datasets and the assumptions on composition adopted in these works, we stress that our conclusion that CI~Tau is optically thin at (sub-)millimetre to centimetre wavelengths is robust: with the exception of highly porous material, all the different compositions tested in \autoref{sec:6} consistently recover optical depths $\tau_\nu\lesssim1$ across all the modelled datasets and the same $\tau_\nu^{\rm abs}$ (see \autoref{sec:app7}). Moreover, at least in those discs, HL~Tau \citep{Carrasco-Gonzalez2019,Zhang2023} and HD~163296 \citep{Guidi2022}, where high-resolution centimetre-wavelength data are available, the much shallower spectral index between $2.1/3.0\,{\rm mm}$ and $8.0/9.1\,{\rm mm}$ than in CI~Tau indicates a genuinely higher optical depth (unless the dust composition differs substantially among these source).

At ALMA wavelengths the (absorption) optical depth trends displayed in \autoref{fig:5} are primarily determined by the shape of the dust surface density. Indeed, as can be noticed in the lower right panel, the total dust opacity hardly changes with radius for $\lambda\leq3.1\,{\rm mm}$. This is due to the maximum grain size radial profile being flat (see \autoref{fig:4}) and the weak dependence of $\chi_\nu$ on $q$ (less than a factor of three at most radii). At $9.1\,{\rm mm}$, instead, since our inferred maximum grain size falls around the Mie interference at $a_{\rm max}\approx\lambda/2\pi$, the total dust opacity rapidly decreases with $q$ (a factor of ten for $3.0\leq q\leq4.0$) and the opacity radial variations affect the shape of the optical depth radial profile the most. These results highlight the importance of the Mie resonance in determining the optical depth as a function of wavelength. While both \citet{Rosotti2019_cliff} and \citet{Tazzari2021} also emphasised the key role of the `opacity cliff' at the Mie resonance, they did it in the context of models where the grain size was a function of radius and where the disc outer radius was therefore set by the location in the disc where the resonance was attained at each wavelength. Instead, in our analysis, where the grains size turns out to be almost constant, there is little variation in opacity with radius, but a pronounced reduction in opacity at wavelengths longward of the Mie resonance (i.e. $\lambda\gtrsim9.1\,{\rm mm}$). A similar argument can also explain the wavelength-dependence of the single-scattering albedo, while its weak radial variation is due mostly to our flat maximum grain size profile.

\subsection{Result comparison for different dust mixtures}\label{sec:5.3}
\begin{figure*}
    \centering
    \includegraphics[width=\textwidth]{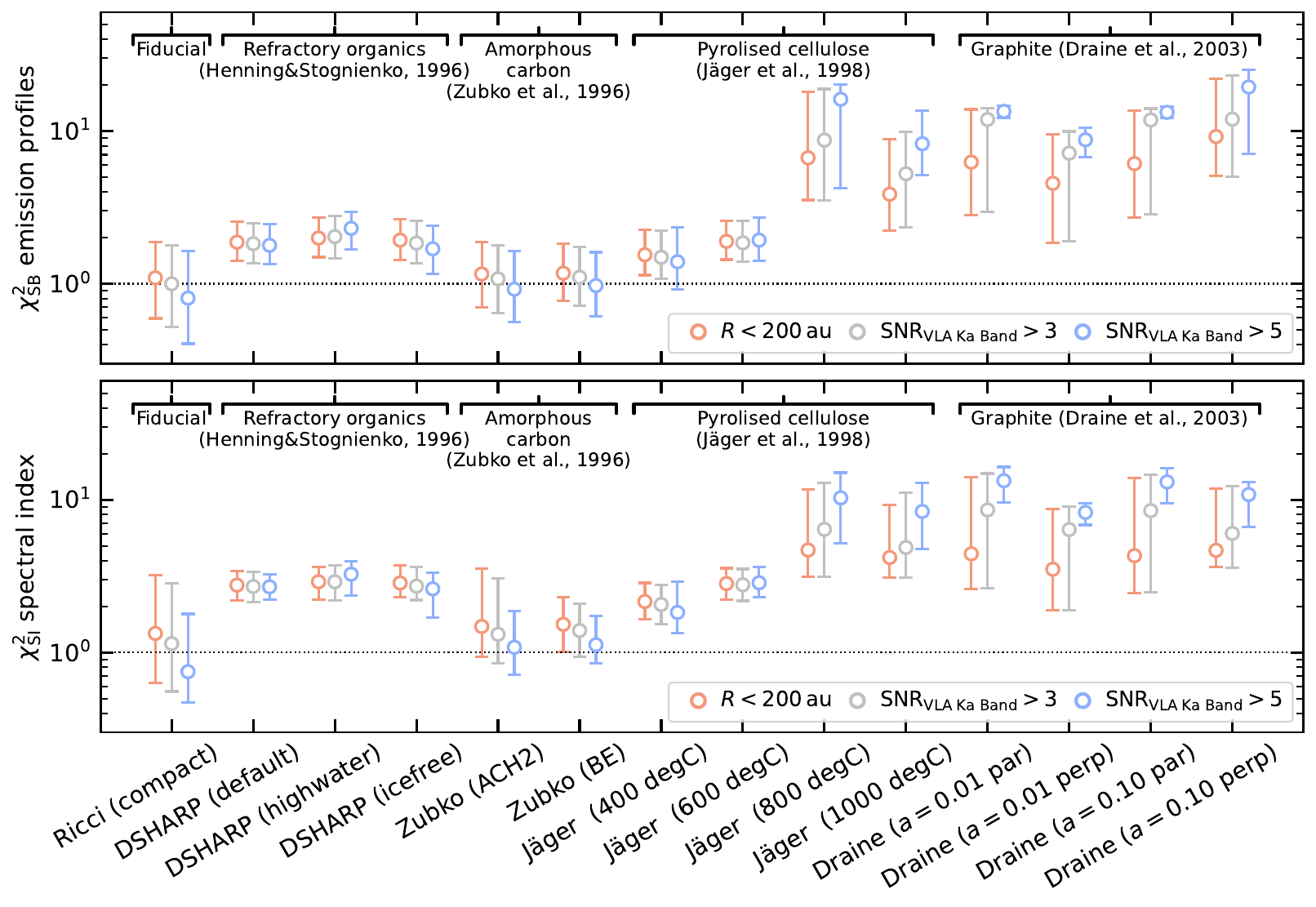}
    \caption{Reduced $\chi^2$ between our model posteriors and the observed surface brightness (upper panel) or spectral index (bottom panel) radial profiles for different dust compositions, colour-coded by the outermost disc radius the $\chi^2$ was averaged over. Mixtures including amorphous carbonaceous grains (e.g. `Ricci (compact)' and `Zubko') can fit our data the best. No constraints on the relative abundance of organics and other materials (especially water ice and silicates) can be drawn.}
    \label{fig:6}
\end{figure*}

To test the dependence of our results on the adopted dust mixture, we fitted CI~Tau's low resolution surface brightness profiles as in \autoref{sec:4.3}, \textit{ex novo} adopting different sets of optical properties, generated using the \texttt{dsharp\_opac} package \citep{Birnstiel2018}. Grains were assumed to be compact and spherical, their mixed dielectric constants were computed following the Bruggeman rule and their optical properties using Mie theory. To avoid artefacts due to poorly fitted temperature profiles, we fixed $T_{\rm dust}$ to be the temperature of a passively irradiated disc (as from \autoref{eq:app5.1}) with flaring angle and luminosity matching the best-fit results of our fiducial case (i.e. $\phi=0.035$ and $L_\star=1.04\,L_\odot$). We evaluated the quality of these fits by comparing the reduced $\chi^2$ between our models and the surface brightness and spectral index radial profiles following the procedure introduced in \autoref{sec:4.4}. \autoref{fig:6} displays the median and $1\sigma$ uncertainty (defined as the range between the 16\textsuperscript{th} and 84\textsuperscript{th} percentiles) of the reduced $\chi^2$ posterior distributions for the different dust compositions we tested, colour-coded by the outermost disc radius $\chi^2_{\rm SB}$ (upper panel) and $\chi^2_{\rm SI}$ (bottom panel) were averaged over. \autoref{tab:app5} summarises the mixtures and dielectric constants considered. 

First of all, for our fiducial dust composition, \autoref{fig:6} shows that $\chi^2\approx1$ with a small uncertainty ($\lesssim1$ for $R\lesssim82\,{\rm au}$), thus confirming (see \autoref{sec:app6}) that `Ricci (compact)' grains can fit remarkably well CI~Tau's surface brightness and spectral index radial profiles at most radii. We compared our fiducial results with those obtained using different dust optical properties, starting from the default mixture adopted by the DSHARP collaboration \citep{Andrews2018,Birnstiel2018}, labelled `DSHARP (default)'. In this case, while the brightness profiles are fitted reasonably well through all the disc ($\chi^2_{\rm SB}\approx2$), $\alpha_{\rm B7-B6}$ is systematically underestimated, leading to a larger $\chi^2_{\rm SI}\approx3$ for the spectral index radial profiles. This difference from our fiducial results does not depend on the different fraction of water ice adopted in the two mixtures, that only marginally affects dust optical properties (as already noticed by \citealt{Birnstiel2018}). Indeed, increasing the dust water ice content to 60\% by volume (as in the `Ricci (compact)' composition) or removing it altogether (compositions labelled `DSHARP (highwater)' and `DSHARP (icefree)', respectively), no clear difference from the `DSHARP (default)' results can be seen in \autoref{fig:6}, neither for the surface brightness, nor for the spectral index radial profiles. 

Instead, the different organic materials (hence dielectric constants) adopted by \citet{Ricci2010} and \citet{Birnstiel2018} impact our results the most. Depending on their synthesis temperature (e.g. \citealt{Jager1998}), interstellar carbonaceous material can come in a mixture of different hybridisation states, such as sp$^2$ (where sheets of carbon atoms are formed, as in graphite), sp$^3$ (where carbon atoms are arranged in tetrahedral geometry, as in diamonds), or even mixed states (where curved structures, such as fullerene, are abundant). The relative fraction of these hybridisation states and the amount of hydrogenation heavily affect the dielectric constants of carbon grains (e.g. \citealt{Zubko1996}), with substantial effects on their opacities. To test the sensitivity of our results to these properties, following \citet{Birnstiel2018}, we generate optical constants for dust mixtures with the same volume fractions as the `DSHARP (default)' ones, but replacing the refractory organics with different carbonaceous materials\footnote{See Appendix~B of \citet{Birnstiel2018}, reference notebook \href{https://github.com/birnstiel/dsharp_opac/blob/master/notebooks/AppendixB.ipynb}{here}.}. First, we tested the dielectric constants of amorphous carbon grains produced by arc discharge in hydrogen atmosphere (ACH2-sample) or burning benzene (BE-sample) proposed by \citet{Zubko1996}, labelled `Zubko (ACH2)' and `Zubko (BE)', respectively. In both cases, \autoref{fig:6} shows that our models are in excellent agreement with the data, with $\chi^2\approx1$ and a small spread ($\lesssim1$ for $R\lesssim82\,{\rm au}$) both for the surface brightness and spectral index radial profiles. Noticeably, as reported in \autoref{sec:4.1} and \autoref{tab:app5}, also our fiducial `Ricci (compact)' opacities were generated using the dielectric constants of \citet{Zubko1996} for ACH2 amorphous carbon, further strengthening the evidence that amorphous carbon grains can fit our data remarkably well. The different fractions of mixing materials adopted in the `Ricci (compact)' and `Zubko (ACH2)' compositions, instead, suggests that (i) small fractions of troilite do not substantially affect our results, and (ii) our observations cannot discriminate between mixtures of the same materials in different fractions (within $\lesssim15\%$), among those tested in \autoref{fig:6} (see \autoref{tab:app5}). 

Instead, when the carbonaceous materials in the mixtures are made mostly by graphite, our models provide poorer fits to the data. This can be seen in \autoref{fig:6} when the refractory organics \citep{Henning&Stognienko1996} in the `DSHARP (default)' mixture are replaced by carbonaceous materials synthesised by pyrolising cellulose at progressively higher temperatures (\citealt{Jager1998}, labelled `J\"{a}ger' followed by the pyrolysis temperature). Indeed, while organics produced by low temperature pyrolysis are made essentially by amorphous carbon and fit our data reasonably well ($\chi^2\approx2$ to 3), at temperature higher than $\geq800\,^\circ$C such grains become dominated by large graphitic areas and fit our data much worse ($\chi^2\gg5$). Such low quality fits were also obtained adopting the graphite optical constants of (\citealt{Draine2003}, labelled `Draine' followed by the carbon grain size and the relative direction of the electric field and the graphite plane our dielectric constants were computed for), confirming our hypothesis that mixtures including organics made by amorphous carbon can fit our data the best.

Such different results for different compositions can be explained by the Mie interference location and magnitude in their absorption opacity spectral index. In the case of `Ricci (compact)', `Zubko', and `DSHARP (default)' compositions, such a resonance leads to a peak-like opacity spectral index profile, with $\beta^{\rm abs}>3$ at $a_{\rm max}\approx\lambda/2\pi$ for $q\leq2.5$. Instead, for `J\"{a}ger (>800 $^\circ$C)' and `Draine' compositions, the opacity spectral index is sigmoid-shaped and $\leq2.5$, regardless of $q$. The number of resonances present varies from none up to a few, but they are low-amplitude and just make the $\beta^{\rm abs}$ profile look more jagged. Since dust emission is optically thin in VLA Ka~band (regardless of the compositions adopted in our analysis), only the former set of opacities can explain a $\beta_{\rm B3-Ka}=\alpha_{\rm B3-Ka}-2>2.5$ for $R\gtrsim30\,{\rm au}$ (see \autoref{fig:3}). New (sub-)centimetre wavelength observations deep enough to detect dust emission with high S/N across the entire disc could provide stronger indications that $\alpha_{\rm B3-Ka}>5$ in the outer disc. This will be useful to further constrain dust composition because different peak magnitudes in the absorption spectral index can be seen between BE- and AC2H-sample amorphous carbon species (e.g. $\max(\beta^{\rm abs})\approx4.3$ and 4.7 for `Zubko (ACH2)' and `Ricci (compact)', while $\max(\beta^{\rm abs})\approx3.3$ for `Zubko (BE)'), but also for different water ice volume fractions (e.g. $\max(\beta^{\rm abs})\approx3.9$ for `DSHARP (icefree)' opacities, while $\max(\beta^{\rm abs})\approx3.1$ for~the `DSHARP (highwater)' ones).

For the `Ricci (compact)', `Zubko (BE)', and `DSHARP (default)' compositions we also performed high angular resolution fits as detailed in \autoref{sec:4.3}. Their marginalised posterior distributions are discussed in \autoref{sec:app7} and indicate that (i) the surface density and maximum grain size retrieved for the `Ricci (compact)' and `Zubko (BE)' mixtures are consistent within a factor of three, and (ii) the `DSHARP (default)' mixture gives less reliable results.

To summarise, CI~Tau's curse of a steepening spectral index between ALMA Band~3 and VLA Ka~band, which makes it difficult to detect and resolve dust emission at wavelengths longer than a few millimetres, turned out to be its blessing because only a few dust mixtures, those with the sharpest Mie resonance transition, such as `Ricci (compact)' and `Zubko', can explain this feature, providing unexpectedly strong constraints on the bulk dust composition in this disc.

\subsection{Result comparison for different porosities}\label{sec:5.4}

\begin{figure*}
    \sidecaption
    \includegraphics[width=\textwidth]{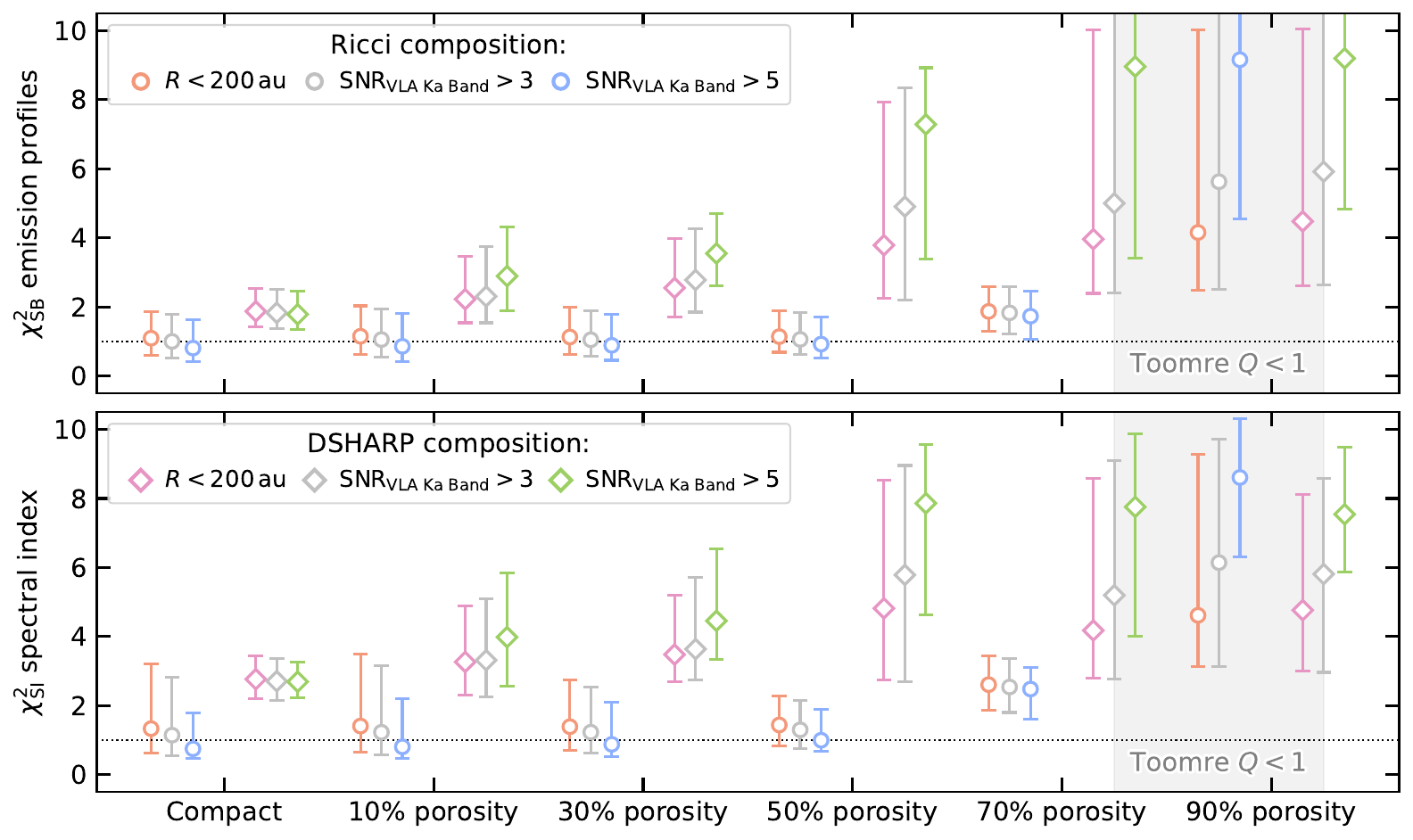}
    \caption{Reduced $\chi^2$ between our model posteriors and the observed surface brightness (upper panel) and spectral index (bottom panel) radial profiles for different porosity fractions, colour-coded by the outermost disc radius the $\chi^2$ was averaged over. Dots and diamonds are used for the `Ricci' and `DSHARP' mixtures, respectively. The shaded grey regions indicate where the inferred dust surface density is high enough for the disc to be in the marginally gravitational unstable regime under the assumption of a standard dust-to-gas ratio $Z=10^{-2}$. The `Ricci' mixture provides better fits than the `DSHARP' one for every porosity level and suggests that dust in CI~Tau can be up to 50\% porous. The `DSHARP' mixture, instead, can explain the data only for compact aggregates.}
    \label{fig:9}
\end{figure*}

We tested the impact of dust porosity on our results fitting the low resolution surface brightness radial profiles with `Ricci' and `DSHARP' dust compositions, but progressively increasing their void fraction from 10\%, 30\%, 50\%, and 70\% up to 90\% by volume\footnote{We focused on the mixtures that best reproduce CI~Tau's observations for compact grains. Since `Ricci' and both `Zubko' mixtures give similar results, hereafter we only discuss the former.}. We used the \texttt{dsharp\_opac} package \citep{Birnstiel2018} to determine the dielectric constants of each mixture with the Bruggeman rule\footnote{Here we followed \citet{Zhang2023}, while \citet{Birnstiel2018} preferred the Maxwell-Garnett rule \citep{Garnett1904} for porous grains, treating the physical mixture as inclusion in a background `void matrix'. \citet{Ricci2010} also used the Bruggeman rule for their default optical properties (i.e. `Ricci' mixture with 30\% porosity). These two methods lead to significant differences in the optical constants, and a generally worse agreement with the data when the Maxwell-Garnett rule is adopted. Furthermore, following \citet{Kataoka2014} and the argument of \citet{Ueda2024}, we treated porous particles as aggregates of compact monomers, mixing first the refractive materials, then adding void as a separate component, as explained \href{https://github.com/birnstiel/dsharp_opac/blob/master/notebooks/porosity.ipynb}{here}.} and their optical properties using Mie theory for spherical grains. As for the comparison of different dust mixtures (\autoref{sec:5.3}), we kept the temperature fixed while fitting for the other parameters. Overall, for both the `Ricci' and `DSHARP' compositions we noticed an increase of the best-fit density and grain size with the porosity fraction, especially for filling factors $>50\%$ and $>30\%$ for the `Ricci' and `DSHARP' grain mixtures, respectively. This trend can be motivated by the reduced absorption opacity and the progressive shift of the Mie resonance to sizes much larger than the reference wavelength for fluffier particles.

To estimate the quality of these fits, we measured the reduced $\chi^2$ between our models and the surface brightness and spectral index radial profiles following the procedure introduced in \autoref{sec:4.4}. The median and $1\sigma$ uncertainty of the reduced $\chi^2$ posterior distributions as a function of the adopted porosity fraction, colour-coded by the outermost disc radius the $\chi^2$ was computed for, are displayed in \autoref{fig:9}. The top and bottom panels summarise our model agreement with the surface brightness and spectral index radial profiles, respectively. Our fits with the `Ricci' mixture (red, grey, and blue dots) can reproduce remarkably well both the brightness and spectral index profiles for compact grains and particles with porosity fractions up to 50\%. In these cases, even though the estimated $\chi^2$ slightly increases with the void fraction, it remains below two. For every chosen porosity, the `DSHARP' mixture (pink, grey, and green diamonds) fits our data worse than the `Ricci' one, with $\chi^2>3$ already for void fractions as small as 10\%. For both `Ricci' and `DSHARP' mixtures, porosities $>50\%$ lead to very poor quality fits. In addition for highly porous particles (with void fractions of 70\% and 90\% or larger, for the `DSHARP' and `Ricci' composition, respectively) our inferred dust density is so high that, for a standard dust-to-gas ratio $Z=10^{-2}$, CI~Tau should be marginally gravitationally unstable (Toomre parameter $Q<1$, grey shaded area in \autoref{fig:9}). Since we observe no clear evidence of instability in the dust and gas emission, we favour lower porosity fractions.

The differences between `Ricci' and `DSHARP' results can be attributed once again to shape of the $\beta^{\rm abs}$ profile. Increasing the porosity fraction in the dust mixture leads to progressively shallower Mie resonances in the absorption opacity spectral index (for a fixed $q$), that almost completely disappears for void fractions $>50\%$. Therefore, the `DSHARP' composition gives worse results than the `Ricci' one because, for any given porosity, its optical properties lead to much shallower $\beta^{\rm abs}$ profiles. They are so shallow that even reducing $q$ (as in \autoref{fig:A12}) does not provide enough flexibility to compensate for the effect of porosity on the Mie resonance. A similar justification applies to the case of `Ricci' opacities with $>70\%$ porosity.

To summarise, dust porosity fractions up to $\approx50\%$ are compatible CI~Tau's observations when considering `Ricci' opacities. The `DSHARP' mixture, instead, provides good quality fits almost only for compact particles. This result suggests that, by robustly assessing the particle porosity (e.g. using full polarisation observations, see \autoref{sec:6.2}), we can tell `Ricci' and `DSHARP' opacity (or ultimately amorphous carbonaceous or refractory organic) models apart.

\section{Discussion}\label{sec:6}

\subsection{Interpretation of the best-fit radial profiles}\label{sec:6.1}
\paragraph{Maximum grain size} A puzzling result from \autoref{sec:5.1} is our smooth and almost constant best-fit grain size radial profile.

As noticed by \citet{Jiang2024}, smooth maximum grain size radial profiles are common to almost all the discs targeted by previous studies similar to ours. Based on the results of dust coagulation and transport simulation, they suggested that, even in the presence of low levels of turbulence ($\alpha_{\rm turb}\approx10^{-4}$), fragile pebble collisions (i.e. with a low fragmentation velocity threshold of $u_{\rm frag}\approx1\times10^2\,{\rm cm}\,{\rm s}^{-1}$) make dust growth fragmentation-limited through the whole disc extent. While this argument qualitatively explains the absence of peaks and troughs in our maximum grain size radial profile, under the commonly adopted assumption that the gas surface density declines more steeply than the temperature radial profile, it requires turbulence to decrease radially. In fact, since, in the fragmentation-dominated regime, $a_{\rm frag}\propto\Sigma_{\rm gas}/(\alpha_{\rm turb}T_{\rm dust})$, where $\Sigma_{\rm gas}$ is the gas surface density \citep{Birnstiel2012}, if the gas surface density and temperature can be expressed as power law functions of the disc radius with exponent $\gamma$ and $p$, where $\gamma<p$, then the turbulence will decrease~radially as $\gamma-p=\gamma+0.5<0$. In \autoref{sec:app8} we discuss more in detail the turbulence levels expected of CI~Tau if $a_{\rm max}=a_{\rm frag}$ and for a standard dust-to-gas ratio, $Z=10^{-2}$. 

Alternatively, the very weak radial dependence of the maximum grain size in CI~Tau could be the consequence of particle bouncing. In fact, since the bouncing velocity threshold depends on the particle mass, the bouncing barrier is almost radially flat (e.g. in the simulations of \citealt{Stammler2023} and \citealt{Birnstiel2024}, $a_{\rm max}\propto (R/10\,{\rm au})^{-3/16}$, in excellent agreement with our results). Furthermore, it scales much more weakly with $\Sigma_{\rm gas}$ and $\alpha_{\rm turb}$ then the fragmentation barrier ($a_{\rm boun}\propto a_{\rm frag}^{1/4}$, \citealt{Dominik&Dullemond2024}), thus requiring less stringent constraints on their radial profiles. However, we caveat against a straightforward interpretation of our results in favour of the bouncing and against the fragmentation barrier for two main reasons. First, despite having been considered only rarely in collisional models \citep[e.g.][]{Windmark2012}, it has been proposed that also the fragmentation velocity threshold \citep{Beitz2011} and the dust tensile strength \citep{SanSebastian2020} depend on the particle mass. Secondly, it could be possible that our fit would not be able to recover a radial decrease of the maximum grain size, even if genuinely present, because of the limited quality of our VLA Ka~band data. Deeper (sub-)centimetre wavelength observations able to confirm that continuum emission is detected out to $\lesssim200\,{\rm au}$ are crucial to support our results. 

Nevertheless, some sources show different trends, with significant increases of $a_{\rm max}$ at the position of some bright rings. Examples are the $100\,{\rm au}$ ring in HD~163296 (\citealt{Sierra2021}, but we note that the analysis of \citealt{Guidi2022}, which is based on higher angular resolution observations over a broader frequency range, disputed this claim), some rings in HL~Tau (\citealt{Carrasco-Gonzalez2019}, but see the updated model of \citealt{Guerra-Alvarado2024}, whose maximum grain size radial profile is less substructured, within the uncertainties\footnote{In the case of HL~Tau, the significant modulations of $a_{\rm max}$ at the location of gaps and rings inferred by \citet{Carrasco-Gonzalez2019} were not recovered by \citet{Guerra-Alvarado2024}, whose best-fit maximum grain size radial profile is roughly constant and $a_{\rm max}\approx1\,{\rm cm}$ in the inner $60\,{\rm au}$, then abruptly decreases and levels off to $a_{\rm max}\approx3\times10^{-2}\,{\rm cm}$ further out. This difference might be physical. Since HL~Tau is still very young, it is possible that dust did not grow enough to reach the fragmentation (or bouncing) threshold, as suggested by \citet{Jiang2024}. This hypothesis is also consistent with the decreasing trend of $a_{\rm max}$ with radius in this disc. There are, however, examples of (self-gravitating) Class 0/I discs where dust growth was proposed to be already in the fragmentation-dominated regime \citep{Xu2023}.}) and TW~Hya \citep{Macias2021}. However, in all of these cases, a fixed power-law density distribution ($q=2.5$ by \citealt{Sierra2021}, and $q=3.5$ by \citealt{Carrasco-Gonzalez2019}, \citealt{Guerra-Alvarado2024} and \citealt{Macias2021}) was adopted.

To better compare these results with ours, we also ran a set of low resolution fits fixing $q=3.5$ (the typical size distribution exponent of the ISM dust \citealt{Mathis1977}, that also lies in the range of values constrained by our fit, as shown in the bottom-right panel of \autoref{fig:4}) and $q=2.5$ (that spans a much wider range of absorption opacity spectral indices, $\beta^{\rm abs}$, thus providing lower limits on $a_{\rm max}$, as noticed by \citealt{Sierra2021}). However, neither option can fit the data well because our prescription is not flexible enough to explain \textit{both} the spectral index at ALMA wavelengths and between ALMA Band~3 and the VLA Ka~band data, that provide a crucial additional constraint when compared with the results of \citet{Macias2021} and \citet{Sierra2021}, based on ALMA-only data\footnote{We notice that, in line with our results of a radially varying $q$, also \citet{Pinte2016} in their radiative-transfer models of ALMA 0.9 to $2.9\,{\rm mm}$ observations of HL~Tau could not find a model able to reproduce the dust surface density at all the wavelengths with a radially constant grain size distribution power-law. Indeed, their best-fit radial profile is closer to $q\approx3.5$ in the inner disc and $q=4.5$ in the outer disc, qualitatively similar to our results in CI~Tau, albeit with the substantial difference that \citet{Pinte2016} found a lower (higher) $q$ in gaps (rings).}. Interestingly, \citet{Macias2021} fitted their data also leaving $q$ as a free parameter. In this case, they obtained much more similar results to ours: an almost constant $a_{\rm max}$ profile with radius, albeit with a larger uncertainty, and a density distribution power-law index radial profile that decreases towards the inner disc and at the position of the bright rings. 

\paragraph{Grain size distribution} In contrast with the maximum grain size radial profile, our dust density distribution decreases towards the inner disc and at the location of the bright rings. We stress that, in the absence of radial variations of $a_{\rm max}$, the dependence of $q$ on the disc radius is essential to explain the radial increase of CI~Tau's spectral indices (see \autoref{fig:3}), which, according to our models, reflects a similar increase of the absorption spectral index, as is show in \autoref{sec:app9}. These features suggest that a larger fraction of the dust distribution is dominated by the largest particles (i.e. those with size $\approx a_{\rm max}$) in the inner disc and the bright rings.

In the outer disc, our best-fit dust size distribution is broadly consistent with the expectations of fragmentation-limited growth ($q=3.5$, \citealt{Birnstiel2024}, see also the more-refined predictions for a fragmentation-coagulation equilibrium of \citealt{Birnstiel2011}). Instead, in the inner disc, $q$ lies between the values expected when the maximum grain size is limited by the fragmentation and the radial drift threshold ($q=2.5$, \citealt{Birnstiel2024}). Such a best-fit dust number density distribution is also not consistent with the results of bouncing-limited dust growth simulations, that predict an almost monodisperse particle size distribution \citep{Stammler2023,Dominik&Dullemond2024,Birnstiel2024}. Nevertheless, such predictions are based on models that did not consider processes, such as erosion by abrasion, that are detrimental to growth and could substantially increase the fraction of small particles, thus widening the dust size distribution and increasing $q$ \citep{Dominik&Dullemond2024}. Dedicated collisional growth and dust transport simulations are needed to interpret our results self-consistently. As for the rings, instead, in the fragile pebble collision scenario discussed above (i.e. where rings are not favourable locations for dust to grow), the lower $q$ could be explained by the preferential confinement of larger grains in pressure bumps co-located with the rings \citep{Pinilla2012}. In fact, smaller and better-coupled solids, that are more easily diffused away or mixed up by turbulence, can filter through a pressure maximum \citep[e.g.][]{Petrovic2024}, locally increasing the fraction of larger particles.

To summarise, our results suggest that turbulent fragmentation or bouncing are the dominant processes halting dust growth in CI~Tau. On top of this global trend, dust confinement in the rings provides a possible explanation for the small-scale modulations of the dust density and grain size power law index radial profiles. Deeper (sub-)centimetre observations, able to detect continuum emission with higher S/N in the outer disc or provide stringent upper limits, are essential to reassess this picture.

\subsection{Dust polarisation and the role of porosity}\label{sec:6.2}
As an alternative to the multi-frequency continuum analysis we presented in this paper, (multi-wavelength) linear polarisation observations have also been used to provide constraints on the properties of dust in planet-forming discs \citep[e.g.][]{Miotello2023}. Over the last decade, linear polarisation fractions of about 1\% have been detected in a number of sources, such as HL~Tau \citep{Stephens2017}, HD~163296 \citep{Dent2019}, HD 142527 \citep{Kataoka2016}, IM~Lup \citep{Hull2018}, CW~Tau and DG~Tau \citep{Bacciotti2018}. At $0.9\,{\rm mm}$, the polarisation vectors are most often oriented along the disc minor axis. This polarisation pattern is commonly interpreted as due to dust self-scattering \citep{Kataoka2015}. However, 1.3 and $3.1\,{\rm mm}$ observations revealed that the polarisation pattern can be frequency-dependent \citep[e.g.][]{Stephens2017}, with polarisation vectors being progressively more azimuthally aligned at longer wavelengths, suggesting that other processes, such as dust grain alignment with the radiation field \citep{Tazaki2017}, become the dominant source of polarisation. The (wavelength dependence of the) polarisation fraction due to self-scattering can be used to provide estimates of the maximum grain size by comparison with model predictions \citep{Kataoka2015}, as in the case of HL~Tau, where the decrease of self-scattering polarisation with wavelength was interpreted by \citet{Kataoka2017} as a strong evidence that only grains of $a_{\rm max}\lesssim100\,\mu$m are present in the disc, in puzzling contrast with the results of the analysis of multi-frequency continuum observations of the same source \citep{Carrasco-Gonzalez2019,Guerra-Alvarado2024}.

Several explanations were put forward to account for this~discrepancy, such as dust vertical settling and optical depth effects (\citealt{Lin2020,Ueda2021}, but see \citealt{Sierra&Lizano2020} for a counter-argument), the different nature of continuum and polarised-light emission (e.g. signal contamination by polarisation due to alignment of oblate dust with the magnetic or radiation field, \citealt{Kirchschlager&Bertrang2020}), or dust composition (\citealt{Yang&Li2020} suggested that adopting dust mixtures including amorphous carbonaceous grains instead of refractory organics might alleviate the tension between the results of polarised-light and continuum observation analyses). Following \citet{Tazaki2019}, who noticed that the range of particle sizes with high self-scattering polarisation fraction is wider for porous grains than compact particles, \citet{Zhang2023} proposed that particles with 70\% to 97\% porosity can explain both the dust continuum and linear polarisation multi-frequency observations of HL~Tau. Similar porosity fractions of 80\% were shown to be in remarkable agreement with scattered light and (sub-)millimetre (continuum and linear polarisation) observations of IM~Lup by \citet{Ueda2024}.

Unfortunately, full polarisation observations of CI~Tau have never been taken, and no information on the dust polarisation fraction in this disc is available. As a consequence, all our constraints on particle porosities are based on our continuum profile analysis in \autoref{sec:5.4}. Our preference for low porosity fractions in CI~Tau is in contrast with the aforementioned results of \citet{Zhang2023}~and \citet{Ueda2024} in HL~Tau and IM~Lup, even more so since they adopted the default `DSHARP' mixture, that provides~good quality fits to our data almost only in the compact case. Instead, our results are consistent with those of \citet{Guidi2022}, whose multi-frequency continuum observations of HD~163296 can be better explained by low particle porosities (25\% instead of 80\%). While both our analysis and that of \citet{Guidi2022} did not consider polarisation observations, and therefore provided more uncertain estimates of the dust porosity, it is likely that~the differences between the porosity fractions of HL~Tau/IM~Lup and CI~Tau/HD~163296 are real. Indeed, while the former sources are known to be very young ($\lesssim1\,{\rm Myr}$, e.g. \citealt{Alcala2017}), the latter are relatively (CI~Tau, $>3\,{\rm Myr}$, \citealt{Gangi2022}) to significantly (HD~163296, $\gtrsim6\,{\rm Myr}$, \citealt{Fairlamb2015}) older. Thus, it can be speculated that time-dependent processes (e.g., dust compaction) might have played a role in determining those differences. 

Both laboratory experiments and numerical simulations suggest that the initial stages of particle growth take place by hit-and-stick, a process that leads to very fluffy aggregates with low fractal dimensions \citep[see e.g.][and references therein]{Testi2014}. Subsequently compaction is thought to take place: the fractal dimension increases and the porosity decreases up to 40 to 60\% due to a combination of bouncing- and fragmentation-driven compression \citep{Michoulier2024}. On the one hand, the high porosity fractions inferred in HL~Tau and IM~Lup are more in line with the models of \citet{Ginski2023}, who showed that the small particles in the upper disc layers probed by scattered light observations are very porous or have small fractal dimensions. This suggests that dust in these young discs is still growing or undergoing compression. On the other hand, our results for CI~Tau and those of \citet{Guidi2022} for HD~163296 are more consistent with the models of \citet{Rosotti2019,Zormpas2022,Delussu2024}, that showed that compact grains can better reproduce the size luminosity correlation detected at (sub-)millimetre wavelengths in nearby 1 to $3\,{\rm Myr}$-old star-formation regions. This being said, the porosity fraction that particles can reach after compaction depend on a number of disc parameters, such as the radial location (because of the different collisional timescale and dust-to-ice ratio, e.g. \citealt{Lorek2018}, and \citealt{Michoulier2024}). Thus the observed differences, if confirmed, might also reflect more intrinsic source differences than a genuine time-dependent process.
 
\subsection{Pebble mass estimates}\label{sec:6.3}

\begin{figure*}
    \centering
    \includegraphics[width=\textwidth]{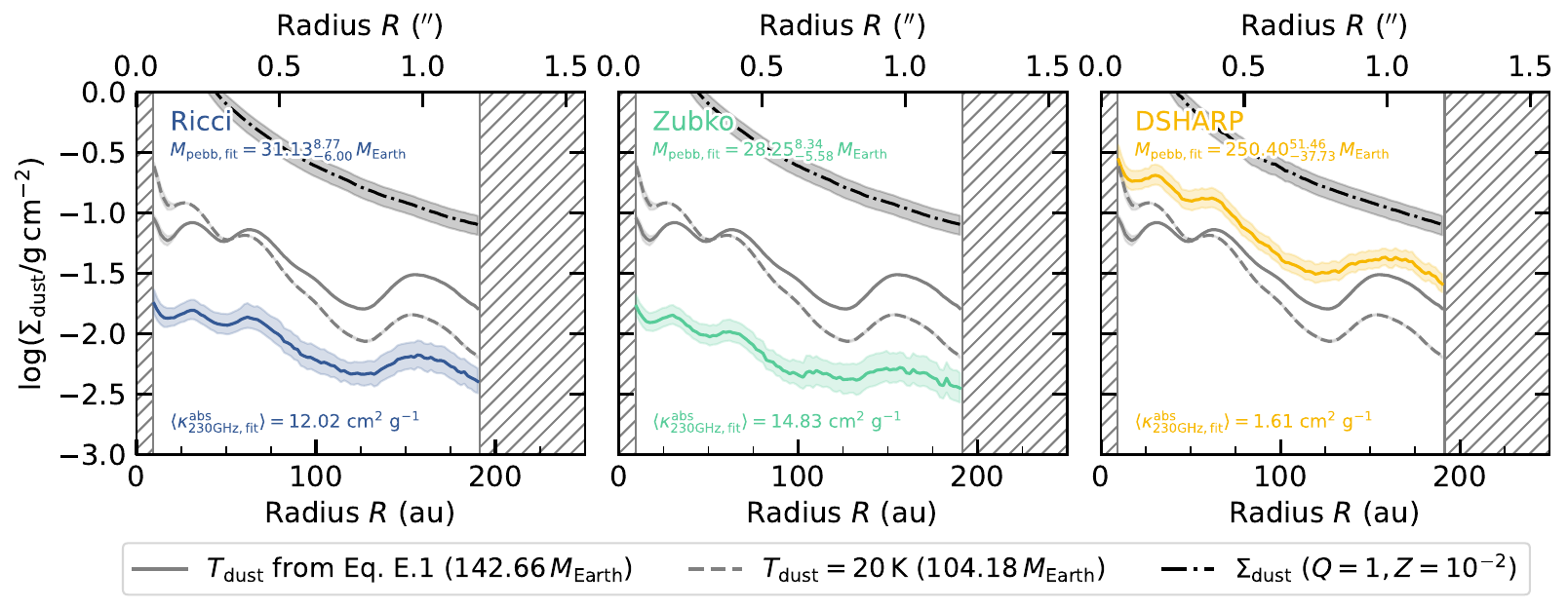}
    \caption{Comparison between the pebble masses (top-left corner, within $R<200\,{\rm au}$) inferred from the dust surface density radial profiles fitted using different compositions (for compact grains only). The solid and dashed grey lines display the expected dust surface density radial profiles obtained by converting the ALMA Band~6 surface brightness radial profile using a constant temperature of $20\,{\rm K}$ and the temperature profile from \autoref{eq:app5.1}, both for $\kappa_{\rm 230GHz}=2.3\,{\rm cm}^2\,{\rm g}^{-1}$. The median absorption opacity from our multi-wavelength fit for different dust mixtures is annotated in the bottom-left corner of each subplot. The dashed-dotted black line shows the upper limit on the dust surface density, corresponding to a Toomre parameter $Q=1$ and a gas-to-dust ratio $Z=10^{-2}$.}
    \label{fig:10}
\end{figure*}

Knowledge of the disc dust mass is crucial to determine the potential of a disc to form (new) planets. Being sensitive to small solids, but blind to the dust sequestered in larger aggregates, such as planetesimals and planetary cores, continuum observations in the (sub-)millimetre have been traditionally used to provide lower limits on the disc dust mass, under the assumption that continuum emission at these wavelengths is optically thin, as \citep{Hildebrand1983}
\begin{equation}\label{eq:7.1}
    M_{\rm dust} \geq M_{\rm pebb}\approx\dfrac{F_\nu d^2}{\kappa_\nu B_\nu(T_{\rm dust})},
\end{equation}
where $F_\nu$ is the disc dust flux density at frequency $\nu$ and $d$ is the source distance, adopting a fiducial opacity ($\kappa_\nu\approx2.3\,{\rm cm}^2\,{\rm g}^{-1}$ at $230\,{\rm GHz}$, see e.g. \citealt{Ansdell2016,Ansdell2018,Pascucci2016,Barenfeld2016}) and dust temperature ($T_{\rm dust}=20\,{\rm K}$, see e.g. \citealt{Ansdell2016,Ansdell2018}, sometimes rescaled by the stellar luminosity, e.g. \citealt{Pascucci2016,Barenfeld2016}). We refer to this lower limit as pebble mass, $M_{\rm pebb}$.

Over the last years, it was realised that the pebble masses inferred from (sub-)millimetre surveys in nearby star-formation regions were not high enough \citep{Manara2018} or barely sufficient \citep{Najita&Kenyon2014,Mulders2021} to form the rocky cores of the currently detected exoplanets. A number of solutions to this `mass budget' problem were proposed, such as the presence of highly optically thick regions capable of hiding large pebbles mass fractions \citep[e.g.][]{Zhu2019,Ribas2020}, the possibility that planetary cores form much earlier than the Class~II stage \citep[e.g.][]{Tychoniec2020}, or, more recently, the replenishment of fresh ISM material \citep[e.g.][]{Winter2024}.

Multi-frequency continuum observations can provide more reliable estimate of the pebble mass in a disc because of their accurate constraints on the dust optical depth and mid-plane temperature. \autoref{fig:10} displays a comparison between the pebble masses inferred for the `Ricci (compact)', `Zubko (BE)', and `DSHARP (default)' compositions in CI~Tau and the results obtained under the traditional optically thin and fixed temperature assumptions. The dust surface densities posterior distributions from our high resolution multi-frequency fits for the aforementioned compositions are displayed in each subplot in blue, turquoise, and yellow, respectively. On top of these profiles, we also plotted the dust surface density estimated from the ALMA~Band~6 surface brightness radial profile, under the assumption of optically thin emission, defined as (\autoref{eq:5.1}, for $\tau_\nu^{\rm abs}\ll1$)
\begin{equation}
    \Sigma_{\rm dust,OT}=\dfrac{I_\nu}{B_\nu}\dfrac{\cos i}{\kappa_\nu}.
\end{equation}
We adopted the same opacity law traditionally used to convert dust flux densities into dust masses: i.e. $\kappa_\nu$ is $2.3\,{\rm cm}^2\,{\rm g}^{-1}$ at $230\,{\rm GHz}$ \citep{Ansdell2016} and scales linearly with frequency \citep{Beckwith1990}. Solid and dashed grey lines were used for a constant temperature of $20\,{\rm K}$ and the temperature profile of a passively irradiated disc (\autoref{eq:app5.1}, with $\phi=0.035$ and $L_\star=1.04\,L_\odot$), respectively. Finally, the dashed-dotted black curves show the dust surface density profile corresponding to a Toomre parameter $Q=1$ \citep{Toomre1964} for a standard dust-to-gas ratio $Z=10^{-2}$:
\begin{equation}
    \Sigma_{\rm dust,Q}=\dfrac{c_{\rm s}\Omega_{\rm K}Z}{\pi G},
\end{equation}
where $c_{\rm s}$ is the locally isothermal sound speed (computed with a mean-molecular mass of $\mu=2.4$), $\Omega_{\rm K}$ is the Keplerian angular velocity (estimated for a central mass of $M_\star=1.29\pm0.45\,M_\odot$ \citealt{Gangi2022}), and $G$ is the gravitational constant.

As is clear from \autoref{fig:10}, different assumptions on dust composition provide different estimates of the pebble mass in the disc. The masses estimated under the traditional optically thin assumption, regardless of the temperature being $20\,{\rm K}$ (grey solid line) or decreasing with radius (grey dashed line), are a factor of two lower than those based on the `DSHARP (default)' opacity fit, but overestimate those measured from the `Ricci (compact)' and `Zubko (BE)' best-fit surface density profiles by a factor of three to five. This naively gives credit to the realisation, dating back to the work of \citet{Beckwith1990}, that knowledge of the dust composition is crucial to provide reliable constraints on the pebble mass. Our marginal preference for `Ricci (compact)' and `Zubko (BE)' opacities over the `DSHARP (default)' ones, extensively motivated in the previous sections, has two main consequences. Firstly, if taken at face value, the pebble mass comparison shown in \autoref{fig:10} suggests that, under standard assumptions on opacity and temperature, \autoref{eq:7.1} does not always underestimate the dust mass, as is often assumed.\footnote{Despite being counter intuitive, this result is a natural consequence of the different adopted opacities. When pebble masses are estimated using \autoref{eq:7.1} and the radially averaged best-fit absorption opacities from our multi-wavelength analysis (see bottom-left corner of each panel in \autoref{fig:10}), $M_{\rm pebb}\approx20.19$, 16.27, and $148.83\,M_{\rm Earth}$ for $T=20\,{\rm K}$ and the `Ricci (compact)', `Zubko (BE)', `DSHARP (default)' mixtures, respectively. These values are lower than the pebble masses measured from our multi-frequency analysis (see top-left corner of each panel in \autoref{fig:10}), confirming that the combination of observations at different wavelengths leads to higher (and generally more reliable) pebble mass estimates by allowing to correct for optical depth effects. This being said, at least in the case of CI~Tau, and presumably other (marginally) optically thin discs (i.e. large sources, where most of the mass is in optically thin outer disc regions), the uncertainty introduced by the adopted opacity law far outweighs the optical depth corrections when measuring $M_{\rm pebb}$. Whether this conclusion can be extrapolated to a demographic level, and specifically to the results of (sub-)millimetre snapshot surveys in nearby star-formation regions, however, is not trivial. On the one hand, most of these surveys observed their targets only at one wavelength and thus cannot provide any constraints on their maximum grain size. This leads to uncertainties up to a factor of 100 in their (sub-)millimetre opacity (for $0.1\leq a_{\rm max}/{\rm mm}\leq 10$; see the top panel of Figure 10 of \citealt{Birnstiel2018}). On the other hand, the bulk of the disc population is made of very compact sources, where optical depth corrections might be significant. In the most unfavourable scenario of marginally self-gravitating discs ($Q=1$) with a reasonable grain size of $a_{\rm max}=1\,{\rm mm}$ (corresponding to $\omega_{\rm 1.3mm}\geq0.8$ for the `DSHARP (default)' mixture), the Monte Carlo Radiative Transfer simulations of \citet{Zhu2019} suggest that, in the presence of efficient dust scattering, the dust mass can be underestimated by a factor of 10 to 100 for discs smaller than 10 to $30\,{\rm au}$. This being said, the strong dependence of the optical depth corrections on the efficiency of dust scattering, and thus on the single-scattering albedo, would also imply that the unknown dust composition might be the dominant source of uncertainty also in the smaller and optically thicker discs.} Secondly, it would suggest that, if giant planet formation occurred in CI~Tau, this process might be already over. This is supported by current estimates of the total mass of heavy elements in Solar System giants, thought to be 8 to $46\,M_{\rm Earth}$ in Jupiter and 16 to $30\,M_{\rm Earth}$ in Saturn (\citealt{Guillot2023} and references therein), and thus in line or higher than the pebble mass in CI~Tau measured for `Ricci (compact)' and `Zubko (BE)' dust mixtures. Clearly, the (still poorly constrained) dust porosity fraction adds to the previously mentioned sources of uncertainty. However, for the `Ricci' mixture, the pebble mass inferred from our multi-wavelength analysis does not exceed $M_{\rm pebb}\approx134.39\,M_{\rm Earth}$ for a $50\%$ porosity fraction. This value does make it more feasible to assemble the core of a new gas giant in the system with a more reasonable, but still high \citep{Drazkowska2023}, efficiency of $\approx 30\%$.

Finally, in \autoref{tab:app6} we compare our results for CI~Tau with similar estimates of the pebble mass from $1.3\,{\rm mm}$ continuum observations and multi-frequency observations in the literature. Our results for `DSHARP (default)' opacities for CI~Tau are in line with those obtained in all the other sources with the exception of IM~Lup, that is more massive and potentially marginally gravitationally unstable, as it might be hinted by the $m=2$ spiral detected in (sub-)millimetre continuum observations \citep{Huang2018} and its high gas mass inferred from a kinematic analysis of its super-Keplerian rotation curve \citep{Martire2024}. This is suggesting that adopting our fiducial `Ricci (compact)' composition also for these sources, will likely decrease their pebble masses and planet-formation potential (see \citealt{Sierra2025} for LkCa~15).

\section{Summary and conclusions}\label{sec:8}
We have introduced, for the first time, deep ALMA Band~3 and VLA Ka, Ku, X, and C~band continuum observations of CI~Tau. We took advantage of the high angular resolution and sensitivity of ALMA Band~3 and VLA Ka~band observations and of combining them with similar quality archival ALMA Band~6 and 7 data to study the properties of dust in this system. Our main results are summarised in the following list:
\begin{enumerate}
    \item At ALMA wavelengths, continuum emission towards CI~Tau displays similar morphologies, characterised by a sequence of three gaps and four rings, as previously reported by \citet{Clarke2018} and \citet{Long2018} at $1.3\,{\rm mm}$ and \citet{Rosotti2021} at $0.9\,{\rm mm}$. Non-parametric modelling of the visibilities (with the code \texttt{frank}) revealed that the innermost bright ring is substructured (i.e. made of two sub-beam rings and a gap in between) and highlighted the presence of an additional gap at $\approx 5\,{\rm au}$ in the $1.3\,{\rm mm}$ data (see also \citealt{Jennings2022b}). 
    \item In contrast, the VLA Ka~band emission is very faint and shows no clear substructures, not even in the best-fit \texttt{frank} profile. When reconstructed with a large-enough beam ($>0\farcs150$), extended emission can be detected, as can also be seen in the best-fit \texttt{frank} profile, albeit at a low ${\rm S/N}\approx3$.
    \item CI~Tau's integrated spectral index between 0.9 and $3.1\,{\rm mm}$ is $2.6\pm0.1$, which is in line with those measured in HL~Tau, TW~Hya, and HD~163296. Between 3.1 and $9.1\,{\rm mm}$, the spectral flux density distribution steepens, with an integrated spectral index of $4.3\pm0.2$, and is thus significantly larger than in the previously mentioned sources. At even longer wavelengths, emission is not dominated by dust but likely by a combination of optically thick gyrosynchrotron emission (at $\lambda\gtrsim2\,{\rm cm}$), as suggested by the high X~band flux density variability and optically thin free-free emission (at $\lambda\lesssim1\,{\rm cm}$). Deeper (sub-)centimetre simultaneous observations (e.g. from C to Q~band) probing variability on longer timescales are needed to conclusively assess the nature of CI~Tau's continuum radio emission.
    \item We developed a two-step Bayesian method to model CI~Tau's surface brightness radial profiles. First, the observations between 0.9 and $9.1\,{\rm mm}$ were fitted at low resolution ($0\farcs195$). These results where then used as priors for a high-resolution ($0\farcs058$) fit only to the ALMA data. This new technique allows one to trade off resolution for sensitivity in centimetre-wavelength observations, where targets are generally very faint and long integration times are required to get good quality data.
    \item When adopting the `Ricci (compact)' bulk composition, our best-fit dust temperature profile is smooth and radially declines as expected for a passively irradiated disc. The dust surface density profile increases, while $q$ (the slope of the dust density distribution, i.e. the fraction of small grains) decreases locally at the position of the bright rings, which is in line with the predictions of dust trapping simulations. Our best-fit maximum grain size, however, is smooth and radially flat, as expected from models of fragile dust collisions in low-turbulence discs or bouncing-limited dust growth. Our analysis revealed that CI~Tau's continuum emission is optically thin between 0.9 and $9.1\,{\rm mm}$, in contrast with most of the sources previously studied with similar quality multi-wavelength observations.
    \item For compact grains, we tested the dependence of our results on dust composition, comparing the $\chi^2$ between the observed and predicted surface brightness and spectral index radial profiles. We showed that the fraction of water ice included in the mixture only negligibly affects the goodness of the fit, while changing the refractive constants of organics leads to substantial differences. In particular, our data disfavour graphite-dominated carbon and marginally prefer amorphous carbonaceous grains \citep{Zubko1996} over refractory organics \citep{Henning&Stognienko1996}. Deeper (sub-)centimetre wavelength data are needed to confirm our results. The different optical depths expected in the case of `Ricci (compact)' and `DSHARP (default)' mixtures provide an additional opportunity to constrain internal composition with independent estimates of the total dust extinction (see \autoref{sec:app7}).
    \item For a fixed particle composition, we tested the dependence~of our results on dust porosity using the same $\chi^2$ comparison method. We showed that the `Ricci' mixture can fit the data well for void fractions up to $50\%$, while for porosities larger than $70\%$, it leads to worse results. This is in contrast with the recent inferences of high porosity fraction (80 to 90\%) in the younger HL~Tau and IM~Lup discs. This difference could be attributed to time-dependent compaction, but dedicated models are needed to confirm this hypothesis. Increasing the void fraction, the `Ricci' mixture performs progressively better than the `DSHARP' one. Deeper (sub-)centimetre continuum and (multi-wavelength) polarisation observations are needed to confirm our results and better constrain particle porosity. Given the increasingly different behaviour of the `Ricci' and `DSHARP' compositions for larger void fractions, knowledge of porosity could also be used to better constrain the bulk dust composition.
    \item For our fiducial `Ricci (compact)' composition, we estimated CI~Tau's pebble mass to be $M_{\rm pebb}=31.13^{+8.77}_{-6.00}\,M_{\rm Earth}$, thus a factor of three to five lower than expected from the dust luminosity conversion traditionally employed in snapshot surveys (\autoref{eq:7.1}). This difference is primarily driven by~the adopted absorption opacities, which are much higher for `Ricci (compact)' grains ($\langle\kappa^{\rm abs}_{\rm 230GHz,fit}\rangle=11.9\,{\rm cm}^2\,{\rm g}^{-1}$, mediated over our $a_{\rm max}$ and $q$ posteriors) than is commonly assumed ($\kappa^{\rm abs}_{\rm 230GHz}=2.3\,{\rm cm}^2\,{\rm g}^{-1}$). Such a low pebble mass is barely enough to account for the heavy element mass in Jupiter or Saturn. This suggests that if giant planet formation is taking place in CI~Tau, the rocky cores of such planets mostly likely have already been assembled. Compared to other more optically thick sources, uncertainties on the bulk composition rather than optical depth corrections dominate our uncertainty on the pebble mass estimate in CI~Tau.
\end{enumerate}
Deeper (sub-)centimetre observations (e.g. with ALMA in Band~1) and follow-up polarisation observations, are needed to confirm our preference for amorphous carbonaceous grains with low porosities. Expanding our analysis to other sources, especially moderate-to-low optical depth discs such as AS~209 or GM~Aur \citep{Sierra2021}, could be ideal to further test different dust composition models and, by comparison with our results for CI~Tau, understand if composition and porosity change from source to source. Future facilities, first and foremost the next generation Very Large Array (ngVLA), will be crucial to obtaining better resolution and sensitivity centimetre-wavelength observations and to extending our analysis to larger samples.

\begin{acknowledgements}
This paper makes use of the following ALMA data:\\ ADS/JAO.ALMA\#2015.1.01207.S,
ADS/JAO.ALMA\#2016.1.01370.S,\\
ADS/JAO.ALMA\#2017.A.00014.S, and
ADS/JAO.ALMA\#2018.1.00900.S,\\
ALMA is a partnership of ESO (representing its member states), NSF (USA), and NINS (Japan), together with NRC (Canada), NSC and ASIAA (Taiwan), and KASI (Republic of Korea), in cooperation with the Republic of Chile. The Joint ALMA Observatory is operated by ESO, AUI/NRAO, and NAOJ. This paper makes use of the following VLA data: VLA/19A-440 and VLA/20A-373. The National Radio Astronomy Observatory (NRAO) is a facility of the National Science Foundation operated under cooperative agreement by Associated Universities, Inc. FZ acknowledges support from STFC and Cambridge Trust for a Ph.D. studentship, is grateful to the Institute for Astronomy, University of Hawai'i at Manoa, for hosting him for a \textit{Dustbusters} secondment when this project came to be. FZ thanks Til Birnstiel, Ted Bergin, Greta Guidi, Oliver Shorttle, Elena Viscardi, and Mark Wyatt for insightful discussions on dust properties and the analysis method. SF is funded by the European Union (ERC, UNVEIL, 101076613), and acknowledges financial contribution from PRIN-MUR 2022YP5ACE. PC acknowledges support by the Italian Ministero dell Istruzione, Universit\`a e Ricerca through the grant Progetti Premiali 2012 – iALMA (CUP C52I13000140001) and by the ANID BASAL project FB210003. A.R. has been supported by the UK Science and Technology Facilities Council (STFC) via the consolidated grant ST/W000997/1. RAB is supported by a University Research Fellowship. GR is funded by the European Union under the European Union’s Horizon Europe Research \& Innovation Programme No.~101039651 (DiscEvol) and by the Fondazione Cariplo, grant no. 2022-1217. Views and opinions expressed are however those of the author(s) only and do not necessarily reflect those of the European Union or the European Research Council. Neither the European Union nor the granting authority can be held responsible for them. This project has received funding from the European Union's Horizon 2020 research and innovation programme under the Marie Sklodowska-Curie grant agreement No. 823823 (Dustbusters RISE project).\\

\textit{Software:} \texttt{CASA} \texttt{v6.4.3.27} and \texttt{6.2.1.7} \citep{McMullin2007,CASAteam2022}, \texttt{analysisUtils} \citep{Hunter2023}, \texttt{gofish} \texttt{v1.4.1} \citep{Teague2019}, \texttt{frank} \citep{Jennings2020}, \texttt{uvplot} \citep{Tazzari2017}, \texttt{corner} \texttt{v2.2.1} \citep{corner}, \texttt{emcee} \texttt{v3.1.2} \citep{Foreman-Mackey2013,Foreman-Mackey2019}, \texttt{numpy} \texttt{v1.24.2} \citep{numpy}, \texttt{scipy} \texttt{v1.10.1} \citep{Virtanen2020}, \texttt{matplotlib} \texttt{v3.7.1} \citep{Hunter2007}, \texttt{astropy} \texttt{v5.2.1} \citep{astropy2013,astropy2018,astropy2022}, \texttt{JupyterNotebook} \texttt{v6.5.2} \citep{jupyternootbok}, \texttt{dsharp\_opac} \texttt{v1.1.4} \citep{Birnstiel2018}. Scripts and data are publicly available on \href{https://github.com/fzagaria/multi-freq_CITau}{\texttt{GitHub}}.
\end{acknowledgements}

\bibliographystyle{aa} % style aa.bst
\bibliography{aa52986-24} % your references Yourfile.bib

\begin{appendix}

\section{ALMA flux density calibration}\label{sec:app1}
Before concatenation, the visibilities of all the ALMA datasets in the same band were deprojected and inspected to identify any mismatches in the amplitude scales. These mismatches were then corrected for by rescaling the flux densities of all the EBs at the same frequency to that of the reference one, using the \texttt{gencal} task. This reference observing block was selected by comparing the observed luminosity of each flux calibrator with that interpolated from their tabulated light curves at the same epoch. Hereafter we discuss how the reference EB was determined for each band. 

\begin{figure*}
    \centering
    \includegraphics[width=\textwidth]{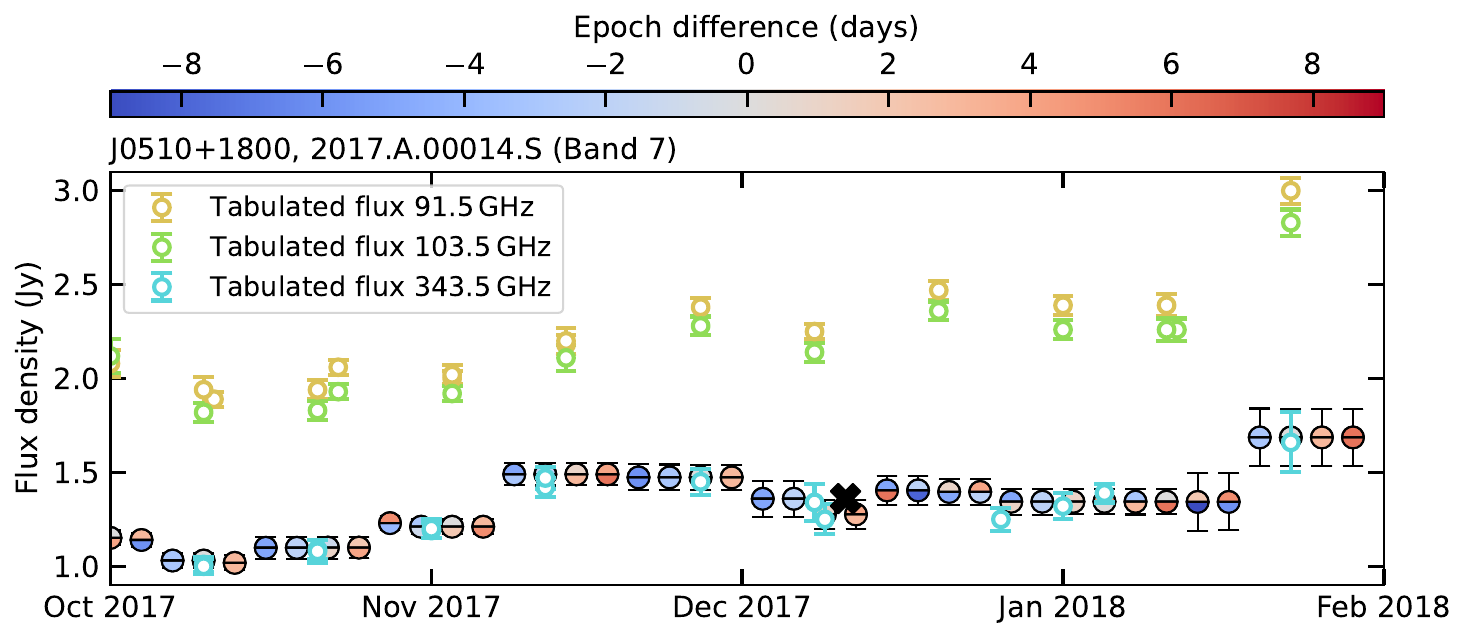}
    \includegraphics[width=\textwidth]{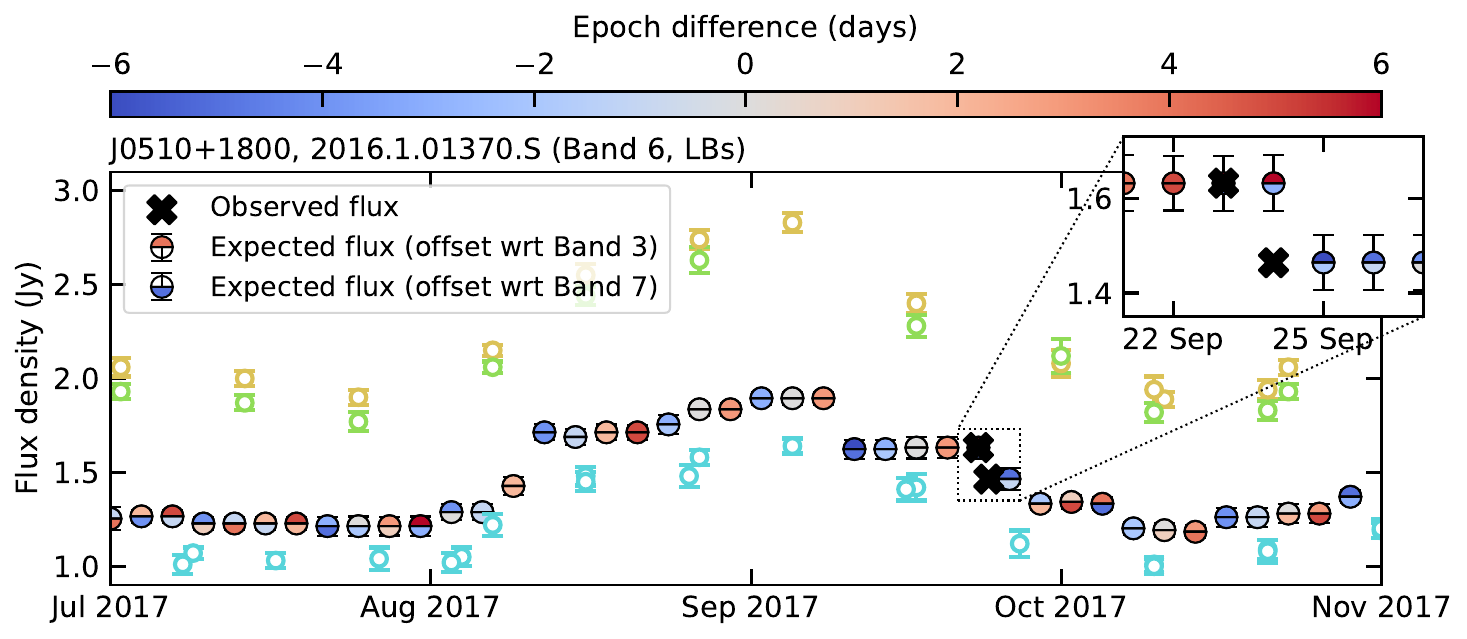}
    \includegraphics[width=\textwidth]{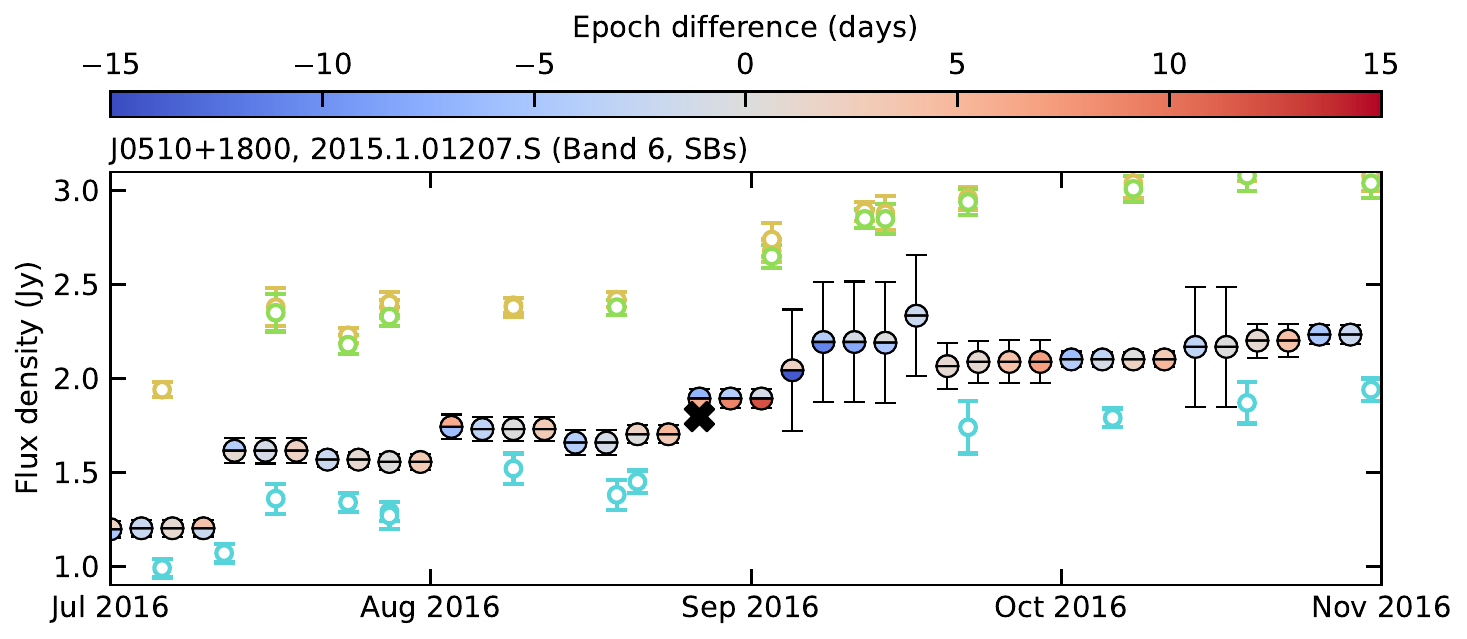}
    \caption{Light curves of the flux calibrators (Band~6 and 7). The expected flux densities were interpolated at 330.7, 242.0, and $234.4\,{\rm GHz}$, respectively. The insert shows the one-day cadence interpolation for the week LB data were taken.}
    \label{fig:A1}
\end{figure*}

\begin{figure*}
    \centering
    \includegraphics[width=\textwidth]{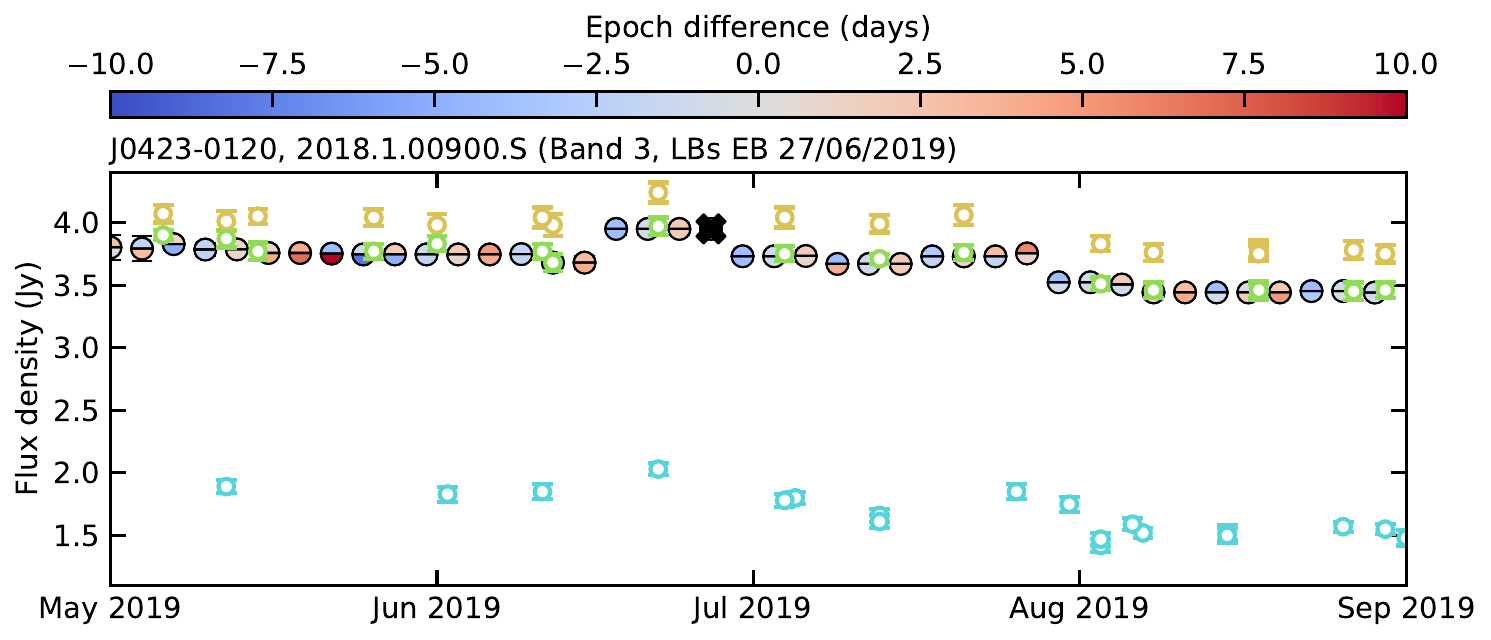}
    \includegraphics[width=\textwidth]{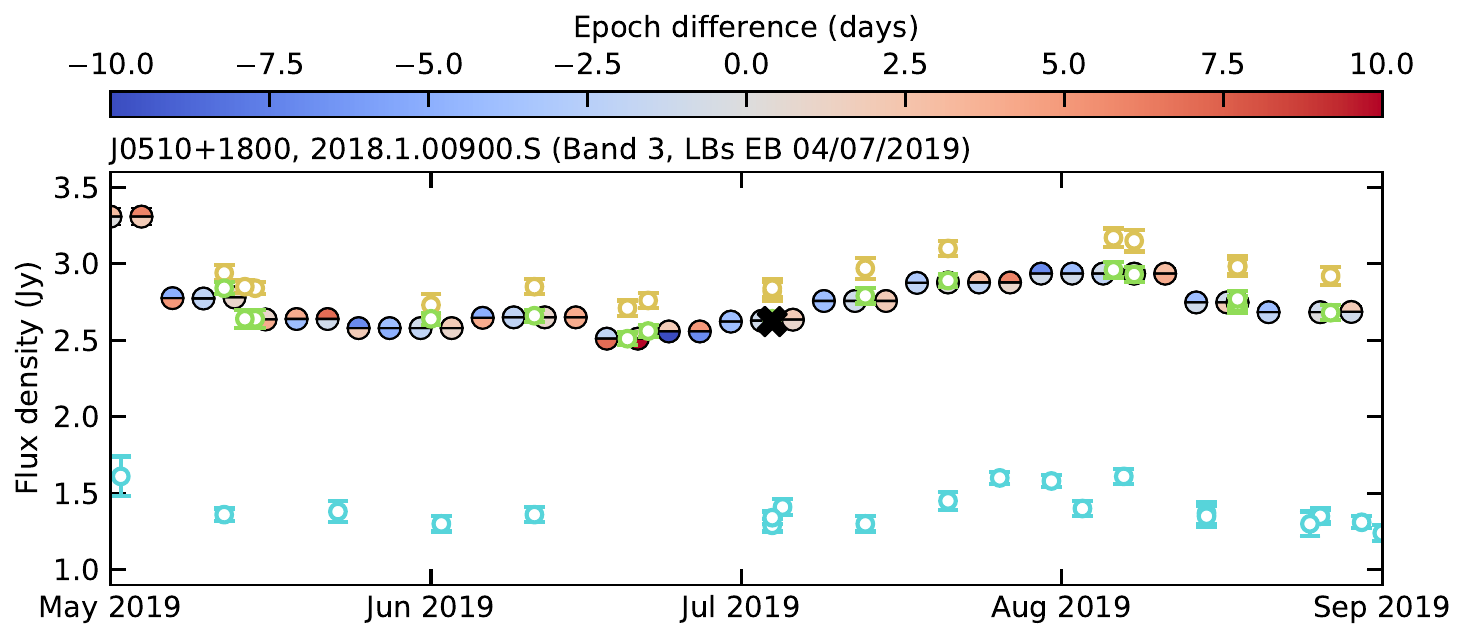}
    \includegraphics[width=\textwidth]{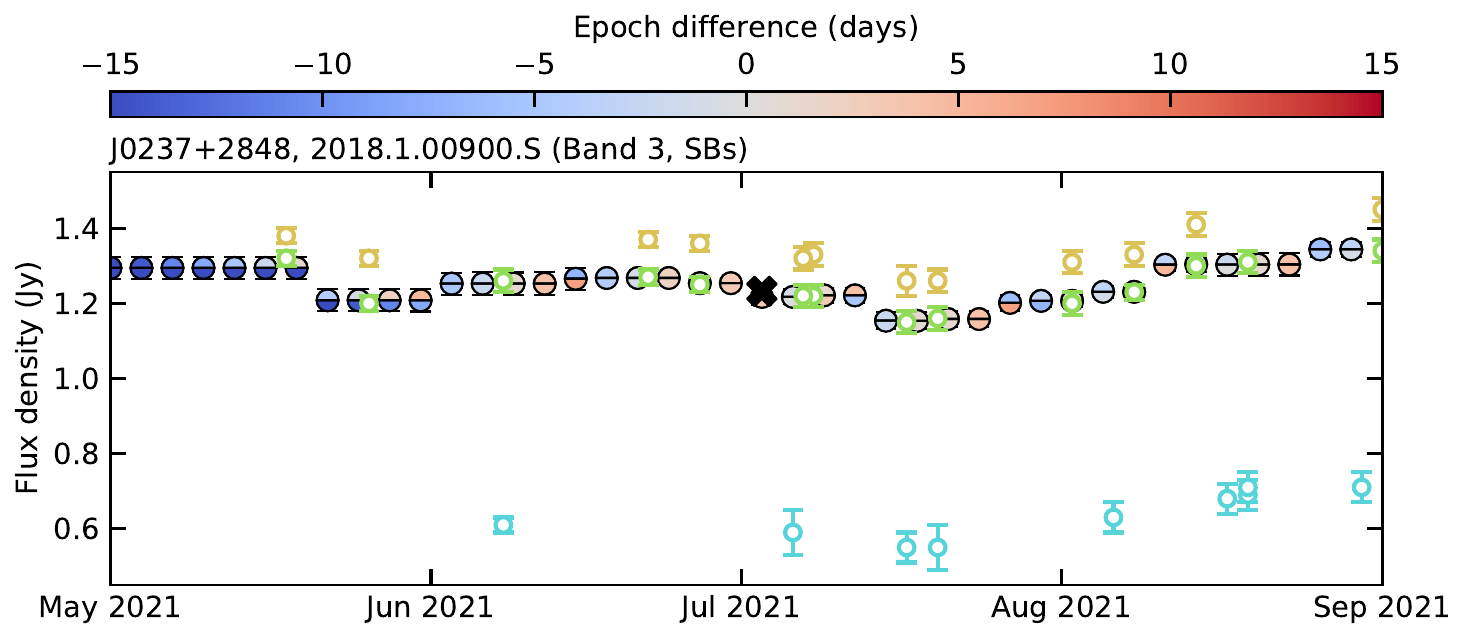}
    \caption{Light curves of the flux calibrators (Band~3). Expected flux densities were interpolated at $104.5\,{\rm GHz}$.}
    \label{fig:A2}
\end{figure*}

\autoref{fig:A1} and \ref{fig:A2} display the light curves (both tabulated and interpolated) of the flux calibrators targeted during ALMA Band~7, 6, and 3 observations. Different subplots refer to a different observing block and hence to a different flux calibrator or observing time. In each subplot, the tabulated flux densities\footnote{As reported in the \href{https://almascience.eso.org/sc/}{ALMA Calibrator Source Catalogue}.} are shown as hollow dots over a timescale of four months around the observation date. Yellow and green colours were used for Band~3 measurements, cyan for the Band~7 ones. The larger, full dots, instead, display the expected luminosities of the flux calibrators at a fixed cadence of three days. These were computed at the frequency of one of the SPWs in the observations (see captions), interpolating between the closest tabulated Band~3 and 7 flux densities, using the \texttt{au.getALMAFlux} task \citep{Hunter2023}. These dots are colour-coded by the epoch offset from (i.e. number of days before or after) the closest tabulated flux density measurements in Band~3 (top-half) and 7 (bottom-half of the dot). Black crosses display the observed luminosity of each flux calibrator in the same SPW the expected luminosities were interpolated at.

\paragraph*{ALMA Band~7 --} The two EBs in Band~7 were observed the same day (11 December 2017). Their flux densities differ by $\approx5\%$ and the luminosities of their flux calibrator (J0510+1800) are within $\lesssim6\%$ of their closest tabulated flux density measurements (see top panel of \autoref{fig:A1}), consistent with the 10\% (2$\sigma$) absolute flux density accuracy reported in the ALMA Technical Handbook\footnote{See Sect.~10.2.6 of the \href{https://almascience.nrao.edu/proposing/technical-handbook/}{ALMA Technical Handbook}.}. Since the two measurements are essentially identical, we selected the latest one as the reference EB.

\paragraph*{ALMA Band~6 --} The long and short baseline observations in Band~6 were taken one year apart (27 August 2016 for the SBs and 23 and 24 September 2017 for the LBs). Their average frequencies are also slightly different ($225.6\,{\rm GHz}$ for the SBs and $233.0\,{\rm GHz}$ for the LBs), complicating the comparison. As for the SBs, the luminosity of its flux calibrator (J0510+1800) is within $\lesssim5\%$ of the expected one interpolated from its closest tabulated measurement in Band~3 and 7 (see bottom panel of \autoref{fig:A1}), consistent with the 10\% ($2\sigma$) absolute flux density accuracy reported in the ALMA Technical Handbook. However, the tabulated measurements were taken 7 days before (Band~3) and 6 days after (Band~7) our observations. Reassuringly, CI~Tau's SB flux density is in excellent agreement (difference $\approx1\%$) with the one reported by \citet{Long2018} using data from the program 2016.1.01164.S (PI: G. Herczeg) at a very similar reference frequency (difference $<1\%$). The 2016.1.01164.S data were taken the same day (27 August 2017) of a tabulated measurement of the luminosity of their flux calibrator (J0510+1800). Their observed and tabulated flux densities agree within $<1\%$. Therefore, we are confident that our SB data were flux calibrated accurately. However, since their average frequency is lower than that of the LB data, using the SBs as our reference EB would systematically underestimate the Band~6 luminosity.

Despite having been observed only one day apart, the flux densities of the two LB execution blocks differ by more than 10\%. The luminosity of the EB observed on 23 September (hereafter LB0) is $175.55\,{\rm mJy}$, while that of the EB observed the day after (hereafter LB1) is $155.63\,{\rm mJy}$. The first would imply an intraband (i.e. with respect to the SB data) spectral index of $6.9\pm4.4$, while the second is consistent with the spectral index between the Band~6 SB and Band~7 data. In addition to these considerations, we chose LB1 as reference EB for its better $uv$-coverage at short baselines, and rescaled the LB0 luminosity to the LB1 one. Curiously, though, the luminosity of the LB0 flux calibrator perfectly matches the expected one, while that of the LB1 differs by more than 10\% (see central panel of \autoref{fig:A1}). This is likely due to the calibrator progressively dimming between September and November 2017, and the closest tabulated measurements of its luminosity being taken 7 days after (Band~3) and 3 days before (Band~7) our observations.

\paragraph*{ALMA Band~3 --} The three EBs in Band~3 were observed more than two years apart (27 June and 4 July 2019, 3 July 2021). Their flux density ratios are $\lesssim4\%$ and the luminosities of their flux calibrators (J0423-0120, J0510+1800, and J0237+2848) are $\lesssim1\%$ of their closest tabulated flux density measurements (see \autoref{fig:A2}), consistent with the 5\% ($2\sigma$) absolute flux density accuracy reported in the ALMA Technical Handbook. Since the same day of our observations conducted on 4 July 2019 a tabulated measurement of the luminosity of the flux calibrator J0510+1800 is available and it agrees within $\lesssim1\%$ with the observed one, we selected this EB as the reference one.

\section{Ku, X, and C band images}\label{sec:app2}
\begin{figure*}
    \centering
    \includegraphics[width=\textwidth]{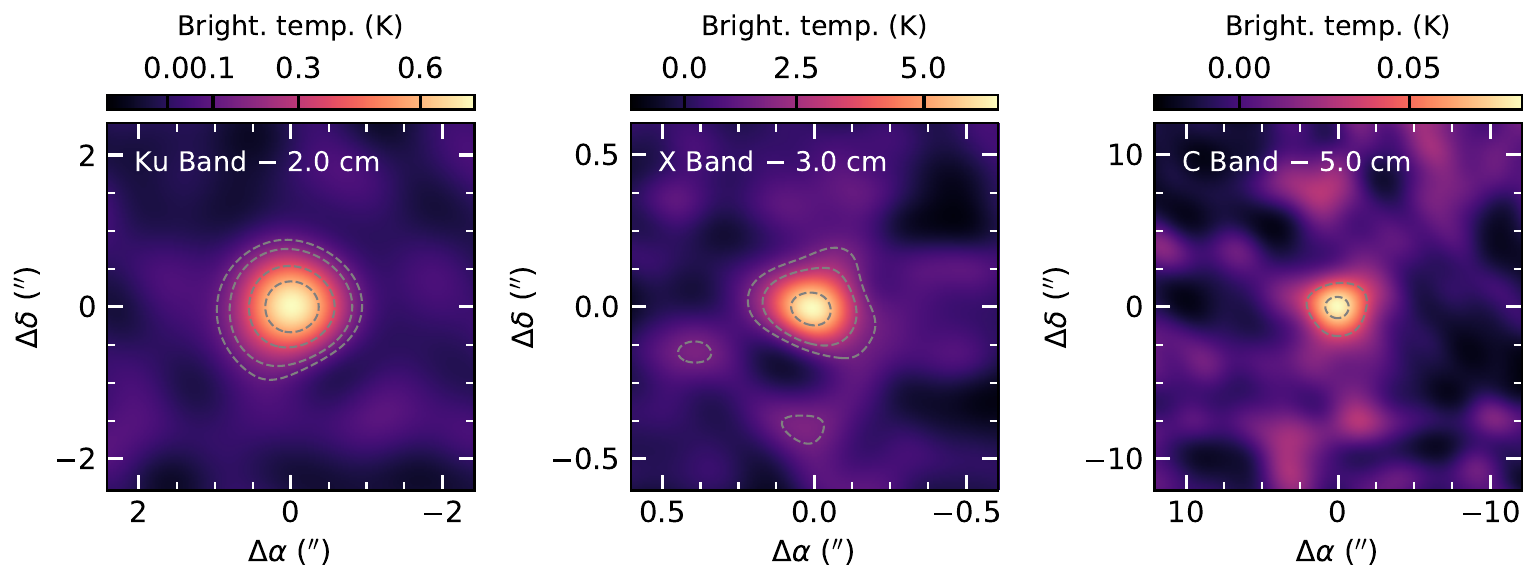}
    \caption{CI~Tau Ku, X, and C~band images, the first one reconstructed with \texttt{uniform} weighting and the others with \texttt{natural} weighting. Disc emission is unresolved at all the wavelengths. Grey contours indicate the $3\sigma$, $5\sigma$, $10\sigma$ (for Ku~band data), and $15\sigma$ (for X~band data) detection levels.}
    \label{fig:A3}
\end{figure*}

VLA Ku, X, and C~band images were reconstructed with the default \texttt{tclean} task parameters. The CLEANing region was determined using auto-masking \citep{Kepley2020}, and, in C~band, where the field of view is polluted by a few background galaxies, additionally setting the subparameter \texttt{noisethreshold=3.65}, to only image the sources at least as bright as CI~Tau at this wavelength. We used a multi-scale multi-frequency synthesis deconvolver with \texttt{nterms=2}, and adopted a set of (Gaussian) deconvolution scales, including a point source and scales corresponding to 8, 16, 30 (for C~band), and 80 (for X and Ku~band) pixels. We adopted the \texttt{natural} and \texttt{uniform} weighting scheme for Ku~band data, while for X and C~band data only the \texttt{natural} weighting was used, due to the lower S/N. CLEANing was performed with a $1\sigma$ noise threshold and $0\farcs04$, $0\farcs008$, and $0\farcs05$ cell sizes for Ku, X, and C~band images, respectively. The image noise was estimated over a circular annulus centred on and larger than the target, with inner and outer radii of $5\farcs0$ and $7\farcs5$ for Ku and X~band, and $10\farcs0$ and $25\farcs0$ for C~band. The images are shown in \autoref{fig:A3} and their parameters are summarised in \autoref{tab:app3}. Disc emission is unresolved at all of these wavelengths.

\section{\texttt{frank} visibility fits}\label{sec:app3}
\begin{figure*}
    \centering
    \includegraphics[width=\textwidth]{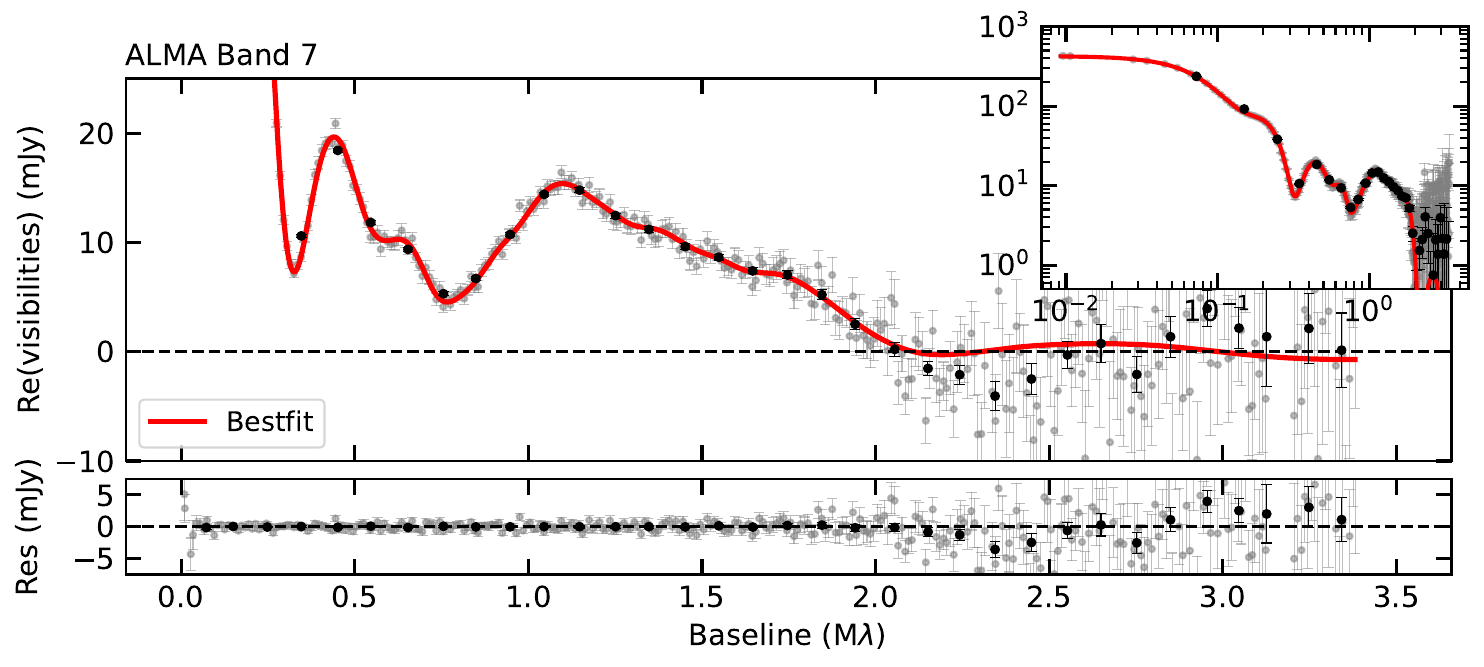}
    \includegraphics[width=\textwidth]{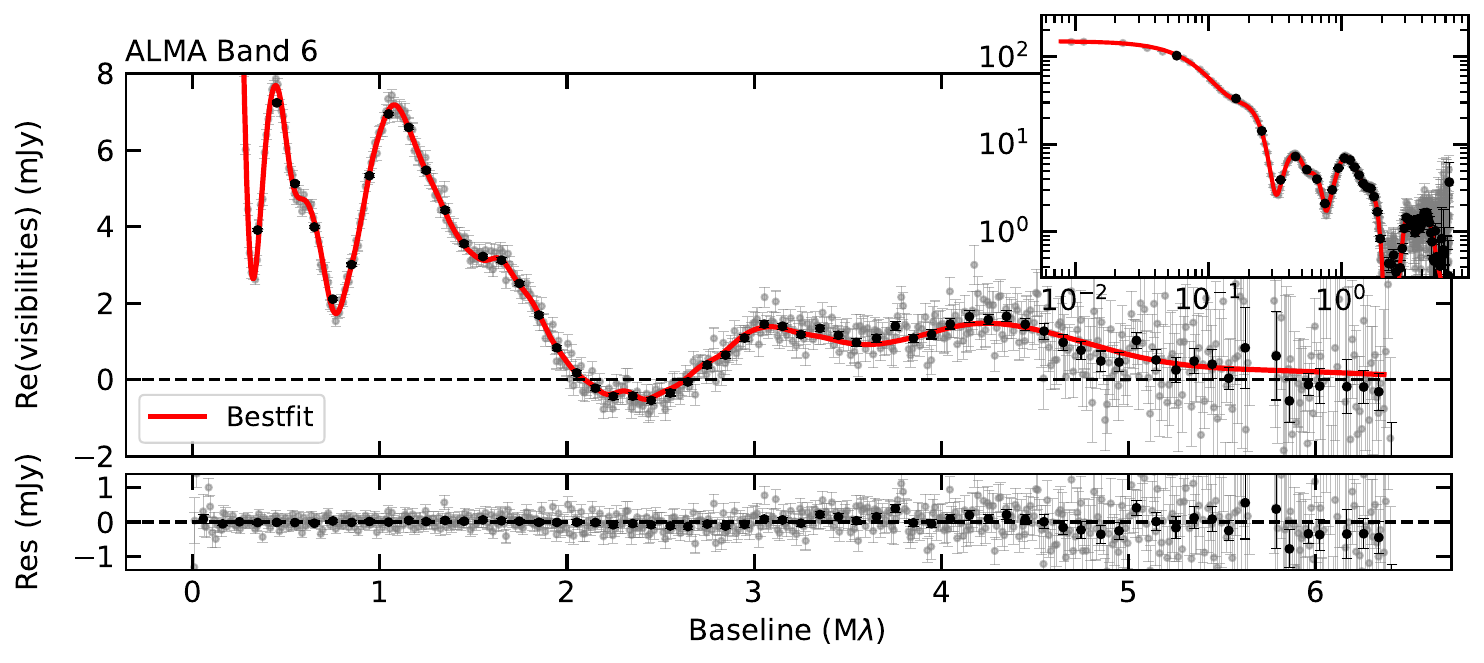}
    \caption{ALMA Band~7 (top) and 6 (bottom) deprojected (real part of the) visibilities, \texttt{frank} fits (red), and residuals. Data and residuals are averaged into bins of equal size ($10^{-2}$ and $10^{-1}\,{\rm M}\lambda$ for grey and black dots, respectively). The inserts show the full visibility range.}
    \label{fig:A4}
\end{figure*}

\begin{figure*}
    \centering
    \includegraphics[width=\textwidth]{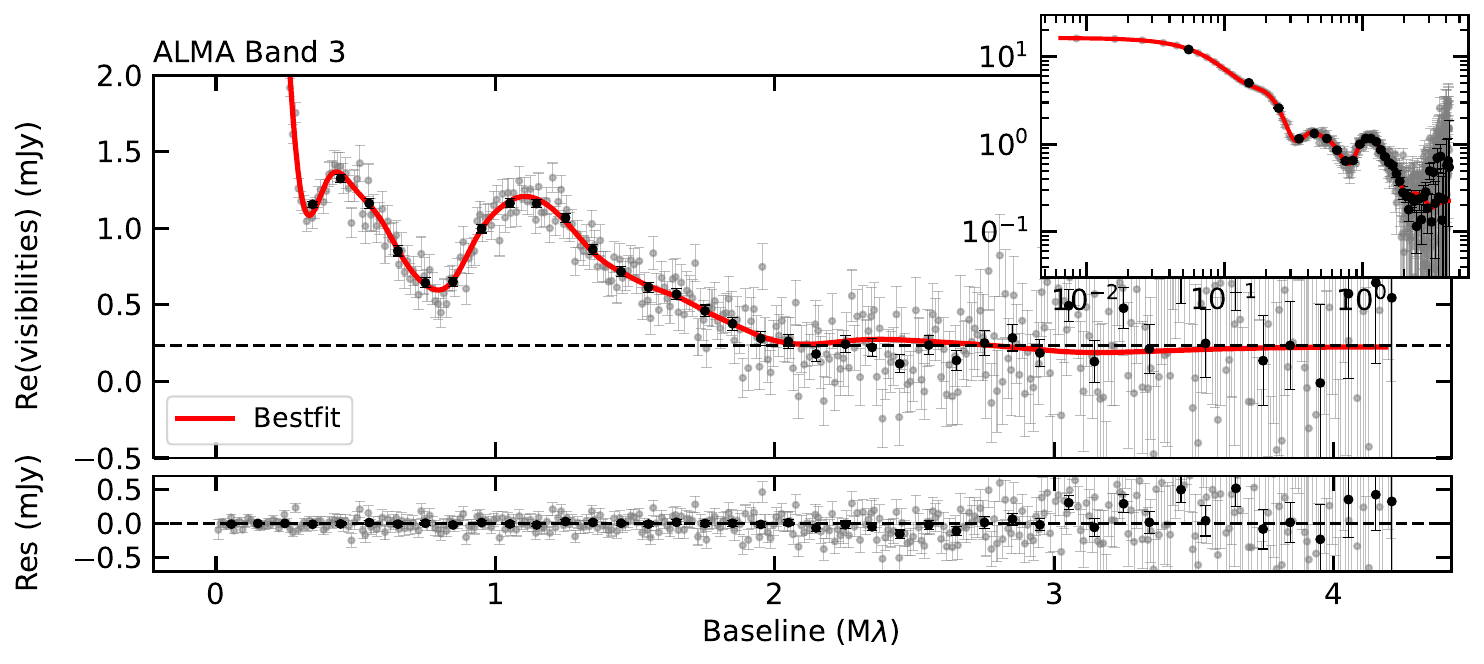}
    \includegraphics[width=\textwidth]{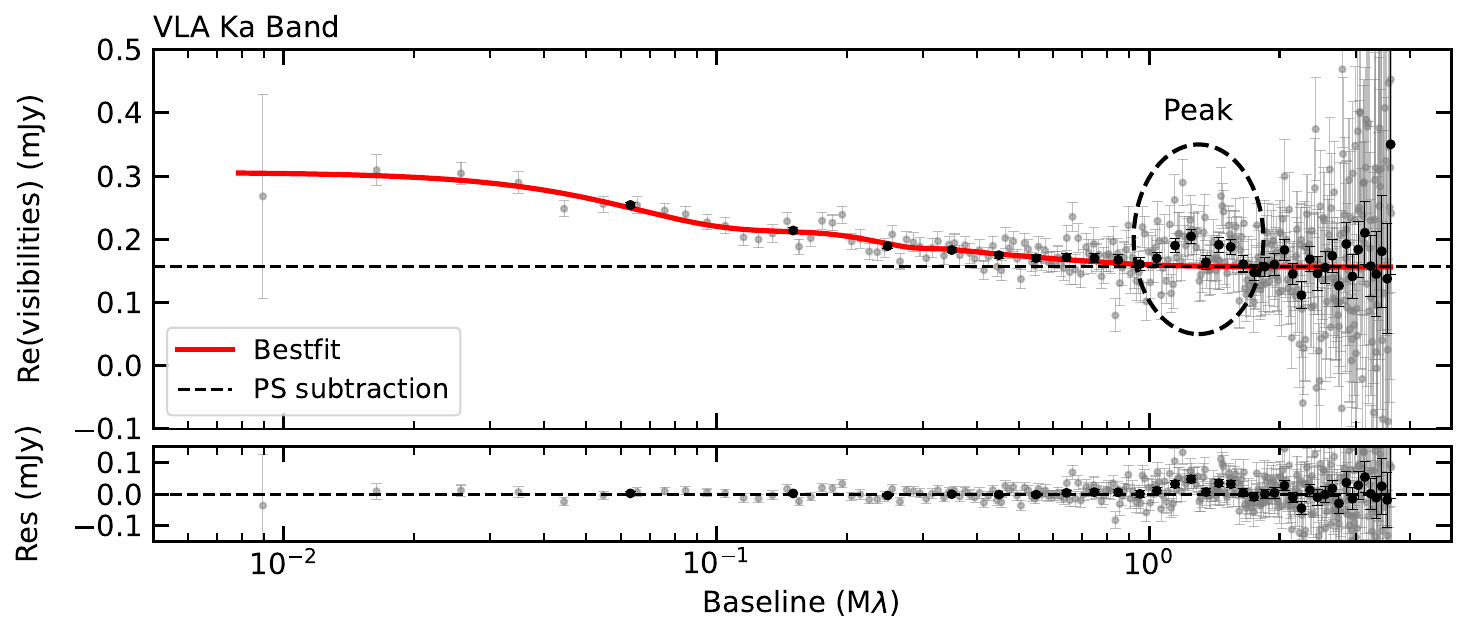}
    \caption{ALMA Band~3 (top) and VLA Ka~band (bottom) deprojected (real part of the) visibilities, \texttt{frank} fits (red), and residuals. Data and residuals are averaged into bins of equal size ($10^{-2}$ and $10^{-1}\,{\rm M}\lambda$ for grey and black dots, respectively). The dashed lines display the non-null flux density offset. The insert shows the full Band~3 visibility range. We plot the VLA Ka~band visibilities in log-scale to highlight that emission is extended.}
    \label{fig:A5}
\end{figure*}

For ALMA data we first (re-)computed and set the weights based on the variance of the data using the \texttt{CASA} task \texttt{statwt}. This step had no visible effect on the images and radial profiles reconstructed using \texttt{tclean}, but significantly improved the residual maps of our visibility fits, especially in the case of band~3 data. CI~Tau visibilities were then extracted from each dataset after applying time averaging to 30 and $60\,{\rm s}$ (for ALMA and VLA Ka~bands, respectively), and spectral averaging to one channel per SPW\footnote{We found identical results when averaging less (i.e. two and four channels per SPW in Band~6 and 3, respectively).}. First, we determined the disc offset from the phase centre ($\Delta\alpha$, $\Delta\delta$), fitting a Gaussian, whose inclination and position angle were fixed to the best-fit values of \citet{Clarke2018}, to the visibility. Compared to a full fit of the disc geometry, this choice reduced the significance of large scale residuals (due to the PA being off by up to $4\,{\rm deg}$ in Band~3 and $30\,{\rm deg}$ in Ka~band). The visibilities where then deprojected and inspected for the presence of flux density offsets. In the case of ALMA Band~3 and VLA~Ka~band data, where such offsets are clearly present, they were subtracted as discussed in \autoref{sec:app4.1}. Subsequently, the resulting visibilities where fitted in log-space, to avoid negative solutions and reduce non-physical wiggles, using the hyper-parameters in Columns (2)--(5) of \autoref{tab:app4}. The deprojected visibilities, our best fits, and the residuals are shown in \autoref{fig:A4} and \ref{fig:A5}, while the best-fit surface brightness profiles are plotted in purple in the central column of \autoref{fig:1}. We measured the 68\% and 95\% flux radius from these profiles using a curve of growth method\footnote{To avoid spurious effects due to our choice of $R_{\rm max}$ and the outer edge of the best-fit profile, we discarded any emission below the $3\sigma$ sensitivity threshold of the azimuthally averaged \texttt{tclean} brightness profiles reconstructed with \texttt{robust=1.5}.}. Size uncertainties were estimated as for the CLEAN image radial profiles, from the uncertainties obtained from the covariance matrix of the visibility fit (see Equation 28 of \citealt{Junklewitz2016}). Our results are summarised in Columns (6)--(7) of \autoref{tab:app4}. In the case of $R_{\rm 68\%}$, that is less sensitive to the profile cut-off, they confirm the trend already seen in the \texttt{tclean} profiles, of a decreasing disc size with wavelength. Except for VLA Ka~band data, where the \texttt{frank} profiles clearly detect extended emission up to $\lesssim1\farcs5$, and due to the point-source subtraction, the dust sizes agree well with those in \autoref{tab:app3}.

\subsection{Point-source subtraction}\label{sec:app4.1}
The ALMA Band~3 and VLA Ka~band deprojected visibilities display a clear flux density offset (i.e. on average are non-null at their longest baselines). Since a point source (PS) in the image plane maps to a constant in the Fourier space, we interpreted this offset as the contribution of a point-like source at the disc centre (e.g. due to free-free from the central star, or a localised, non-resolved inner disc wind, or gyrosynchrotron emission, see \autoref{sec:3.2} for a discussion). Especially in the case of VLA data, where it accounts for about 50\% of the emission, this flux density offset induces high-amplitude spurious oscillations in the fitted profile \citep[e.g.][]{Jennings2022a,Sierra2024}. For this reason, we subtracted it from the deprojected visibilities and only considered the PS-subtracted visibilities in our \texttt{frank} fits. 

We estimated the PS flux density by fitting a constant to the real~part of the unbinned deprojected visibilities at the longest baselines. It goes without saying that our estimated PS flux density is considerably affected by the choice of the longest baseline we fit beyond of. In this paper, this is defined as the minimum baseline where the visibilities flatten out and determined as follows. First, we used the Levenberg-Marquardt algorithm implemented in \texttt{scipy.optimize.curve\_fit} \citep{Virtanen2020} to fit a linear function to the real part of the unbinned deprojected visibilities over a range of baselines $q\in\left[q_0,q_0+q_{\rm incr}\right]$, where $q_0$, the minimum baseline, and $q_{\rm incr}$, the baseline width, are varied as follows. We chose $1.0\,{\rm M}\lambda\leq q_0\leq4.0\,{\rm M}\lambda$ (3.0 for VLA Ka~band data), progressively increasing by $0.1\,{\rm M}\lambda$, and for each $q_0$ we varied $0.4\,{\rm M}\lambda\leq q_{\rm incr}\leq1.0\,{\rm M}\lambda$, every $0.1\,{\rm M}\lambda$. Then, for each chosen baseline width, we inspected the best-fit angular incline as a function of $q_0$ and identified the minimum baseline that minimises the incline. Since, in principle, this can change as a function of the baseline width, we selected as the longest baseline to fit beyond of that consistently minimising the incline over most values of $q_{\rm incr}$. Then we used the Levenberg-Marquardt algorithm implemented in \texttt{scipy.optimize.curve\_fit} to determine the PS flux density. These PS flux densities are displayed as dashed lines in both panels of \autoref{fig:A5}.

In the case of ALMA Band~3 data, the minimum angular incline is attained for $2.0\leq q_0/{\rm M}\lambda\leq 2.4$ regardless of $q_{\rm incr}$. For these values of $q_0$, the PS flux density is $0.23\pm0.03\,{\rm mJy}$. For shorter minimum baselines, the angular incline rapidly increase, while for longer ones, it considerably depends on $q_{\rm incr}$ and the PS flux density is highly uncertain. In the case of VLA Ka~band data, instead, the picture is less clear because of the lower S/N. On the one hand, as can be seen in the bottom panel of \autoref{fig:A5}, the visibilities locally peak between 1.0 and $1.6\,{\rm M}\lambda$, driving the PS flux density fictitiously up. On the other hand, as in the case of Band~3, at longer minimum baselines the PS flux rapidly becomes very uncertain. Thus, we chose $1.7\,{\rm M}\lambda$ as longest baseline to fit beyond of, leading to a PS flux density of $0.16\pm0.01\,{\rm mJy}$. For this value of $q_0$ the minimum angular incline is also roughly null. 

\section{Radio emission variability}\label{sec:app4}

To explore the short-term continuum emission variability in the Ku and X~band observations, we split our data scan by scan and measured their flux densities. Ku~band data were taken two weeks apart (on February 19 and March 04, 2020), in 3 scans $\approx7\,{\rm min}$-long for each execution block, while X~band data were taken the same day (August 03, 2019), in 12 scans $\approx7\,{\rm min}$-long. For both Ku and X~band data, for each single-epoch dataset, we reconstructed continuum emission images and estimated their noise using the same parameters as in \autoref{sec:app2}, \texttt{natural} weighting, and a fixed circular mask of $5\farcs0$ and $1\farcs0$ radius for Ku and X~band data, respectively (almost identical to those generated by auto-masking in \autoref{sec:app2}). Emission was detected in all the images with peak ${\rm S/N}>20$ for Ku~band data and $\gtrsim3$ for X~band data. Since those detection levels are rather low, especially in X~band, we opted to measure flux densities fitting the continuum visibilities for each single-epoch dataset with a PS model adopting a Bayesian approach similar to the one described in \autoref{sec:4.2}.

\begin{figure*}
    \centering
    \includegraphics[width=\textwidth]{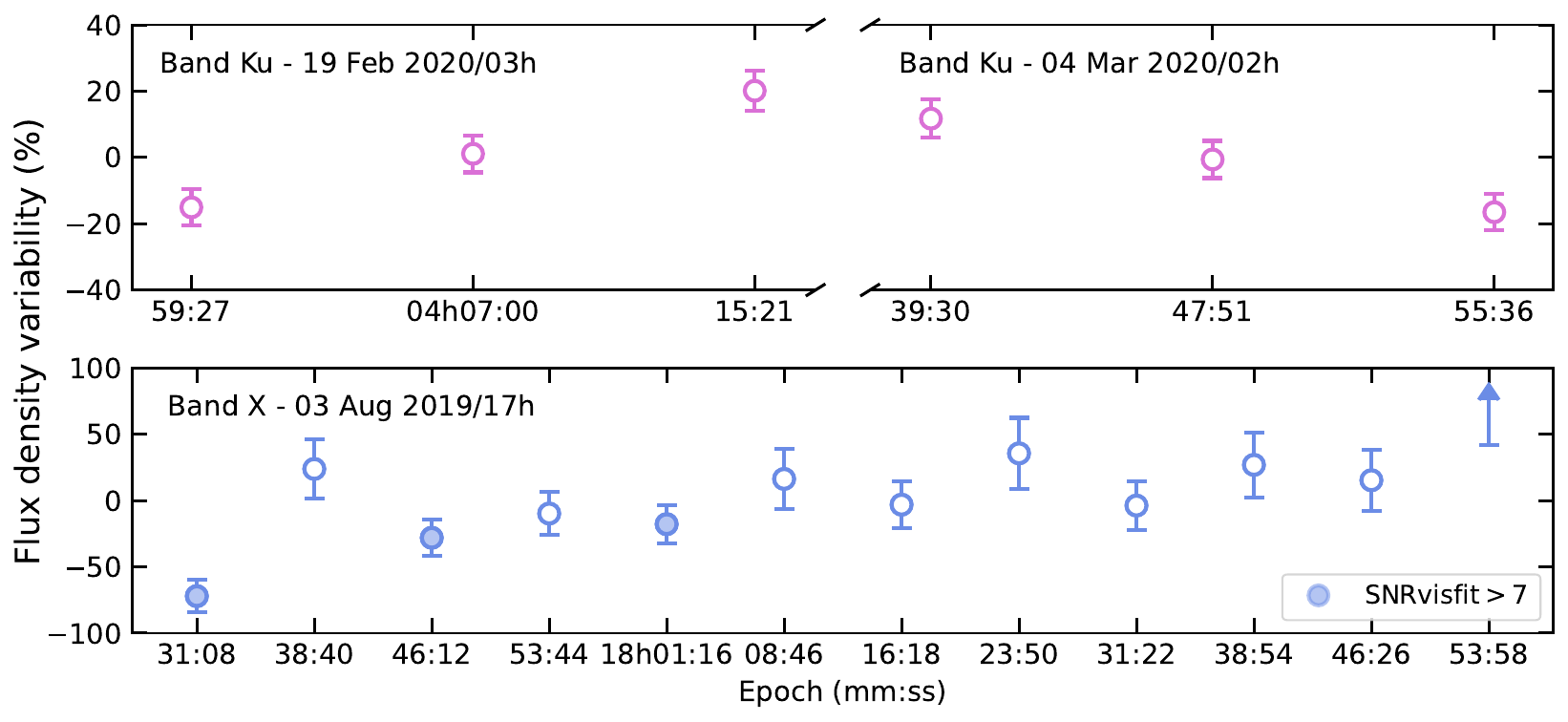}
    \caption{CI~Tau VLA Ku (top) and X~band (bottom) flux density variability is $\lesssim30\%$ over a timescale of minutes to hours (X~band) and weeks (Ku~band).}
    \label{fig:A8}
\end{figure*}

\autoref{fig:A8} displays the time variability of the VLA Ku and X~band emission, estimated as $100 \times (1 - F_{\rm scan}/F_{\rm conc})$, where $F_{\rm scan}$ is the flux density of each single-scan dataset and $F_{\rm conc}$ is the flux density of the concatenated dataset, measured in the visibility plane as for each scan. These flux densities are consistent, with those inferred by integrating the peak intensity of each \texttt{tclean} image over a synthesised beam within their $1\sigma$ RMS noise. The uncertainties were propagated from the confidence interval of each fit and, for Ku~band single-epoch data, a 5\% flux calibration uncertainty was also considered because observations were taken in different execution blocks. As is clear from \autoref{fig:A8}, while the continuum flux density changes by $\lesssim30\%$ over a timescale of minutes to weeks in the Ku~band, in the X~band it dims by more than 50\%, almost flares back in $\approx15\,{\rm min}$, and then remains steady in the subsequent hour. No emission is detected in the last scan.

\section{Temperature prior}\label{sec:app5}
In this Section we discuss the temperature prior used in our fit. For a passively irradiated disc, the mid-plane temperature profile can be approximated as \citep{Kenyon&Hartmann1987}
\begin{equation}\label{eq:app5.1}
    T=\left(\dfrac{\phi L_\star}{8\pi R^2\sigma_{\rm SB}}\right)^{1/4},
\end{equation}
where $\phi$ is the disc flaring angle, $L_\star$ is the stellar luminosity, $R$ is the disc radius and $\sigma_{\rm SB}$ is the Stefan-Boltzmann constant. Following \citet{Macias2021}, we assumed that the stellar luminosity is distributed as a Gaussian, with mean $L_\star=1.04\,L_\odot$\footnote{CI~Tau is an actively accreting young star, with accretion luminosity, $\log(L_{\rm acc}/L_\odot)=0.02\pm0.29$ (inferred by \citealt{Gangi2022} averaging over 19 emission lines in the optical and near IR), comparable to its stellar luminosity. We chose not to include accretion luminosity in our temperature prior since we expect it to change our estimates by $\lesssim20\%$.} and standard deviation $0.48 L_\odot$ \citep{Gangi2022}, and that the flaring angle is uniformly distributed between 0.01 and 0.06.

We computed analytically the probability density function associated with the disc temperature considering the more general case of a random variable $Z=(cXY)^k$, i.e. the power-law of the product of two random variables and a constant. The probability density function of this random variable can be written as
\begin{equation}
    \begin{gathered}
        F_Z(z):=\mathbb{P}(Z\leq z)=\mathbb{P}\Bigl((cXY)^k\leq z\Bigr)=\mathbb{P}\left(XY\leq z^{1/k}/c\right)=\\
        = \mathbb{P}\left(Y\leq z^{1/k}/Xc,X\geq0\right)+
        \mathbb{P}\left(Y\geq z^{1/k}/Xc,X\leq0\right)=\\
        \hspace{-15ex}=\int_0^\infty f_X(x)\int_{-\infty}^{z^{1/k}/xc}f_Y(y)dydx+\\
        \hspace{+15ex}+\int^0_{-\infty} f_X(x)\int^\infty_{z^{1/k}/xc}f_Y(y)dydx
    \end{gathered}
\end{equation}
by the fundamental theorem of calculus and the Leibniz rule
\begin{equation}
    f_Z(z):=\dfrac{d}{dz}F_Z(z)=\int_{-\infty}^\infty f_X(x)f_Y\left(\dfrac{z^{1/k}}{cx}\right)\dfrac{z^{1/k-1}}{kc\lvert x\rvert}dx.
\end{equation}
If we take $X$ to be distributed as a Gaussian with mean $\mu$ and standard deviation $\sigma$, and $Y$ to be uniformly distributed between $y_{\rm min}$ and $y_{\rm max}$, then
\begin{equation}\label{eq:app5.2}
    \begin{gathered}
        \hspace{-15ex}f_Z(z)=\dfrac{1}{y_{\rm max}-y_{\rm min}}\dfrac{1}{\sqrt{2\pi\sigma^2}}\times\\
        \hspace{+15ex}\times\int_{z^{1/k}/cy_{\rm max}}^{z^{1/k}/cy_{\rm min}}\exp\left\{-\dfrac{(x-\mu)^2}{2\sigma^2}\right\}\dfrac{z^{1/k-1}}{kcx}dx.
    \end{gathered}
\end{equation}
When $z=T$, $x=L_*$, $y=\phi$, $c=(8\pi R^2\sigma_{\rm SB})^{-1}$ and $k=0.25$, \autoref{eq:app5.2} gives the probability density function corresponding to our definition of the temperature profile in \autoref{eq:app5.1}.

\section{Data and model comparison}\label{sec:app6}

\begin{figure*}
    \centering
    \includegraphics[width=\textwidth]{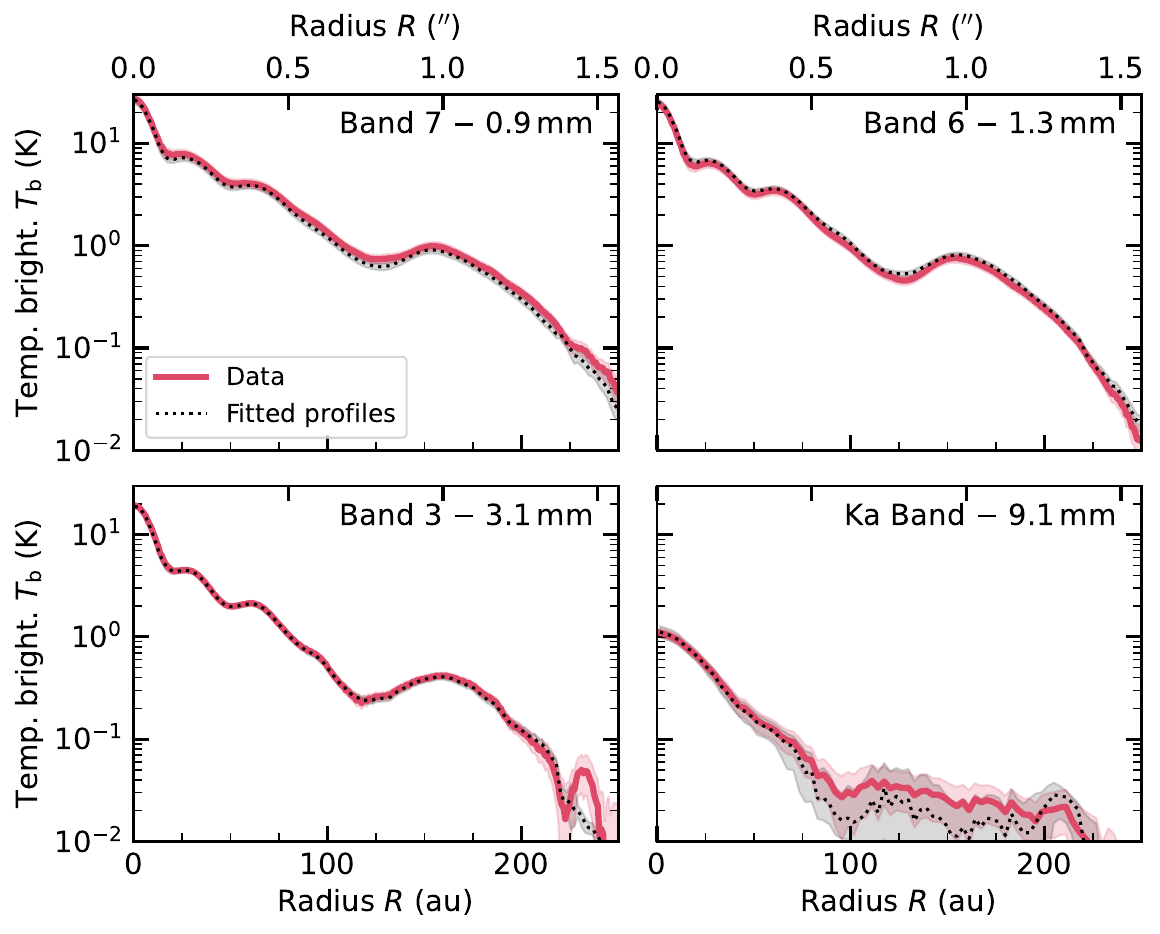} 
    \caption{Comparison between the azimuthally averaged ALMA Band~7, 6, 3, and VLA Ka~band surface brightness radial profiles (purple solid line and shaded region, including the calibration uncertainty) with the posterior distribution of our fit (black dashed line and shaded region, for the median and the area between the 16\textsuperscript{th} and 84\textsuperscript{th} percentiles). Results from the high (low) resolution fit are displayed for ALMA (VLA) data. An excellent agreement between models and observations can be seen.}
    \label{fig:A10}
\end{figure*}

\begin{figure*}
    \centering
    \includegraphics[width=\textwidth]{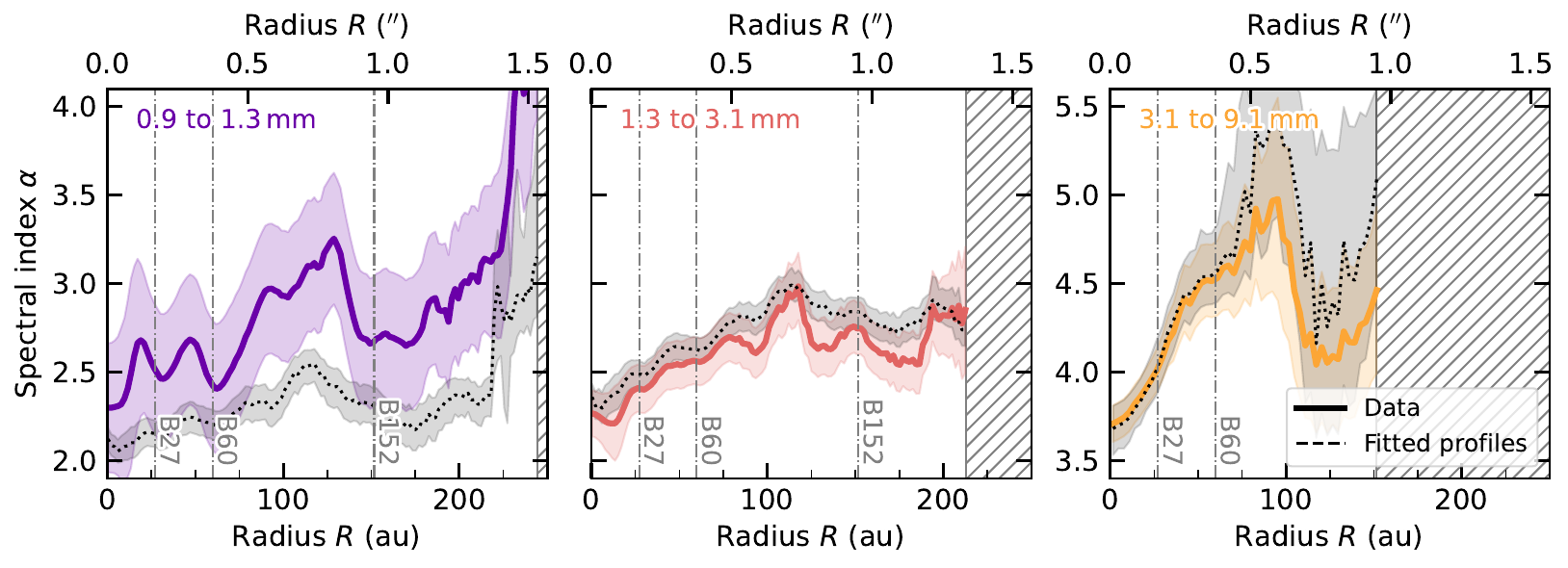} 
    \caption{Same as in \autoref{fig:A10}, but for the spectral index radial profiles discussed in \autoref{sec:3.3} (see \autoref{fig:3}). Results from the high (low) resolution fit are displayed for $\alpha_{\rm B7-B6}$ and $\alpha_{\rm B6-B3}$ ($\alpha_{\rm B3-Ka}$) data.}
    \label{fig:A11}
\end{figure*}

As a sanity-check, in this section we show that the posterior distribution of our fitting parameters can recover the input surface brightness and spectral index radial profiles. To do so, we used our MCMC samples to determine the posterior distributions of the model intensity, $S_\nu$, at 0.9, 1.3, 3.1, and $9.1\,{\rm mm}$, and the spectral indices ($\alpha_{\rm B7-B6}$, $\alpha_{\rm B6-B3}$, and $\alpha_{\rm B3-Ka}$), radius by radius.

\autoref{fig:A10} displays the azimuthally averaged ALMA Band~7, 6, 3 ($0\farcs058$ resolution) and VLA Ka~band ($0\farcs195$ resolution) surface brightness radial profiles as purple solid lines and their uncertainty (the quadrature sum of the error of the mean due to deprojection and the absolute flux density calibration uncertainty) as a shaded region of the same colour. The median and $1\sigma$ spread of our (high resolution for ALMA bands and low resolution for VLA Ka~band) model (defined as the region between the 16\textsuperscript{th} and 84\textsuperscript{th} percentiles of the posterior distribution) surface brightness radial profiles are shown as black dashed lines and shaded areas of the same colour. A remarkable agreement can be seen between the data and model profiles at all the wavelengths.

Similarly, \autoref{fig:A11} displays the spectral index radial profiles extracted as is \autoref{sec:3.2} in violet (for $\alpha_{\rm B7-B6}$), purple (for $\alpha_{\rm B6-B3}$), and yellow (for $\alpha_{\rm B3-Ka}$). Their uncertainties were obtained propagating those on the azimuthally averaged surface brightness radial profiles and include the absolute flux density calibration error. The median and $1\sigma$ spread of our model spectral indices (defined as for the surface brightness radial profile posteriors above) are shown as black dashed lines and shaded areas of the same colour. Models and observations are in remarkable agreement, with the exception of the $\alpha_{\rm B7-B6}$ profiles. This is because the model surface brightness radial profile slightly overestimates (underestimates) the observed ALMA Band~6 (7) one by $\approx 10\%$ ($5\%$). Increasing the absolute flux calibration uncertainty does not improve the agreement between models and data. 

To further investigate the origins of this discrepancy, we performed a low resolution fit with an additional free parameter: a brightness scale factor for the $1.3\,{\rm mm}$ data that can be interpreted as a flux calibration offset. We adopted this additional degree of freedom only for the ALMA Band~6 surface brightness radial profile because, as discussed in \autoref{sec:app1}, its flux calibration uncertainty is the largest among our data due to the light curve variability of the LB flux calibrator. Although, by construction, this scale factor is allowed to be different at each radius, its posterior distributions are similar at all the radii as long as the VLA Ka~band surface brightness radial profile is detected with ${\rm S/N}>5$. In particular, its median is radially constant, and it perfectly matches the flux density offset between the two EBs of our Band~6 LB data (1.128). This suggests that LB0 (rather than LB1; see \autoref{sec:app1}) should be used as a reference EB when setting the Band~6 LB flux scale. This conclusion is also in line with the recently published SMA photometry of CI~Tau \citep{Chung2024}. We also re-ran our fitting procedure after re-scaling our Band~6 surface brightness radial profile by the flux density offset determined in the previous step. Our results show that, in this case, the quality of the fit improves and the mismatch between $\alpha_{\rm B7-B6}$ and our models completely vanishes. The marginalised posterior distributions of the four fitted parameters are perfectly consistent with those shown in \autoref{fig:4}.

\section{Best-fit radial profiles}\label{sec:app7}
\begin{figure*}
    \centering
    \includegraphics[width=\textwidth]{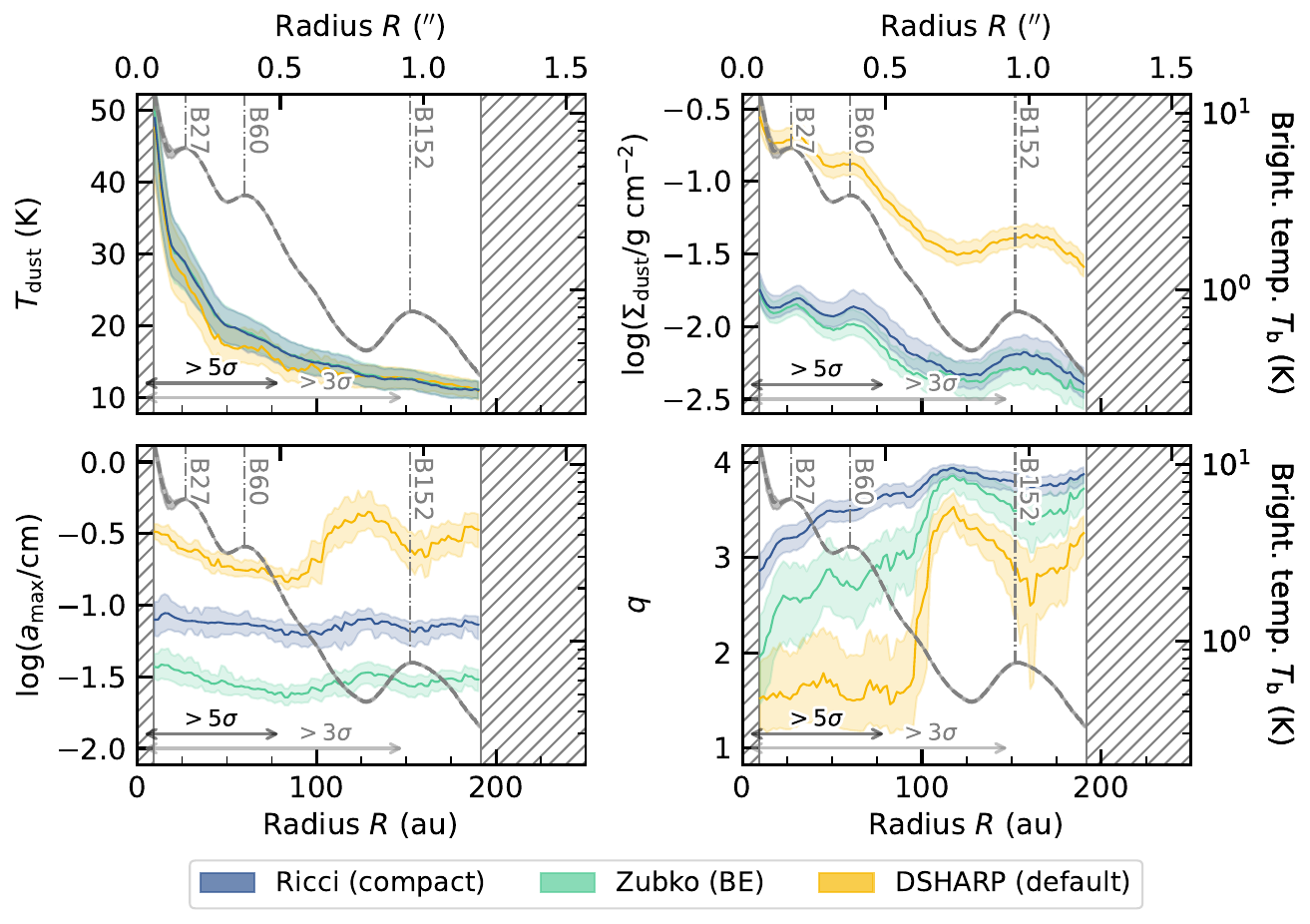}
    \caption{Same as in \autoref{fig:4} but showing a comparison among the best-fit results for different compositions: `Ricci (compact)' in blue, `Zubko (BE)' in turquoise, and `DSHARP (default)' in yellow. The highest quality (i.e. lowest $\chi^2$) fits, obtained adopting the `Ricci (compact)' and `Zubko (BE)' optical properties, provide similar marginalised posterior distributions for all the parameters (within a factor of 3 for $a_{\rm max}$ and 2 for $\Sigma_{\rm dust}$).}
    \label{fig:A12}
\end{figure*}

For the three sets of compositions (`Ricci (compact)', `Zubko (BE)', and `DSHARP (default)') that provide the best agreement with our data, we also performed high angular resolution fits\footnote{`Zubko (ACH2)' opacities give almost identical results to those obtained for the `Zubko (BE)' ones and will not be discussed further. Similar considerations apply to the results obtained with `DSHARP (highwater)' and `DSHARP (icefree)' optical properties in comparison with the standard `DSHARP (default)' ones.} as explained in \autoref{sec:4.3}. Corner plots comparing the posterior distributions obtained for such opacities at the position of the three bright rings in the system can be found in \autoref{sec:app10}, while their marginalised posterior distributions
are displayed in \autoref{fig:A12} in blue, turquoise, and yellow, respectively. 

\begin{figure*}
    \centering
    \includegraphics[width=\textwidth]{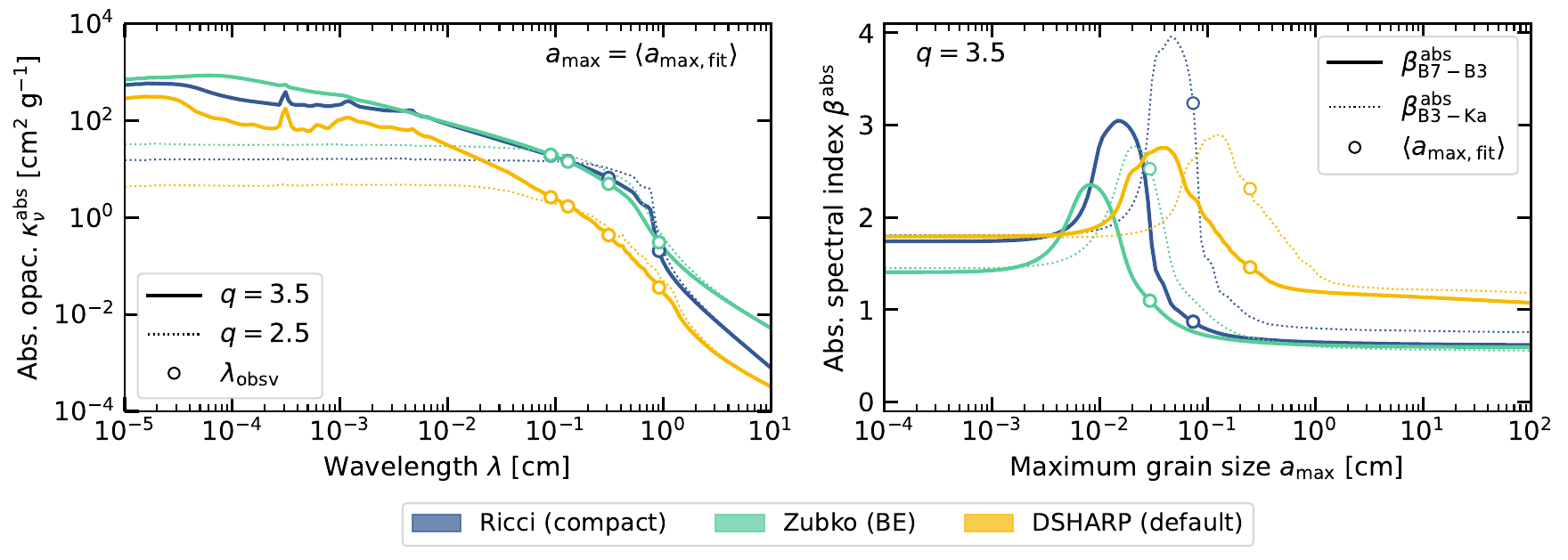}
    \caption{Left: Absorption opacity, $\kappa_\nu^{\rm abs}$, averaged over different dust size distribution ($q=3.5$, solid lines, and $q=2.5$, dotted lines) and the radially averaged best-fit maximum grain size from \autoref{fig:A12}, $\langle a_{\rm max,fit}\rangle$, plotted as a function of $\lambda$ for the `Ricci (compact)' in blue, `Zubko (BE)' in cyan, and `DSHARP (default)' composition in turquoise. Dots of the same colours mark the wavelengths our high-resolution observations were taken at. Right: Absorption spectral index, $\beta^{\rm abs}$, at different wavelengths ($\beta^{\rm abs}_{\rm B7-B3}$, solid lines, and $\beta^{\rm abs}_{\rm B3-Ka}$, dotted lines), for $q=3.5$ as a function of the maximum grain size for the same dust mixture in the left panel. Dots of the same colours mark the radially averaged best-fit maximum grain size from \autoref{fig:A12}.}
    \label{fig:A13}
\end{figure*}

As is clear from the upper-left panel, the temperature profile is consistent with that of a passively irradiated disc regardless of the assumed dust composition. Instead, the marginalised posteriors for the other fitting parameters show substantial differences. The dust surface density profiles, displayed in the upper-right panel of \autoref{fig:A12}, all decrease towards the outer disc and locally peak at the position of the bright rings. However, while the posterior distributions obtained for `Ricci (compact)' and `Zubko (BE)' opacities are very similar (within less than a factor of two), the `DSHARP (default)' composition prefers a much denser disc (more than a factor of six, increasing up to fifteen in the inner disc). This discrepancy is due to the different absorption opacity normalisation among these sets of optical properties (and more specifically the different dielectric constants of amorphous carbon and refractory organics), as can be seen in the left panel of \autoref{fig:A13}, clearly showing that the `DSHARP (default)' mixture is the less absorbing one among those tested. Moving on to the bottom-left panel of \autoref{fig:A12}, the `Ricci (compact)' and `Zubko (BE)' mixtures also lead to similarly flat maximum grain size radial profiles with comparable magnitudes ($a_{\rm max}\approx7.4\times10^{-2}\,{\rm cm}$ and $2.9\times10^{-2}\,{\rm cm}$, respectively). Instead, the best-fit profile obtained for the `DSHARP (default)' opacities is not flat, but peaks at the location of the outermost dust gap. We do not consider this feature to be physical, but more likely an artefact of the low S/N of our VLA Ka~band data at these locations, combined with the lower flexibility of the optical properties of this mixture. The `DSHARP (default)' opacities also favour the presence of much larger grains in this system, with $a_{\rm max}\approx2.5\times10^{-1}\,{\rm cm}$ (i.e. more than three times larger than for `Ricci (compact)' grains). This discrepancy is primarily due to the different location of the Mie resonance in the $\beta^{\rm abs}$ profile for the optical properties we considered (see the right panel of \autoref{fig:A13}). Finally, as can be seen from the bottom-right panel of \autoref{fig:A12}, the `Ricci (compact)' and `Zubko (BE)' opacities display similar trends in the dust density distribution too. Noticeably, in the outer disc these profiles are also in line with those inferred adopting the `DSHARP (default)' composition. However, in this case the density distribution sharply decreases and flattens to $q\approx1.5$ for $R\leq100\,{\rm au}$. Once again, this behaviour can be motivated by the significantly higher minimum absorption spectral index (for any fixed dust density distribution) for the `DSHARP (default)' mixture compared with the `Ricci (compact)' and `Zubko (BE)' ones, as can be seen in the right panel of \autoref{fig:A13} for $a_{\rm max}\gtrsim1\,{\rm mm}$. Thus, since for steeper dust density distributions the Mie resonance is sharper, the only possibility of fitting both the large $\alpha_{\rm B3-Ka}$ and the small $\alpha_{\rm B6-B3}$ with `DSHARP (default)' grains is to prefer a smaller $q$. Instead, the sharp transition between the inner and the outer disc is due to the increase in $\alpha_{\rm B6-B3}$ and the lower VLA Ka~band S/N, that makes $\alpha_{\rm B3-Ka}$ more easy to fit because of its larger uncertainty.

Our results for the three dust compositions considered in this section agree that CI~Tau's continuum emission is (marginally) optically thin between 0.9 and $9.1\,{\rm mm}$, but their best-fit optical depths show remarkable differences. In fact, while for both the `Ricci (compact)' and `Zubko (BE)' mixtures $\tau_\nu\leq0.4$ even at $0.9\,{\rm mm}$, for the `DSHARP (default)' composition we inferred significantly higher optical depths, with $\tau_\nu\geq0.1$ already at $9.1\,{\rm mm}$. This difference is primarily driven by the single-scattering albedo. Indeed, because of their higher absorption opacity (left panel of \autoref{fig:A13}), for the mixtures including amorphous carbon (i.e. `Ricci (compact)' and `Zubko (BE)') $\omega_\nu\leq0.5$. Instead, when the `DSHARP (default)' composition is considered, our results suggest that $\omega_\nu\geq0.6$ (and $\omega_\nu\geq0.8$ for $\lambda\geq3.1\,{\rm mm}$), leading to a more opaque disc. However, in the (marginally) optically thin limit, these two effects compensate each other: a higher optical depth increases continuum emission (second term in the curly brackets of \autoref{eq:4.1}), but a larger albedo reduces it (third term in the curly brackets of \autoref{eq:4.1}), and indeed, unsurprisingly, the absorption optical depth (i.e. $\tau_\nu^{\rm abs}=\Sigma_{\rm dust}\kappa_\nu^{\rm abs}$), is, in fact, almost identical for the three compositions we tested (at each wavelength). As a final remark, we stress that these optical depth differences come as an additional opportunity to tell what dust mixtures can explain CI~Tau's observations the best, for example by comparison with an independent estimate of the total dust extinction. In this regard, a promising direction is the direct measurement of the line-of-sight dust extinction based on the attenuation of the CO brightness from the back side of the disc at the position of bright dust rings, as successfully demonstrated by \citet{Isella2018} for HD~163296 and \citet{Guzman2018} for AS~209. Although comparable-quality CO observations are also available for CI~Tau \citep{Rosotti2021}, they are significantly affected by cloud absorption, and emission from the back side of the disc is detected with too low S/N to perform a similar analysis.

To summarise, in addition to the larger $\chi^2$, already discussed in \autoref{sec:5.3}, the maximum grain size and dust density distribution profiles shown in \autoref{fig:A12} also provide clear indications (e.g. the $a_{\rm amax}$ and $q$ radial trends) that, compared with our fiducial `Ricci (compact)' opacities, the `DSHARP (default)' mixture gives less reliable results in the case of CI~Tau. However, better data, such as deeper CO observations, are needed to confirm our statement. The `Ricci (compact)' and both `Zubko' mixtures, instead, fit our data better, converging to similar best-fit profiles. We are thus confident that the uncertainties on our results for $\Sigma_{\rm dust}$ and $a_{\rm amax}$ displayed in \autoref{fig:4} are at most a factor of 3 and 2, respectively, when also considering different dust mixtures including amorphous carbonaceous grains.

\section{Considerations on turbulence}\label{sec:app8}

In the scenario proposed by \citet{Jiang2024}, the smooth maximum grain size radial profile inferred from our data can be interpreted as the consequence of turbulent fragmentation of fragile pebbles. Under the assumption of a standard (and homogeneous) dust-to-gas ratio, $Z=10^{-2}$, we can use our fiducial dust temperature, density, and grain size posterior distributions to estimate the (posterior probability of the) turbulence as a function of the disc radius in CI~Tau as (see Equation 8 of \citealt{Birnstiel2012})
\begin{equation}\label{eq:app7.1}
    \alpha_{\rm turb}=\dfrac{2}{3\pi}\dfrac{\Sigma_{\rm dust}}{\rho_{\rm s}a_{\rm max}}\left(\dfrac{u_{\rm frag}}{c_{\rm s}Z}\right)^2,
\end{equation}
where $\rho_{\rm s}$ is the dust material density ($1.70\,{\rm g}\,{\rm cm}^{-3}$ for our fiducial `Ricci (compact)' composition), $u_{\rm frag}$ is the fragmentation velocity threshold ($1\times10^2\,{\rm cm}\,{\rm s}^{-1}$ for fragile pebbles, see the discussion of \citealt{Pinilla2021} and \citealt{Jiang2024}, and the references therein), and $c_{\rm s}\propto T_{\rm dust}^{1/2}$ is the locally isothermal sound speed.

\begin{figure}
    \centering
    \includegraphics[width=\columnwidth]{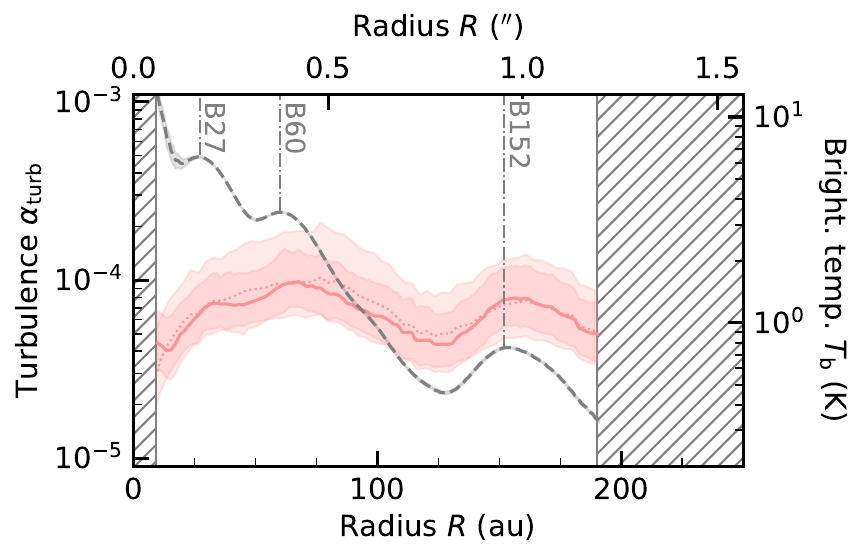} 
    \caption{The pink solid lines and shaded regions display the median and $1\sigma$ uncertainty of the turbulence radial profile estimated for our fiducial “Ricci (compact)” composition. For comparison, the dotted lines show the results of the low resolution fit.}
    \label{fig:A14}
\end{figure}

\autoref{fig:A14} displays the posterior distribution of the turbulence radial profile for our fiducial “Ricci (compact)” dust mixture. This profile is approximately radially constant owing to the shallow dependence of the dust surface density on the disc radius (see \autoref{fig:4}) and shows only low amplitude modulations at the location of the substructures in the continuum. Its magnitude ($\alpha_{\rm turb}\approx7.5\times10^{-5}$) is comparable with that obtained under the same assumptions by \citet{Jiang2024} in TW~Hya, AS~209, HD~163296, and GM~Aur (in this latter case, the turbulence radial profile is also qualitatively similar to CI~Tau's one).

\section{Absorption spectral index}\label{sec:app9}

\begin{figure}
    \centering
    \includegraphics[width=\columnwidth]{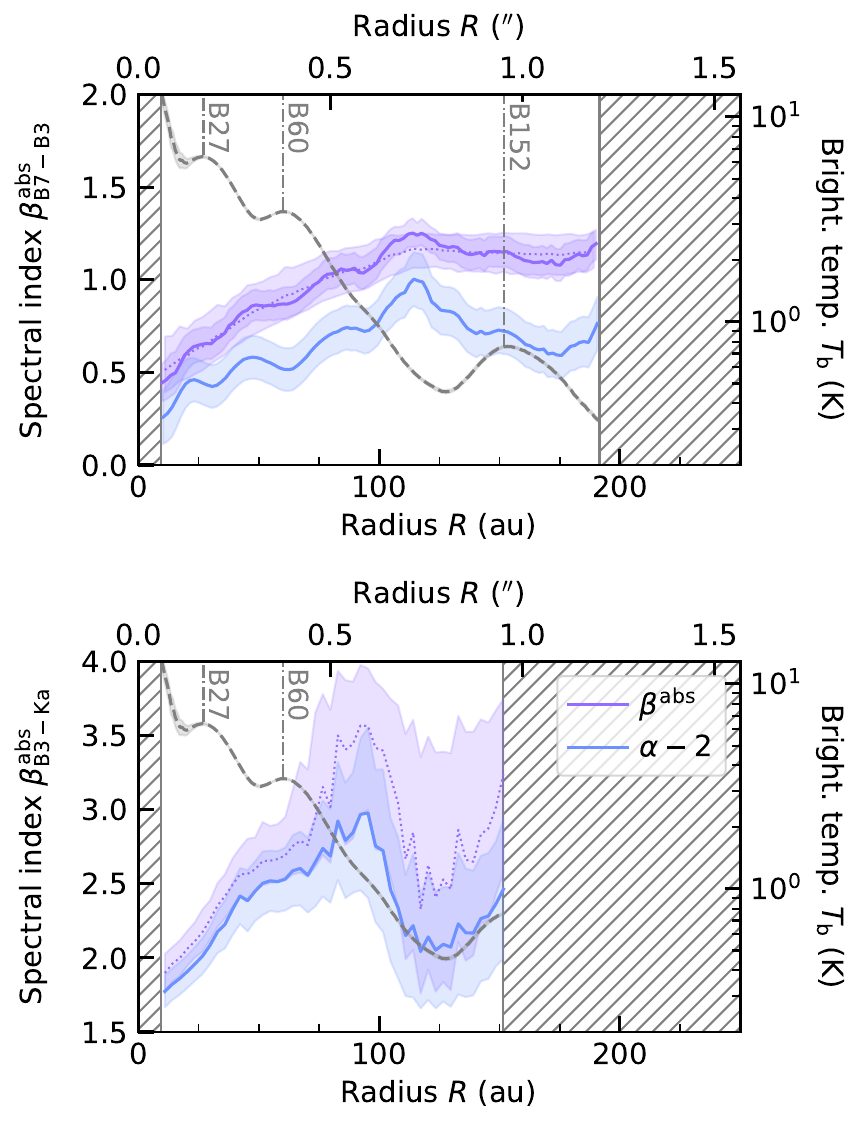} 
    \caption{Absorption spectral index ($\beta^{\rm abs}$) posterior distributions for our fiducial “Ricci (compact)” composition in lilac and the spectral index ($\alpha-2$) radial profile in blue between ALMA Band~7 and 3 (left), and ALMA Band~3 and VLA Ka~band (right). The solid lines and shaded regions display the median and $1\sigma$ uncertainty of each parameter. For comparison, the dotted lines show the results of the low resolution fit.}
    \label{fig:A15}
\end{figure}

\autoref{fig:A15} displays the posterior distribution of the absorption spectral index ($\beta^{\rm abs}$) radial profile between ALMA Band~7 and 3 (left), and ALMA Band~3 and VLA Ka~band (right) for our fiducial “Ricci (compact)” composition in lilac. The absorption spectral index inferred from the spectral index radial profiles at the same wavelengths (\autoref{fig:3}) in the optically thin assumption and Rayleigh-Jeans approximation (i.e. $\beta^{\rm abs}=\alpha-2$) is also plotted for comparison. In both panels, the two profiles show almost identical radial trends (e.g. both $\beta^{\rm abs}_{\rm B7-B3}$ and $\alpha_{\rm B7-B3}-2$ peak at the position of the dark rings in the continuum) with little offsets, that can be explained as follows. The marginal optical depth of our ALMA Band~7 and 3 data in the inner disc, and the low best-fit temperature ($T_{\rm dust}<20\,{\rm K}$) in the outer disc, impose corrections to the assumption that dust emission is completely optically thin and in the Rayleigh-Jeans approximation for $\beta^{\rm abs}_{\rm B7-B3}$. Instead, the discrepancies between $\beta^{\rm abs}_{\rm B3-Ka}$ and $\alpha_{\rm B3-Ka}-2$ are mostly due to our fit preference for a fainter VLA Ka~band~emission profile than the median azimuthal average (see comparison in \autoref{fig:A10}). As a final remark, we stress that our $\beta^{\rm abs}_{\rm B7-B3}$ radial profile increases towards the outer disc. While this trend is qualitatively consistent with the results of \citet{Guilloteau2011}, who analysed together IRAM PdBI 1.3 and $2.7\,{\rm mm}$ observations of CI~Tau prescribing different functional forms for the absorption spectral index radial profile, it is, however, in contrast with the results of \citet{Tazzari2021}, who interpreted the almost constant dust disc sizes between 0.9 and $3.1\,{\rm mm}$ of a sample of 26 bright discs in Lupus as the consequence of a radially constant $\beta^{\rm abs}_{\rm B7-B3}$ radial profile. This difference suggests that, rather than a flat absorption spectral index, outer disc substructures might the reason behind the nearly wavelength-invariant dust disc size at ALMA wavelengths inferred by \citet[][see the case of CX~Tau, that shows no substructured down to a $0\farcs040\approx5\,{\rm au}$ resolution, \citealt{Facchini2019}, and whose dust sizes decline with wavelength as predicted by radial drift models, \citealt{Curone2023}]{Tazzari2021}.

\section{Corner plot comparisons}\label{sec:app10}
\begin{figure}
    \centering
    \includegraphics[width=\columnwidth]{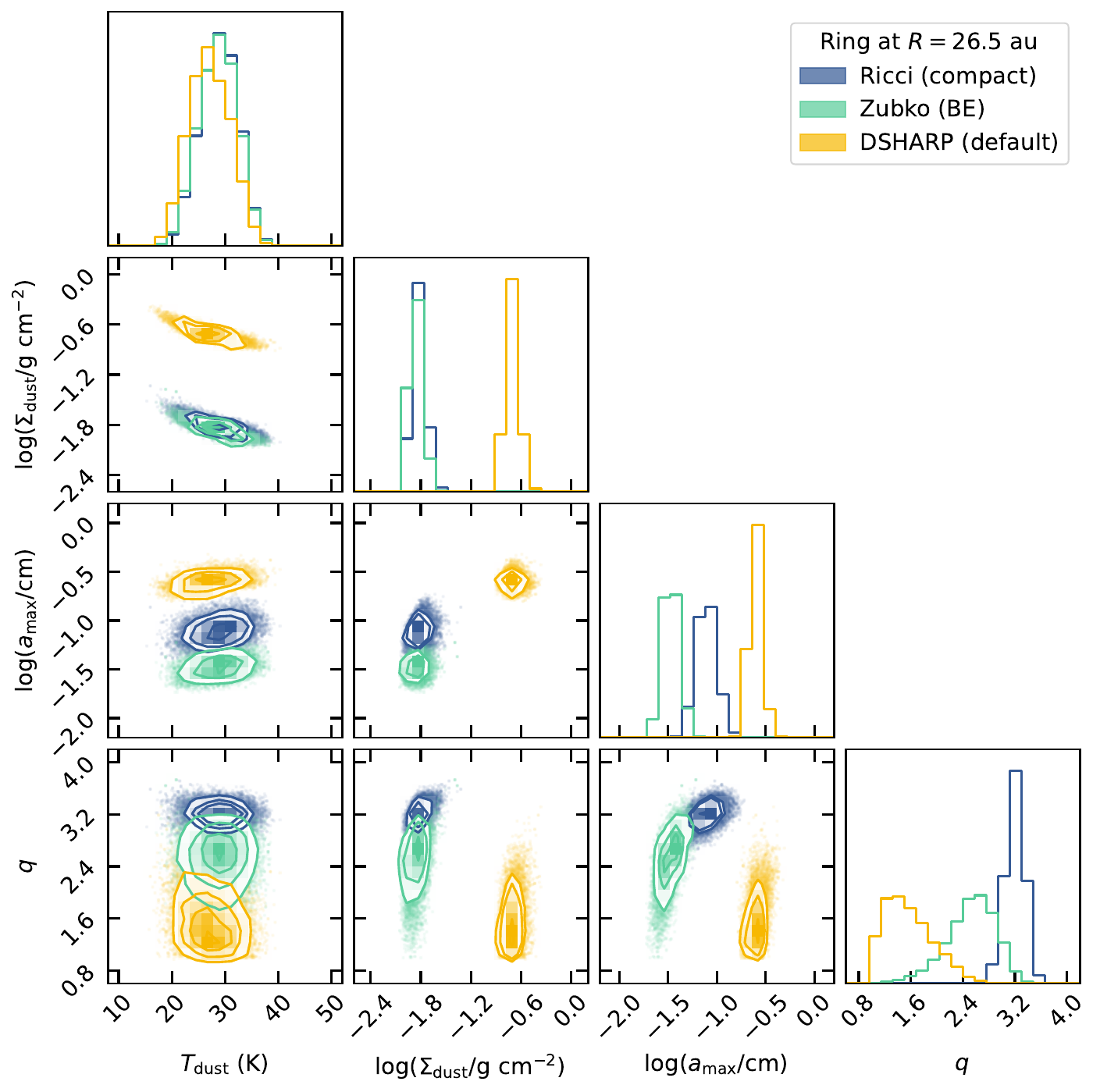} 
    \caption{Comparison of the MCMC posteriors for different assumptions on dust optical properties at the position of the inner ring ($R=26.5\,{\rm au}$).}
    \label{fig:A16}
\end{figure}

\begin{figure}
    \centering
    \includegraphics[width=\columnwidth]{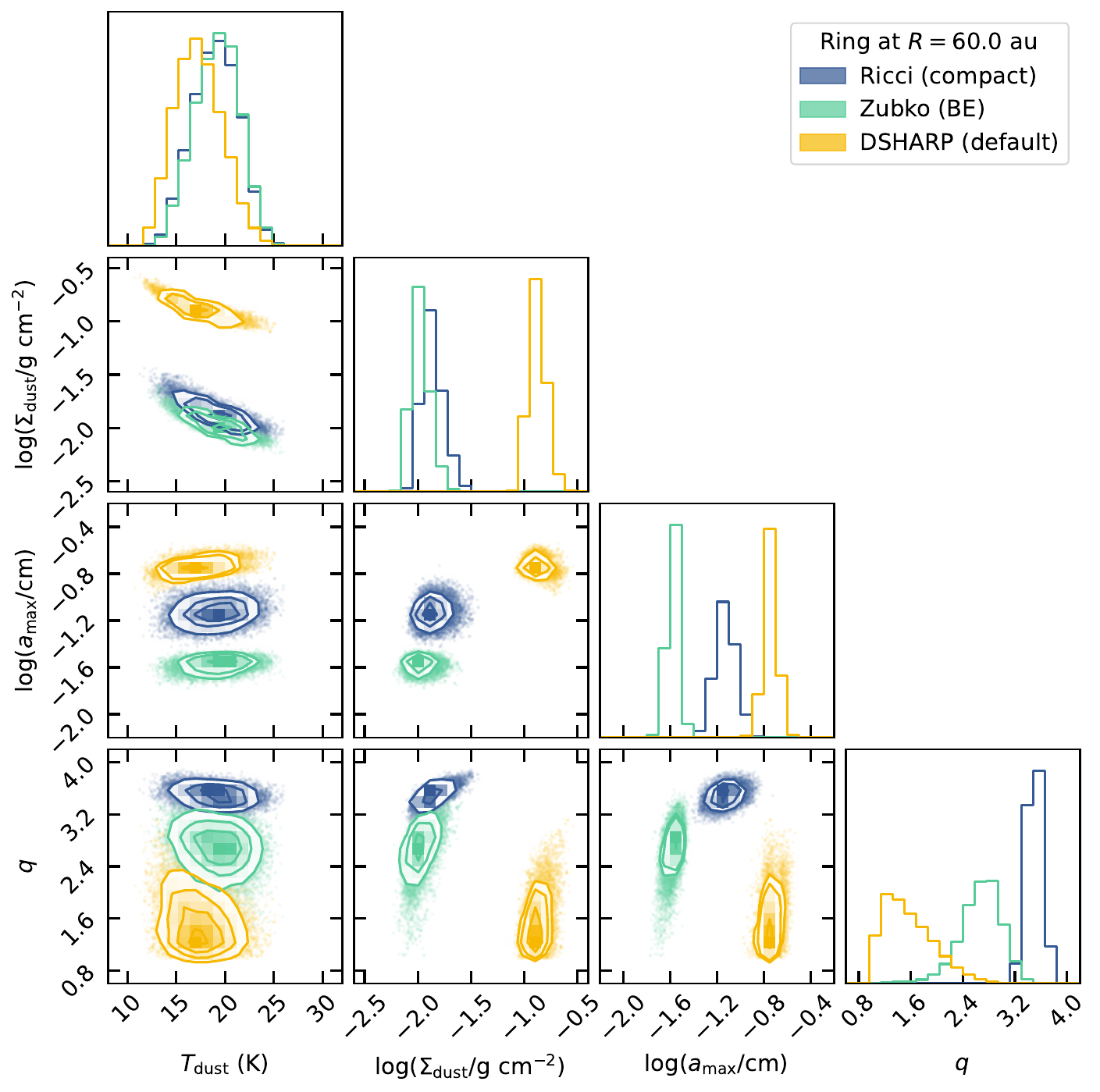} 
    \caption{Same as in \autoref{fig:A16} at the position of the central ring ($R=60.0\,{\rm au}$).}
    \label{fig:A17}
\end{figure}

\begin{figure}
    \centering
    \includegraphics[width=\columnwidth]{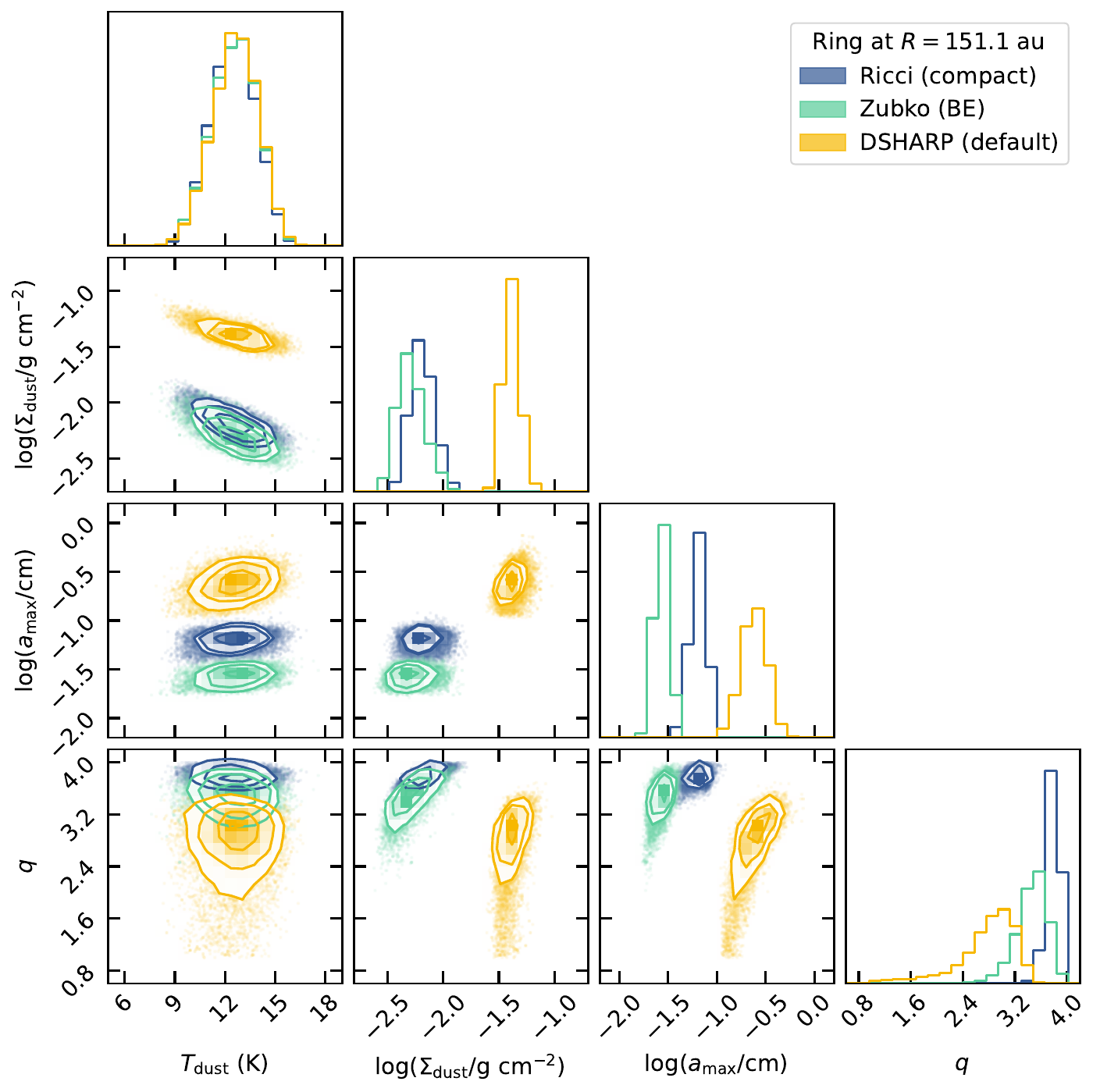} 
    \caption{Same as in \autoref{fig:A16} and \ref{fig:A17} at the position of the outer ring ($R=151.1\,{\rm au}$).}
    \label{fig:A18}
\end{figure}

\autoref{fig:A16}, \ref{fig:A17}, and~\ref{fig:A18} show a comparison of the posterior distributions of our high resolution fits for different assumptions on the dust optical properties: `Ricci (compact)' in blue, `Zubko (BE)' in turquoise, and `DSHARP (default)' in yellow.

\section{Tables}

\FloatBarrier

\begin{sidewaystable*}
    \caption{Summary of CI~Tau's stellar properties.}
    \centering
    \renewcommand{\arraystretch}{1.25}
    \begin{tabular}{lllllllllll}
     \hline
     \multirow{2}{*}{Name} & \multirow{2}{*}{2MASS name} & \multirow{2}{*}{Region} & $d$ & \multirow{2}{*}{SpT} & $T_{\rm eff}$ & $\log L_\star$ & $\log M_\star$ & $t_\star$ & $\log\dot M_{\rm acc}$ & \multirow{2}{*}{References} \\
     & & & (pc) & & (K) & ($L_\odot$) & ($M_\odot$) & (yr) & ($M_\odot\,{\rm yr}^{-1}$) & \\
     (1) & (2) & (3) & (4) & (5) & (6) & (7) & (8) & (9) & (10) & (11) \\
      \hline
      \hline
     CI~Tau & J04335200+2250301 & Taurus & $159.6\pm0.7$ & K3 & $4559\pm114$ & $0.02\pm0.20$ & $0.11\pm0.15$ & $>3\times10^6$ & $-7.28\pm0.40$ & 1, 2 \\
      \hline
    \end{tabular}
    \tablefoot{Column 1: Source name. Column 2: 2MASS designation. Column 3: Star-formation region. Column 4: \textit{Gaia} DR3 distance. Column 5: Spectral type. Column 6: Effective temperature. Column 7: Stellar luminosity. Column 8: Stellar mass. Column 9: Stellar age. Column 10: Mass accretion rate. Column 11: References. \\
    \textbf{References:} (1) \citet{Gaia_DR3} for the source distance; (2) \citet{Gangi2022} for all the other stellar parameters.}
    \label{tab:app1}
    \vspace{50pt}
    \caption{Summary of the new and archival datasets (re-)analysed in this paper.}
    \centering
    \renewcommand{\arraystretch}{1.25}
    \begin{tabular}{lllllllll}
     \hline
     \multirow{2}{*}{Band} & \multirow{2}{*}{Project code} & \multirow{2}{*}{PI} & Date & Time on source & Max. baseline & \multirow{2}{*}{Flux calibrator} & PWV & MRS \\
     & & & (yy/mm/dd) & (min) & (m) & & (mm) & (arcsec)\\
     (1) & (2) & (3) & (4) & (5) & (6) & (7) & (8) & (9) \\
      \hline
      \hline
     Band~7  & 2017.A.00014.S & Rosotti & 2017/12/11 & 76.810 & 3320.96 & J0510+1800 & 0.791 & 1.839 \\
      \hline
     \multirow{2}{*}{Band~6} & 2016.1.01370.S & Clarke  & 2017/09/23 -- 09/24 & 64.713 & 12145.22 & J0510+1800 & 0.622 & 0.700 \\
     & 2015.1.01207.S & Nomura  & 2016/08/27 & 9.072 & 1604.86 & J0510+1800 & 1.864 & 5.511 \\
      \hline
     \multirow{2}{*}{Band~3} & 2018.1.00900.S & Tazzari & 2019/06/27 -- 07/04 & 86.486 & 16196.31 & J0423--0120, J0510+1800 & 0.664 & 1.058 \\
     & 2018.1.00900.S & Tazzari & 2021/07/03 & 37.800 & 2386.15 & J0237+2848 & 0.371 & 7.971 \\
      \hline
     \multirow{2}{*}{Ka~band} & VLA/19A-440 & Tazzari & 2019/03/17 -- 2020/10/08 & 586.57 & 11128.05 & 3C147 & - & 5.3 \\
     & VLA/19A-440 & Tazzari & 2019/09/13 -- 10/11 & 516.67 & 36623.09 & 3C147 & - & 1.6 \\
      \hline
     Ku~band & VLA/20A-373 & Tazzari & 2020/02/19 -- 03/04 & 44.45 & 3387.48 & 3C147 & - & 97 \\
      \hline
     X~band  & VLA/19A-440 & Tazzari & 2019/08/03 & 75.70 & 36623.09 & 3C147 & - & 5.3 \\
      \hline
     C~band  & VLA/20A-373 & Tazzari & 2020/02/25 & 14.92 & 3387.48 & 3C147 & - & 240 \\
      \hline
    \end{tabular}
    \tablefoot{Column 1: ALMA/VLA band. Column 2: Project code. Column 3: Principal Investigator (PI). Column 4: Beginning of each execution block. Column 5: Integration time on the science target. Column 6: Maximum baseline length. Column 7: The quasars observed for bandpass and amplitude scale calibration. Column 8: Precipitable water vapour (PWV) level (not provide by the VLA for their data). Column 9: Maximum recoverable scale (MRS).}
    \label{tab:app2}
\end{sidewaystable*}

\begin{sidewaystable*}
    \caption{Imaging results.}
    \centering
    \renewcommand{\arraystretch}{1.25}
    \begin{tabular}{lllllllllll}
     \hline
     \multirow{2}{*}{Band} & Frequency & \multirow{2}{*}{Robust} & Beam size & Beam PA & RMS noise & Peak intensity & \multirow{2}{*}{Peak S/N} & Flux density & $R_{68}$ & $R_{95}$ \\
     & (GHz) & & ($''^2$) & (deg) & (${\rm mJy}\ {\rm beam}^{-1}$) & (${\rm mJy}\ {\rm beam}^{-1}$) & & (mJy) & ($''$) & ($''$) \\
     (1) & (2) & (3) & (4) & (5) & (6) & (7) & (8) & (9) & (10) & (11) \\
     \hline
     \hline
     \multirow{6}{*}{Band~7} & \multirow{6}{*}{338.228} & $-1.0$ & 
         $0.056\times0.086$ & $-33.20$ & $1.05\times10^{-1}$ & 12.55 & 119.48 & $410.16\pm0.82$ & $0.77_{-0.01}^{+0.01}$ & $1.22_{-0.01}^{+0.01}$ \\
     & &  $-0.5$ & $0.059\times0.088$ & $-33.09$ & $7.30\times10^{-2}$ & 13.02 & 178.27 & $413.80\pm0.60$ & $0.78_{-0.01}^{+0.01}$ & $1.24_{-0.01}^{+0.01}$ \\
     & &   0.0  & $0.065\times0.096$ & $-32.94$ & $5.15\times10^{-2}$ & 14.32& 278.00 & $417.51\pm0.82$ & $0.79_{-0.01}^{+0.01}$ & $1.26_{-0.01}^{+0.02}$ \\
     & &   0.5  & $0.077\times0.111$ & $-33.69$ & $4.14\times10^{-2}$ & 16.77 & 404.76 & $416.02\pm0.69$ & $0.79_{-0.01}^{+0.01}$ & $1.27_{-0.01}^{+0.01}$ \\
     & &   1.0  & $0.092\times0.133$ & $-34.67$ & $3.95\times10^{-2}$ & 19.94 & 505.03 & $418.27\pm0.91$ & $0.81_{-0.01}^{+0.01}$ & $1.28_{-0.01}^{+0.01}$ \\
     & &   1.5  & $0.100\times0.145$ & $-34.90$ & $4.12\times10^{-2}$ & 21.71 & 526.90 & $418.98\pm1.36$ & $0.81_{-0.01}^{+0.01}$ & $1.29_{-0.01}^{+0.01}$ \\
     \hline
     \multirow{6}{*}{Band~6} & \multirow{6}{*}{229.612} & $-1.0$ & 
         $0.021\times0.033$ & 7.95 & $2.89\times10^{-2}$ & 1.73 &  59.74 & $146.88\pm0.89$ & $0.71_{-0.04}^{+0.03}$ & $1.17_{-0.02}^{+0.01}$ \\
     & & $-0.5$ & $0.022\times0.034$ &  7.94 & $2.05\times10^{-2}$ & 1.80 &  88.09 & $146.90\pm0.57$ & $0.71_{-0.02}^{+0.02}$ & $1.18_{-0.01}^{+0.01}$ \\
     & &   0.0  & $0.025\times0.037$ &  7.51 & $1.45\times10^{-2}$ & 2.05 & 141.19 & $147.09\pm0.48$ & $0.73_{-0.02}^{+0.02}$ & $1.19_{-0.01}^{+0.01}$ \\
     & &   0.5  & $0.032\times0.044$ &  8.74 & $1.10\times10^{-2}$ & 2.66 & 241.03 & $147.30\pm0.47$ & $0.73_{-0.01}^{+0.01}$ & $1.21_{-0.01}^{+0.01}$ \\
     & &   1.0  & $0.043\times0.062$ & 29.99 & $9.79\times10^{-3}$ & 3.84 & 392.22 & $147.42\pm0.32$ & $0.73_{-0.01}^{+0.01}$ & $1.21_{-0.01}^{+0.01}$ \\
     & &   1.5  & $0.049\times0.070$ & 35.61 & $9.61\times10^{-3}$ & 4.44 & 462.12 & $147.66\pm0.33$ & $0.74_{-0.01}^{+0.01}$ & $1.22_{-0.01}^{+0.01}$ \\
     \hline
     \multirow{6}{*}{Band~3} & \multirow{6}{*}{97.504} & $-1.0$ &
       $0.034\times0.050$ & $-11.22$ & $1.44\times10^{-2}$ & 0.60 &  41.51 & $15.85\pm0.07$ & $0.57_{-0.06}^{+0.04}$ & $1.09_{-0.02}^{+0.01}$ \\
     & & $-0.5$ & $0.037\times0.054$ & $-10.25$ & $1.04\times10^{-2}$ & 0.66 &  63.36 & $16.20\pm0.07$ & $0.59_{-0.04}^{+0.04}$ & $1.10_{-0.01}^{+0.01}$ \\
     & &   0.0  & $0.044\times0.064$ &  $-6.22$ & $7.55\times10^{-3}$ & 0.79 & 104.85 & $16.42\pm0.08$ & $0.62_{-0.03}^{+0.04}$ & $1.15_{-0.02}^{+0.01}$ \\
     & &   0.5  & $0.055\times0.081$ &     3.21 & $6.26\times10^{-3}$ & 0.97 & 155.14 & $16.41\pm0.14$ & $0.64_{-0.03}^{+0.03}$ & $1.17_{-0.01}^{+0.01}$ \\
     & &   1.0  & $0.066\times0.137$ &    22.34 & $5.66\times10^{-3}$ & 1.25 & 220.42 & $16.35\pm0.20$ & $0.65_{-0.02}^{+0.02}$ & $1.18_{-0.02}^{+0.02}$ \\
     & &   1.5  & $0.077\times0.173$ &    27.84 & $5.50\times10^{-3}$ & 1.43 & 260.64 & $16.33\pm0.25$ & $0.66_{-0.01}^{+0.01}$ & $1.20_{-0.01}^{+0.02}$ \\
     \hline
     \multirow{6}{*}{Ka~band} & \multirow{6}{*}{33.0} & $-1.0$ & 
         $0.042\times0.045$ & $-66.48$ & $5.07\times10^{-3}$ & 0.18 &  34.86 & $0.32\pm0.06$ & $<0.04$ & $0.05_{-0.01}^{+0.01}$ \\
     & & $-0.5$ & $0.046\times0.053$ & $-47.58$ & $3.37\times10^{-3}$ & 0.18 & 52.86  & $0.32\pm0.03$ & $<0.05$ & $0.07_{-0.01}^{+0.01}$ \\
     & &   0.0  & $0.057\times0.068$ & $-45.25$ & $2.39\times10^{-3}$ & 0.18 & 76.05  & $0.33\pm0.02$ & $<0.06$ & $0.08_{-0.01}^{+0.01}$ \\
     & &   0.5  & $0.086\times0.100$ & $-49.15$ & $1.77\times10^{-3}$ & 0.19 & 105.83 & $0.31\pm0.02$ & $<0.09$ & $0.17_{-0.01}^{+0.01}$ \\
     & &   1.0  & $0.136\times0.150$ & 82.64    & $1.58\times10^{-3}$ & 0.19 & 120.29 & $0.31\pm0.02$ & $0.14_{-0.02}^{+0.02}$ & $0.40_{-0.02}^{+0.02}$ \\
     & &   1.5  & $0.156\times0.174$ & 75.85    & $1.56\times10^{-3}$ & 0.19 & 122.88 & $0.31\pm0.03$ & $0.17_{-0.02}^{+0.02}$ & $0.40_{-0.02}^{+0.02}$ \\
     \hline
     \multirow{2}{*}{Ku~band} & \multirow{2}{*}{15.0} & (\texttt{natural}) & $1.643\times1.832$ & 15.97 & $2.83\times10^{-3}$ & 0.16 & 57.62 & $0.16\pm0.01$ & -- & -- \\
     &  & (\texttt{uniform}) & $0.979\times1.005$ & 68.20 & $6.88\times10^{-3}$ & 0.13 & 19.42 & $0.15\pm0.01$ & -- & -- \\
     \hline
     X~band  & 10.0 & (\texttt{natural}) & $0.217\times0.277$ & 58.95 & $2.64\times10^{-3}$ & 0.03 & 11.83 & $0.03\pm0.01$ & -- & -- \\ 
     \hline
     C~band  &  6.0 & (\texttt{natural}) & $3.794\times4.215$ & 45.69 & $6.60\times10^{-3}$ & 0.04 &  5.54 & $0.03\pm0.01$ & -- & -- \\
     \hline
    \end{tabular}
    \tablefoot{Column 1: ALMA/VLA band. Column 2: Average frequency. Column 3: Robust parameter for \texttt{briggs} weighting. For the VLA Ku~band data both the \texttt{natural} and \texttt{uniform} weighting schemes were adopted, while for the X and C~band data only the \texttt{natural} one was considered (see \autoref{sec:app2}). Column 4 and 5: Synthesised beam axes $(\theta_{\rm min}\times\theta_{\rm maj})$ and position angle (PA). Column 6: RMS noise. Column 7: Peak intensity. Column 8: Peak S/N. Column 9: Disc-integrated continuum flux density. Column 10 and 11: 68\% and 95\% flux radius (from the CLEAN image radial profiles).}
    \label{tab:app3}
\end{sidewaystable*}

\begin{table*}[t!]
    \renewcommand{\arraystretch}{1.25}
    \centering
    \caption{\texttt{frank} fit hyper-parameters and disc sizes.}
    \begin{tabular}{lllllll}
    \hline
    Band    & $N$ & $R_{\rm max}$ & $\alpha$ & $\log(w_{\rm smooth}$) & $R_{68}$ ($''$) & $R_{95}$ ($''$)\\
    (1) & (2) & (3) & (4) & (5) & (6) & (7) \\
    \hline
    \hline
    Band~7  & 400 & 2.0 & 1.05 & $1$  & $0.81\pm0.01$ & $1.27\pm0.01$\\
    Band~6  & 500 & 2.0 & 1.50 & $2$  & $0.74\pm0.01$ & $1.23\pm0.01$\\
    Band~3  & 400 & 2.0 & 1.05 & $-3$ & $0.71\pm0.03$ & $1.28\pm0.02$\\
    Ka~band & 300 & 2.0 & 1.05 & $0$  & $0.51\pm0.05$ & $1.09\pm0.01$\\
    \hline
    \end{tabular}
    \tablefoot{Column 1: ALMA/VLA band. Column 2: Number of fitting points. Column 3: Outer disc radius. Column 4 and 5: $\alpha$ and $w_{\rm smooth}$ parameters. Column 6 and 7: 68\% and 95\% flux radius.}
    \label{tab:app4}
\end{table*}

\begin{sidewaystable*}
    \caption{Summary of the volume fractions and complex refractive indices adopted to generate the optical properties tested in \autoref{fig:6}.}
    \centering
    \renewcommand{\arraystretch}{1.25}
    \begin{tabular}{llllllll}
    \hline
    Name & \multicolumn{7}{c}{Component volume fractions} \\ \cline{2-8}
     & (Astro)silicates$^{(1)}$ & Water ice$^{(2)}$ & Troilite$^{(3)}$ & Refractory organics$^{(3)}$ & Amorphous carbon$^{(4)}$ & Pyrolised cellulose$^{(5)}$ & Graphene$^{(1)}$ \\
     & ($3.30\,{\rm g}\,{\rm cm}^{-3}$) & ($0.92\,{\rm g}\,{\rm cm}^{-3}$) & ($4.83\,{\rm g}\,{\rm cm}^{-3}$) & ($1.50\,{\rm g}\,{\rm cm}^{-3}$) & ($2.50\,{\rm g}\,{\rm cm}^{-3}$) & ($1.84\,{\rm g}\,{\rm cm}^{-3}$) & ($2.26\,{\rm g}\,{\rm cm}^{-3}$) \\
     (1) & (2) & (3) & (4) & (5) & (6) & (7) & (8) \\
    \hline
    \hline
    Ricci (compact)                & 0.10 & 0.60 & -- & -- & 0.30 (ACH2) & -- & -- \\
    DSHARP (default)               & 0.17 & 0.36 & 0.03 & 0.44 & -- & -- & -- \\
    DSHARP (highwater)             & 0.10 & 0.60 & 0.02 & 0.28 & -- & -- & -- \\
    DSHARP (icefree)               & 0.26 & 0.00 & 0.04 & 0.70 & -- & -- & -- \\
    Zubko (ACH2)                   & 0.17 & 0.36 & 0.03 & -- & 0.44 (ACH2) & -- & -- \\
    Zubko (BE)                     & 0.17 & 0.36 & 0.03 & -- & 0.44 (BE)      & -- & -- \\
    J\"{a}ger (400$\,^\circ$C)     & 0.17 & 0.36 & 0.03 & -- & -- & 0.44 (400$\,^\circ$C) & -- \\
    J\"{a}ger (600$\,^\circ$C)     & 0.17 & 0.36 & 0.03 & -- & -- & 0.44 (600$\,^\circ$C) & -- \\
    J\"{a}ger (800$\,^\circ$C)     & 0.17 & 0.36 & 0.03 & -- & -- & 0.44 (800$\,^\circ$C) & -- \\
    J\"{a}ger (1000$\,^\circ$C)    & 0.17 & 0.36 & 0.03 & -- & -- & 0.44 (1000$\,^\circ$C) & -- \\
    Draine  ($a=0.01,\parallel$)   & 0.17 & 0.36 & 0.03 & -- & -- & -- & 0.44 ($a=0.01,\parallel$) \\
    Draine ($a=0.01,\perp$)        & 0.17 & 0.36 & 0.03 & -- & -- & -- & 0.44 ($a=0.01,\perp$) \\
    Draine ($a=0.1,\parallel$)     & 0.17 & 0.36 & 0.03 & -- & -- & -- & 0.44 ($a=0.1,\parallel$) \\
    Draine ($a=0.1,\perp$)         & 0.17 & 0.36 & 0.03 & -- & -- & -- & 0.44 ($a=0.1,\perp$) \\
    \hline
    \end{tabular}
    \tablefoot{Slightly different bulk densities for (astro)silicates ($3.50\,{\rm g}\,{\rm cm}^{-3}$) and water ($1.00\,{\rm g}\,{\rm cm}^{-3}$) are used in \texttt{dsharp\_opac} \citep{Birnstiel2018} to generate the optical properties for the `Ricci (compact)' mixture. \citet{Zubko1996} do not report any bulk density for their amorphous carbonaceous grains; a density of $2.50\,{\rm g}\,{\rm cm}^{-3}$ is adopted in \texttt{dsharp\_opac} \citep{Birnstiel2018} following \citet{Ricci2010}. We note, however, that \citet{Woitke2016} and \citet{Dominik2021} use $1.80\,{\rm g}\,{\rm cm}^{-3}$ for the BE-sample. This difference has negligible effects on the optical properties computed with \texttt{dsharp\_opac}. Bulk dust mixture densities are: $1.70\,{\rm g}\,{\rm cm}^{-3}$ for `Ricci (compact)', $1.68\,{\rm g}\,{\rm cm}^{-3}$ for `DSHARP (default)', $2.11\,{\rm g}\,{\rm cm}^{-3}$ for `DSHARP (icefree)', $1.40\,{\rm g}\,{\rm cm}^{-3}$ for `DSHARP (highwater)', $2.12\,{\rm g}\,{\rm cm}^{-3}$ for `Zubko (ACH2)' and `Zubko (BE)', $1.83\,{\rm g}\,{\rm cm}^{-3}$ for `J\"{a}ger', $2.01\,{\rm g}\,{\rm cm}^{-3}$ for `Draine' opacities.}
    \tablebib{(1) \citet{Draine2003}; (2) \citet{Warren&Brandt2008}; (3) \citet{Henning&Stognienko1996}; (4) \citet{Zubko1996}; (5) \citet{Jager1998}.}
    \label{tab:app5}
\end{sidewaystable*}

\begin{sidewaystable*}
    \caption{Comparison between pebble masses measured in the optically thin approximation and from multi-frequency continuum observations for the best studied sources in the literature and CI~Tau.}
    \centering
    \renewcommand{\arraystretch}{1.25}
    \begin{tabular}{lllllllll}
    \hline
    \multirow{2}{*}{Name} & Flux density (1.3$\,$mm) & Distance & Pebble mass (OT) & Pebble mass (MF) & Outer radius & \multirow{2}{*}{Composition} & \multirow{2}{*}{Comments} & \multirow{2}{*}{References} \\
     & (mJy) & (pc) & ($M_{\rm Earth}$) & ($M_{\rm Earth}$) & (au) & & & \\
    (1) & (2) & (3) & (4) & (5) & (6) & (7) & (8) & (9) \\
    \hline
    \hline
    HL~Tau & 744.1 & 147 & 441.89 & 332.95 & 100 & Pollack & $q=3.5$ & 1, 2, 3 \\
    \hline
    \multirow{2}{*}{TW~Hya} & \multirow{2}{*}{575.4} & \multirow{2}{*}{60.1} & \multirow{2}{*}{57.19} & 250 to 330 & \multirow{2}{*}{70} & \multirow{2}{*}{DSHARP (default)} & Large grains & \multirow{2}{*}{4, 5, 4} \\
     & & & & 210 to 310 & & & Large grains, $q=3.5$ & \\
    \hline
    IM~Lup & 253   & 153.8 & 153.62 & $1211.92^{+662.56}_{-469.45}$ & 150 & DSHARP (default) & Large grains, $q=2.5$ & 7, 6, 8 \\
    \hline
    GM~Aur & 264.1 & 155.0 & 129.13 & $246.38^{+113.20}_{-43.28}$ & 175 & DSHARP (default) & -- & 9, 6, 8 \\
    \hline
    AS~209 & 288   & 120.4 & 107.16 & $249.71^{+239.72}_{-69.92}$ & 125 & DSHARP (default) & Large grains, $q=2.5$ & 7, 6, 8 \\
    \hline
    \multirow{2}{*}{HD~163296} & \multirow{2}{*}{715} & \multirow{2}{*}{103.7} & \multirow{2}{*}{197.46} & $276.35^{+449.48}_{-59.93}$ & \multirow{2}{*}{120} & DSHARP (default) & Large grains, $q=2.5$ & 7, 6, 8 \\
     & & & & 265 & & DIANA & -- & 7, 5, 10 \\
    \hline
    MWC~480 & 268  & 156.2 & 179.74 & $396.21^{+589.31}_{-96.55}$ & 100 & DSHARP (default) & Large grains, $q=2.5$ & 11, 5, 8 \\
    \hline
    \multirow{2}{*}{LkCa~15} & \multirow{2}{*}{418.1} & \multirow{2}{*}{154.8} & \multirow{2}{*}{292.05} & 253.12 & \multirow{2}{*}{120} & DSHARP (default) & \multirow{2}{*}{$q=3$} & \multirow{2}{*}{12, 6, 12} \\
     & & & & 9.99 & & Ricci (compact) & & \\
    \hline
    \multirow{3}{*}{CI~Tau} & \multirow{3}{*}{147.7} & \multirow{3}{*}{160.3} & \multirow{3}{*}{107.50} & $31.13^{+8.77}_{-6.00}$ & \multirow{3}{*}{200} & Ricci (compact) & -- & \multirow{3}{*}{13, 6, 13} \\
     & & & & $28.25^{+8.34}_{-5.58}$ & & Zubko (BE) & -- & \\
     & & & & $250.40^{+51.46}_{-37.73}$ & & DSHARP (default) & -- & \\
    \hline
    \end{tabular}
    \tablefoot{Column 1: Source name. Column 2: ALMA Band~6 flux density. Column 3: Source distance. Column 4 and 5: Pebble mass in the optically thin approximation (\autoref{eq:7.1}, OT) and the surface density profile from multi-frequency observations (MF). Column 6: Outermost disc radius the surface density was integrated over. Column 7 and 8: Adopted composition and fitting assumptions. DIANA opacities \citep[][labelled `DIANA']{Min2016,Woitke2016}, are made of 60\% amorphous silicates \citep{Dorschner1995} and 15\% amorphous carbonaceous materials \citep[][BE-sample]{Zubko1996}, with 25\% porosity by volume. The mixture was generated using the Bruggeman rule, with bulk density $\rho_{\rm s}=2.08\,{\rm g}\,{\rm cm}^{-3}$. \citet{Carrasco-Gonzalez2019} generated a set of opacities adopting the optical constant of \citet[][labelled `Pollack']{Pollack1994}. Their mixture is made of 26\% silicates, 31\% organics, and 43\% water ice by volume, with bulk density $\rho_{\rm s}=2.05\,{\rm g}\,{\rm cm}^{-3}$.}
    \tablebib{Flux density, distance, and pebble mass from: (1) \citet{ALMAPartnership2015}; (2) \citet{Galli2018}; (3) \citet{Carrasco-Gonzalez2019}; (4) \citet{Macias2021}; (5) \citet{Gaia_EDR3}; (6) \citet{Gaia_DR3}; (7) \citet{Andrews2018}; (8) \citet{Sierra2021}; (9) \citet{Huang2020}; (10) \citet{Guidi2022}; (11) \citet{Liu2019}; (12) \citet{Sierra2025}; (13) this paper.}
    \label{tab:app6}
\end{sidewaystable*}

\end{appendix}

\end{document}